\documentclass[unnumsec,webpdf,modern,small]{oup-authoring-template}

\graphicspath{{Fig/}}
\usepackage[english]{babel}
\usepackage{parskip}

\usepackage{amsmath,amssymb}
\usepackage{graphicx,subfigure}
\usepackage{rotating}
\usepackage{booktabs}  
\usepackage{multirow}   
\usepackage{caption,rotating}   
\usepackage{xcolor}
\usepackage{booktabs} 
\usepackage{array}
\usepackage{caption}
\usepackage{colortbl}  
\usepackage{xcolor}   
\usepackage{adjustbox}
\usepackage{siunitx}  
\usepackage{multirow}
\theoremstyle{thmstyleone}%
\theoremstyle{thmstyletwo}%
\theoremstyle{thmstylethree}%

\newcommand{\BPV}{\operatorname{BPV}}
\newcommand{\RV}{\operatorname{RV}}

\begin{document}

\journaltitle{Arxiv}
\DOI{DOI HERE}
\copyrightyear{2022}
\pubyear{2019}
\access{Advance Access Publication Date: Day Month Year}
\appnotes{Paper}

\firstpage{1}


\title[AI for Volatility Forecasting]{Foundation Time-Series AI Model for Realized Volatility Forecasting}

\author[1,$\ast$]{Anubha Goel}
\author[2]{Puneet Pasricha}
\author[3]{Martin Magris}
\author[1]{Juho Kanniainen}

\authormark{Anubha Goel et al.}

\address[1]{%
  \orgdiv{Research Group of Financial Computing and Data Analytics}, 
  \orgname{Tampere University}, 
  \orgaddress{%
    \street{Korkeakoulunkatu 7}, 
    \postcode{33720}, 
    \state{Tampere}, 
    \country{Finland}
  }
}
\address[2]{\orgdiv{Department of Mathematics}, \orgname{Indian Institute of Technology Ropar}, \orgaddress{\street{Rupnagar}, \postcode{140001}, \state{Punjab}, \country{India}}}
\address[3]{\orgdiv{Department of Statistics}, \orgname{Instituto Tecnológico Autónomo De México (ITAM)}, \orgaddress{\street{Río Hondo 1, Altavista, Álvaro Obregón}, \postcode{01080}, \state{Ciudad de México}, \country{Mexico}}}

\corresp[$\ast$]{Corresponding author. \href{email:anubha.goel@tuni.fi}{anubha.goel@tuni.fi}}

\received{Date}{0}{Year}
\revised{Date}{0}{Year}
\accepted{Date}{0}{Year}



\abstract{Time series foundation models (FMs) have emerged as a popular paradigm for zero-shot multi-domain forecasting. These models are trained on numerous diverse datasets and claim to be effective forecasters across multiple different time series domains, including financial data.  In this study, we evaluate the effectiveness of FMs, specifically the TimesFM model, for volatility forecasting, a core task in financial risk management. We first evaluate TimesFM in its pretrained (zero-shot) form, followed by our custom fine-tuning procedure based on incremental learning, and compare the resulting models against standard econometric benchmarks. While the pretrained model provides a reasonable baseline, our findings show that incremental fine-tuning, which allows the model to adapt to new financial return data over time, is essential for learning volatility patterns effectively. Fine-tuned variants not only improve forecast accuracy but also statistically outperform traditional models, as demonstrated through Diebold-Mariano and Giacomini-White tests. These results highlight the potential of foundation models as scalable and adaptive tools for financial forecasting—capable of delivering strong performance in dynamic market environments when paired with targeted fine-tuning strategies.}
\keywords{Volatility Forecasting, Realized Variance, Foundation Models, Time Series Analysis, Artificial Intelligence}


\maketitle
\section{Introduction}

Volatility forecasting is the cornerstone of financial modeling and plays a crucial role in risk management, derivative pricing, and investment strategies. The literature on volatility forecasting has grown substantially in recent decades, marked by the development of numerous approaches aimed at improving forecasting accuracy. Historically, econometric models have long been employed to predict market volatility. For instance, the widely recognized Generalized Autoregressive Conditional Heteroscedasticity (GARCH) family of models has been extensively used in modeling volatility and risk over the past several decades. These stochastic econometric time series models have the advantage of being grounded in rigorous statistical theory; however, their effectiveness is often constrained by rigid parametric structures and limited adaptability to evolving market conditions primarily due to the stationarity assumption of the explanatory variables.

Recently, ML approaches such as penalized linear models, tree-based models or neural networks, have been utilized to integrate predictor information, such as macroeconomic and financial variables, to enhance the informational depth of volatility forecasts. Several studies have shown that the prediction performances of Machine Learning (ML) models are significantly better than the benchmark econometric models. While ML models have the power to approximate nonlinear functions and fewer restrictions, they thrive in data-rich environments. That is, with so many parameters to learn, they require massive training data and are computationally costly to train. Recent studies have demonstrated the advantages of various ML models and techniques for financial forecasting \cite[see, for example][]{christensen2023machine,passalis2019deep,tran2019data,zhang2019deeplob}, but these approaches often struggle to adapt to evolving market dynamics without extensive retraining. In highly volatile and rapidly changing financial environments, the need for models that can seamlessly adjust to new market conditions is more pressing than ever. Incremental learning and transfer learning have emerged as promising strategies to refine pre-trained models using new financial data, enabling them to maintain predictive accuracy over time with minimal additional training.

This perspective aligns with our approach, where we leverage foundation models—a new class of deep learning models pre-trained on large-scale, diverse time series datasets—to enhance forecasting performance. These models offer several key advantages over traditional ML and econometric approaches. First, they excel in zero-shot settings, delivering strong predictive performance even on tasks and datasets they were not explicitly trained for. Second, they are highly adaptable across different forecasting tasks and domains. Finally, with minimal task-specific data, foundation models can be fine-tuned efficiently to achieve significant performance improvements, making them both flexible and resource-efficient. Despite these advancements, their application in financial time series, particularly for volatility prediction, remains relatively underexplored.

One of the most prominent examples of the emerging class of foundation time-series models is Google’s TimesFM \cite{das2024decoder}. Developed using a decoder-only transformer architecture and pre-trained on a vast corpus of real-world time series data, TimesFM has demonstrated competitive, and often state-of-the-art, results across diverse forecasting tasks. Benchmark evaluations, including those presented in the GIFT-Eval framework\footnote{\url{https://huggingface.co/spaces/Salesforce/GIFT-Eval}}, consistently position TimesFM as one of the top-performing models, making it a strong candidate for complex forecasting applications, including financial volatility prediction.

TimesFM is a pre-trained model designed to generate point forecasts based on a specified historical context length and prediction horizon. Unlike traditional statistical methods, which require extensive feature engineering and manual tuning, TimesFM automatically captures temporal dependencies in financial data. Recently, \cite{goel2024time} applied time-series foundation models to Value-at-Risk (VaR) forecasting, demonstrating that their approach achieves performance comparable to or better than the best econometric models. Their results underscore the potential of pre-trained architectures to rival established financial forecasting methods, further motivating the exploration of foundation models in volatility prediction. While some models achieve competitive results, their effectiveness can vary significantly depending on the domain, highlighting the need for continued research in adapting and fine-tuning these models for diverse time series applications.

Our paper makes several contributions. While previous research (\cite{bucci2020realized,ge2022neural,christensen2023machine}) has demonstrated the effectiveness of ML models, the use of time series foundation models for volatility forecasting remains underexplored. We bridge this gap and establish a link between two significant and growing areas, time series foundation models in ML and volatility forecasting in financial economics. Thus, by leveraging a pre-trained, adaptable model for volatility forecasting, this study contributes to the growing literature on data-driven financial modeling. 

Second, as mentioned earlier, the foundation models serve as a versatile starting point, and could be fine-tuned on task specific datasets, volatility forecasting in our case. To align TimesFM with our forecasting objective, we implement a fine-tuning procedure using incremental learning, where the model is progressively updated with the most recent data. Specifically,
we employ transfer learning to iteratively fine-tune the model on new data, ensuring that it adapts dynamically to evolving market conditions. Specifically, the pre-trained model is first fine-tuned on an initial subset of data and then iteratively refined on new observations. This method ensures that the model
remains adaptive without overfitting to short-term noise. The fine-tuning process is optimized using gradient clipping, a cosine learning rate schedule, and an exponential moving average decay, providing stability and efficiency in training. We configure the model to predict one-day-ahead realized variance, experimenting with varying historical context lengths (64, 128, and 512 points) to assess its adaptability to different market conditions.

Third, our empirical analysis confirms the insights from the growing literature that ML methods excel in predicting volatility forecasting. In particular, our findings underscore the potential of foundation models in enhancing the accuracy and flexibility of volatility predictions, paving the way for more robust risk assessment and portfolio management strategies. Concretely, we conduct an extensive empirical analysis, investigating 21 major global stock market indices over a period that encompasses several significant events over the period January 2000 to December 2021.\footnote{Refer Section on dataset description for more details.} We also benchmark the performance of TimesFM against standard econometric models (HAR, ARFIMA, RGARCH) and assess the impact of incremental learning on forecast accuracy. Our results demonstrate that TimesFM consistently outperforms traditional volatility models, offering improved predictive accuracy while maintaining computational efficiency. Moreover, the ability to fine-tune and incrementally update the model highlights the advantages of foundation models in financial forecasting applications. In sum, our paper confirms the great potential of time-series foundation models for forecasting volatility, thus, bridging the gap between econometric volatility forecasting and modern time series modeling techniques, contributing to both financial research and practical forecasting applications.

The paper is organized as follows. Section 2 gives the literature review on volatility forecasting. Section 3 discusses the preliminary concepts, the benchmark models and the data used, while Section 4 dicusses the evaluation metrics. Section 5 contains our empirical analysis. Section 6 concludes. 

\section{Literature Review Realized Volatility Forecasting}
The accurate forecasting of financial market volatility has been a longstanding challenge in financial econometrics. The origins of this literature are commonly attributed to the work of \cite{engle1982autoregressive,bollerslev1986generalized,taylor1982financial} 
who introduced discrete-time GARCH and stochastic volatility models to better characterize autoregressive conditional heteroskedasticity in time series data. \cite{corsi2009har} argued that standard volatility models are incapable of reproducing the stylized facts of financial markets and introduced the heterogeneous autoregressive (HAR) model. This model integrates non-parametric RV computed across multiple frequencies with a parametric autoregressive framework in order to emulate the long-memory feature of volatility and capture the impact of varied market participant types. After the success of the HAR model, a lot of variations and extensions of the baseline HAR were proposed to overcome the limitations that arise due to its parsimonious nature, making it inadequate in some directions. Notable extensions of the HAR model include HAR-J (HAR with jumps) and CHAR (Continuous HAR) as introduced by \cite{andersen2007roughing,corsi2009har}; SHAR (Semivariance-HAR), proposed by \cite{patton2015good}, which distinguishes between the effects of positive and negative returns on subsequent realized volatility; and the HARQ model by \cite{bollerslev2016exploiting}, which refines volatility forecasts by accounting for measurement error using realized quarticity.


Despite being successful, these models rely on strong parametric assumptions and may struggle with capturing nonlinearities and regime shifts in financial markets. To address these limitations, recent research has explored the application of ML and time series models to improve volatility forecasting accuracy, primarily due to their ability to model complex, nonlinear dependencies. Several studies have demonstrated the effectiveness of ML models in predicting realized volatility. For instance, \cite{audrino2016lassoing,audrino2020impact,caporin2017building} utilized LASSO, \cite{luong2018forecasting} used random forests (RFs) and  \cite{mittnik2015stock} employed component-wise gradient boosting. Further, artificial neural networks are found to be particularly useful in this context. \cite{christensen2023machine} investigated the application of various ML techniques, including regression trees and neural networks, and found that these models consistently outperformed HAR models, particularly for long-term volatility predictions. Additionally, \cite{rahimikia2020machine} explored the predictive power of ML models in forecasting realized volatility using limit order book (LOB) data and news sentiment analysis. Their findings demonstrated that ML models outperformed HAR models in 90\% of out-of-sample forecasts, particularly when market volatility was not extreme. \cite{bucci2020realized} evaluated feedforward and recurrent neural networks (RNNs) against classical models such as HAR, finding that ML models consistently achieved superior forecast accuracy, particularly during periods of heightened volatility. Adopting a different approach, \cite{liu2023trading} examined the role of trading volume in volatility forecasting, showing that decomposing trading activity into short- and long-run components improves the predictive accuracy of HAR-based models. Long Short-Term Memory (LSTM) networks have been shown to outperform traditional models by capturing long-range dependencies in financial time series \cite{liu2019novel}. Similarly, Neural Basis Expansion Analysis with Exogenous Variables (NBEATSx) has been applied to realized volatility forecasting, demonstrating superior accuracy over traditional statistical methods \cite{souto2024introducing}.

Neural networks have also been used in integration with GARCH models. For instance, an early contribution by \cite{hamid2004using} assessed the forecasting performance of neural networks utilizing implied and realized volatility as inputs and found that neural networks outperformed. \cite{hu1999combining} combined forecasts from four conditional volatility models within a neural network architecture and demonstrated that ANNs produced accurate predictions, especially during crisis periods. Building on this, \cite{arneric2014garch} adopted a Jordan neural network based on squared innovations from a GARCH model and observed that neural network based models delivered superior prediction accuracy when compared to alternative linear and non-linear models. Other articles in this direction include \cite{kristjanpoller2014volatility} and \cite{fernandes2014modeling}, the former applied neural networks to forecast monthly realized volatility for Latin American stock market indices, the latter proposed a neural network-based HAR model that incorporates exogenous variables, thus improving the forecasting of implied volatility.

\section{Preliminaries}

\subsection{Realized Volatility}

Consider a filtered probability space, $(\Omega,(\mathcal{F}_t)_{t\geq 0},\mathcal{F},\mathbb{P})$ and
assume that $P_t$ denotes the price process of a financial asset. Further, let the log-price $\log(P_t)$ can be represented as follows,
\begin{equation*}
\log(P_t) = \log(P_0) + \int_0^t \mu_s \, \mathrm{d}s + \int_0^t \sigma_s \, \mathrm{d}W_s + \sum_{s=1}^{N_t} J_s, \quad t \geq 0,
\end{equation*}
where \( P_0 \) is \( \mathcal{F}_0 \)-measurable, \( \mu = (\mu_t)_{t \geq 0} \) denotes the drift term (a continuous function), \( \sigma = (\sigma_t)_{t \geq 0} \) is the volatility process stochastic (a c\`adl\`ag
function), and $(W_t)_{t\geq0}$ is a standard Brownian motion. In addition, let $(J_k)_{k\geq1}$ be the sequence of jump sizes, where $N=(N_t)_{t\geq0}$ is the counting process that records the number of jumps in the log‑price up to time $t$.

The quadratic variation over the interval $[t-h,t]$, is defined as
\[
\text{QV}_{t,h} = \int_{t-h}^{t} \sigma_s^2 \, \mathrm{d}s + \sum_{k=N_{t-h}+1}^{N_t} J_k^2,
\]
where $h>0$ is the look-back horizon (e.g., 30 minutes or one day). Because the quadratic variation is not directly observable, in practice, the realized variance is commonly used as its empirical counterpart. 

Let \(h>0\) be a fixed horizon and fix an integer \(M\ge1\).
Define the regular grid
\[
t_i := t-h+i\delta, \qquad
\delta := h/M, \quad i=0,\dots,M,
\]
and the corresponding log‑returns
\[
r_i := \Delta_i\log P
     := \log P_{t_i} - \log P_{t_{i-1}}, \qquad i=1,\dots,M.
\]
The realized variance over \([t-h,t]\) based on \(M\) observations is
\[
\RV^{(M)}_{t,h}
    \;=\;
    \sum_{i=1}^{M} r_i^{\,2}
    \;=\;
    \sum_{i=1}^{M} \bigl(\Delta_i \log P\bigr)^2.
\]
As the sampling interval shrinks (\(\delta\to0\), equivalently \(M\to\infty\)),
\(\RV^{(M)}_{t,h}\) converges in probability to the quadratic variation \(\operatorname{QV}_{t,h}\) \cite{barndorff2002estimating}.
Building on this result, our study focuses on forecasting realised volatility to improve predictive modelling in financial markets. 
Throughout this paper, we set \(h = 1\) day and \(\delta = 5\) minutes; hence, \(M\) is interpreted as the number of intraday returns (equal to the number of hours the market is open, multiplied by 12).
Consequently, we abbreviate
\(\RV^{(M)}_{t,1}\) and \(\operatorname{QV}_{t,1}\) by \(\RV_t\) and \(\operatorname{QV}_t\), respectively.

\subsection{Rolling Window Forecasting}

For forecasting realized volatility, we adopt the rolling window scheme, a widely used framework in time series prediction (\cite{nie2022time}). The underlying idea of the rolling window approach is to iteratively shift the training data forward in time to include more recent data and exclude the oldest data points so as to maintain a fixed window size. More specifically, consider a time series represented as a sequence of fixed-dimensional vectors, $\mathbf{x}_t$, indexed by discrete time steps $t$. In each iteration, the model is trained using the most recent $L$ observations, $\mathbf{x}_{t-L+1}, \dots, \mathbf{x}_t$, where $L$ denotes the length of the historical context. The trained model is then used to generate forecasts for the subsequent $H$ time steps, where $H$ denotes the forecast horizon. In the next iteration, the window is shifted forward by $H$ steps, and the process is repeated: the model is retrained on the new $L$-length context, $\mathbf{x}_{t+H-L+1}, \dots, \mathbf{x}_{t+H}$, and used to predict the next $H$ values. This procedure continues until the end of the dataset is reached. This yields a series of predictions, which are then compared with the actual values to assess the predictive accuracy of the model. Throughout this study, we set $H=1$.

\subsection{Dataset Description}

This study employs high-frequency realized volatility data from the Oxford-Man Institute’s Realized Library for 21 major global stock market indices. The indices included in the study are AEX, AORD, BFX, BVSP, DJI, FCHI, FTSE, GDAXI, HSI, IBEX, IXIC, KS11, KSE, MXX, N225, RUT, SPX, SSEC, SSMI, STI, and STOXX50E. The dataset spans from January 2000 to December 2021 for most indices, with the exception of the Nikkei 225 (N225), which begins in February 2000 (see Table \ref{tab:data_summary}). This coverage ensures a comprehensive analysis of market fluctuations over more than two decades across major global financial indices. The sample encompasses several significant events, including the COVID-19 Pandemic, the global financial crisis, the European sovereign debt crisis, multiple negotiations over the U.S. debt ceiling limit, and the flash crashes that occurred on May 6, 2010, and August 24, 2015. The dataset provides daily observations, including key realized measures of market volatility. Specifically, it includes measures such as open price, close price, 5-minute sub-sampled realized variance and bipower variation. This dataset offers a comprehensive foundation for assessing the predictive performance of econometric models and TimesFM, enabling a detailed evaluation of realized volatility forecasting across a diverse set of global financial markets.

  \begin{table}[htbp]
    \centering
    \caption{Start and end dates (yyyy-mm-dd) of the Indices and the total number of observations}
    \label{tab:data_summary}
    \resizebox{.7\textwidth}{!}{%
    \begin{tabular}{lllrr}
        \toprule
        \textbf{Index Name} & \textbf{Symbol} & \textbf{Start Date} & \textbf{End Date} & \textbf{Number of Observations} \\ 
        \midrule
        AEX Index & .AEX & 2000-01-03 & 2021-12-31 & 5735 \\ 
        All Ordinaries & .AORD & 2000-01-04 & 2021-12-31 & 5676 \\ 
        Bell 20 Index & .BFX & 2000-01-03 & 2021-12-31 & 5733 \\ 
        BVSP BOVESPA Index & .BVSP & 2000-01-03 & 2021-12-31 & 5531 \\ 
        Dow Jones Industrial Average & .DJI & 2000-01-03 & 2021-12-31 & 5630 \\ 
        CAC 40 & .FCHI & 2000-01-03 & 2021-12-31 & 5737 \\ 
        FTSE 100 & .FTSE & 2000-01-04 & 2021-12-31 & 5668 \\ 
        DAX & .GDAXI & 2000-01-03 & 2021-12-31 & 5698 \\  
        HANG SENG Index & .HSI & 2000-01-03 & 2021-12-31 & 5505 \\ 
        IBEX 35 Index & .IBEX & 2000-01-03 & 2021-12-31 & 5700 \\ 
        Nasdaq 100 & .IXIC & 2000-01-03 & 2021-12-31 & 5637 \\ 
        Korea Composite Stock Price Index (KOSPI) & .KS11 & 2000-01-04 & 2021-12-31 & 5532 \\ 
        Karachi SE 100 Index & .KSE & 2000-01-03 & 2021-12-31 & 5478 \\ 
        IPC Mexico & .MXX & 2000-01-03 & 2021-12-31 & 5637 \\ 
        Nikkei 225 & .N225 & 2000-02-02 & 2021-12-31 & 5460 \\ 
        Russell 2000 & .RUT & 2000-01-03 & 2021-12-31 & 5634 \\ 
        S\&P 500 Index & .SPX & 2000-01-03 & 2021-12-31 & 5635 \\ 
        Shanghai Composite Index & .SSEC & 2000-01-04 & 2021-12-31 & 5427 \\ 
        Swiss Stock Market Index & .SSMI & 2000-01-04 & 2021-12-31 & 5635 \\ 
        Straits Times Index & .STI & 2000-01-03 & 2021-12-31 & 3691 \\ 
        EURO STOXX 50 & .STOXX50E & 2000-01-03 & 2021-12-31 & 5721 \\ 
        \bottomrule
    \end{tabular}}
\end{table}

\section{Time Series Foundation models}

In recent years, the success of large language models (LLMs) in fields like natural language processing and generative AI has inspired new approaches to time series forecasting. Models such as Generative Pre-trained Transformers (GPTs) have demonstrated impressive capabilities in tasks like text generation and question answering, often without requiring task-specific fine-tuning. This flexibility comes from their decoder-only architecture, which processes sequential data directly and learns generalizable patterns from large and diverse datasets \cite{fu2016using, yenduri2024gpt}.
Recognizing the parallels between language modeling and time series forecasting - both of which deal with sequential, temporally dependent data - researchers have begun applying similar architectural ideas to improve forecast accuracy and model flexibility. Traditional forecasting models, including popular ML techniques and encoder-decoder deep learning architectures, often require extensive data and frequent fine-tuning to adapt to new tasks \cite{pourpanah2022review}. In contrast, decoder-only models offer a simpler and more adaptable alternative, capable of generating accurate forecasts directly from historical data without costly retraining.

One of the most significant advancements in this area is Google’s TimesFM \cite{das2024decoder}, a foundation model specifically designed for time series forecasting. Inspired by the success of LLMs like GPT, TimesFM is built on a decoder-only transformer architecture and pre-trained on a vast and diverse corpus of over 6 billion time series data points. This dataset includes real-world sources such as Google Trends, Wikipedia Pageviews, and financial benchmarks like M4 \cite{makridakis2020m4} and ETT \cite{zhou2021informer}. By combining 80\% real-world data with 20\% synthetic data, TimesFM learns to generalize across a wide range of domains, temporal granularities, and patterns, making it highly effective even in zero-shot scenarios where no task-specific fine-tuning is performed.\footnote{In recent years, numerous zero-shot FMs have been introduced for time series forecasting. These models generally follow a common design paradigm: (a) they employ transformer-based architectures similar to large language models (LLMs), and (b) they undergo extensive pretraining on vast datasets, often comprising billions of time points \cite{rasul2023lag, chen2024visionts, liang2024foundation}. Examples of such models include Chronos \cite{ansari2024chronos}, Moirai \cite{woo2024unified}, and TimesFM \cite{das2024decoder}, all of which align with this conventional FM framework. However, an emerging class of FMs deviates from this standard by adopting alternative architectures or pretraining strategies. Some models, such as TTM \cite{ekambaram2024tiny}, VisionTS \cite{chen2024visionts}, and Mamba4Cast \cite{bhethanabhotla2024mamba4cast}, eschew transformer architectures in favor of other neural network structures. Notably, Mamba4Cast is exclusively trained on synthetic time series data, whereas VisionTS employs an ImageNet-pretrained masked autoencoder \cite{he2022masked} instead of training on time series data. Beyond these developments, newer models, such as TimeDiT \cite{cao2024timedit}, introduce further innovations. TimeDiT employs a diffusion transformer instead of a standard transformer, demonstrating potential advantages in handling multivariate time series (i.e., capable of handling covariates) across multiple forecasting, imputation, and anomaly detection tasks. 
Despite these advancements, the performance of zero-shot FMs in real-world forecasting remains an active area of study. }

Benchmark evaluations, including those presented in the GIFT-Eval framework, consistently position TimesFM as one of the top-performing foundation models across diverse forecasting tasks. In particular, it has achieved state-of-the-art performance, outperforming leading supervised models such as DeepAR \cite{salinas2020deepar} and N-BEATS \cite{oreshkin2019n} by up to 25\% in scaled MAE on the Monash archive dataset \cite{godahewa2021monash}. Importantly, TimesFM delivers these results while maintaining a relatively modest parameter size, ensuring both accuracy and computational efficiency. Given these strengths, foundation models like TimesFM offer a promising new direction for financial forecasting applications, including volatility prediction. This naturally leads to the broader discussion of recent advancements in zero-shot foundation models for time series forecasting.


\subsection{TimesFM}

TimesFM is originally trained on a large number of public and synthetic datasets and exhibits robust out-of-the-box zero-shot performance in comparison to the accuracy of various state-of-the-art forecasting models specific to individual datasets under consideration. At inference time, the model has enough flexibility to forecast with different input lengths, prediction lengths, and time granularities.
\vspace{-1.5cm}


\begin{figure}[ht!]
    \centering
    \includegraphics[width=0.7\linewidth,angle=-90]{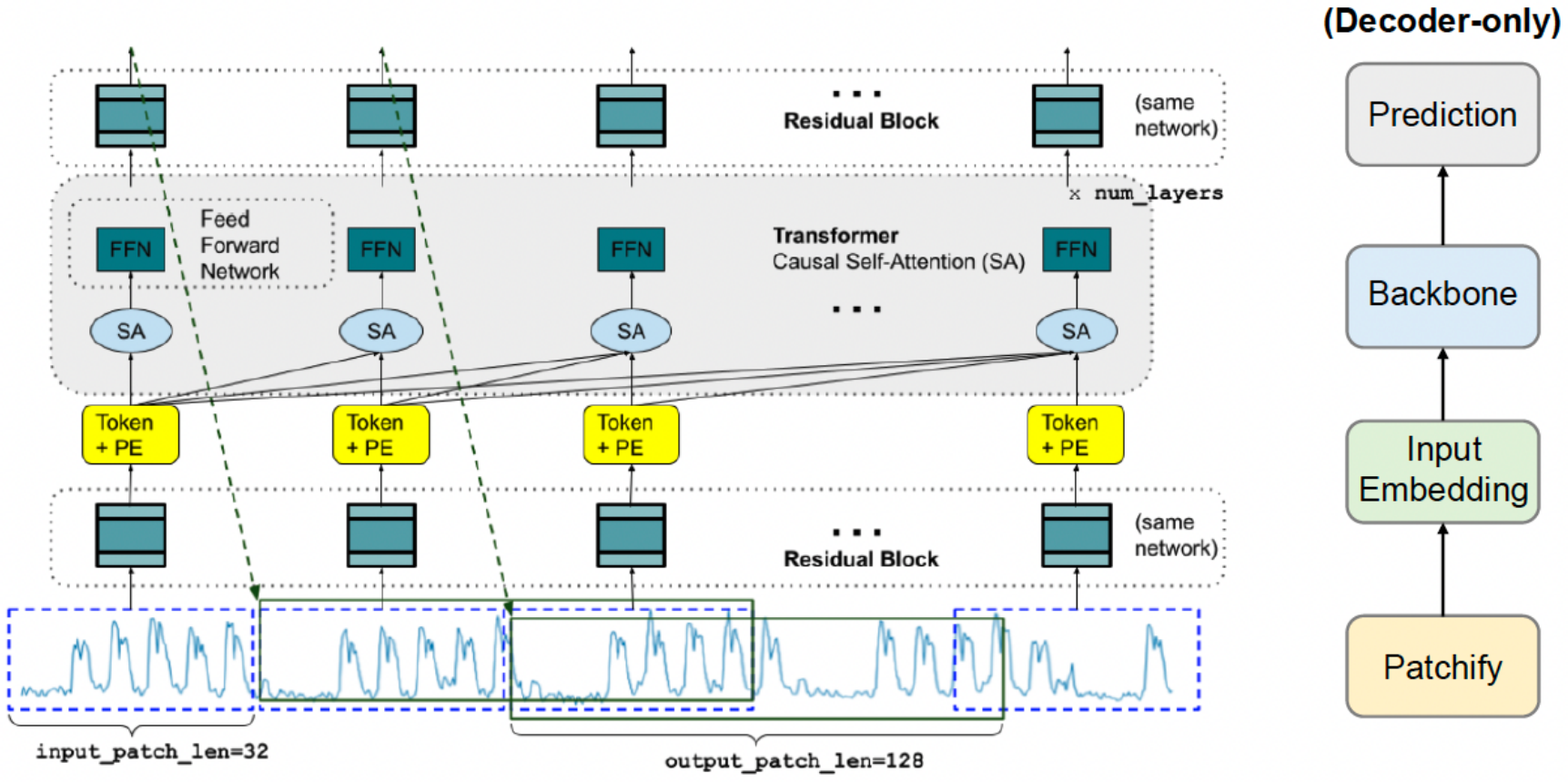}
    \caption{TimesFM structure Note. From \cite{das2024decoder}. A Decoder-Only Foundation Model for Time-Series Forecasting. Proceedings of the Forty-First International Conference on Machine Learning (ICML). Copyright 2024 by the Authors. The figure presents the architecture of TimesFM, a time-series foundation model built on a decoder-only transformer framework. }
    \label{fig:enter-label}
\end{figure}

We utilized the latest TimesFM checkpoint, released on December 30, 2024, which supports context lengths of up to 2048 time points—four times longer than its predecessor. Additionally, it accommodates any horizon length, offering greater flexibility for time-series forecasting. The developers of TimesFM report that this new version improves performance by up to 20\% compared to TimesFM v1.0. Given these advancements, we employed TimesFM v2.0 for predicting realized variance. Furthermore, TimesFM v2.0 has been integrated into GIFT-Eval, one of the most comprehensive time-series benchmarking frameworks. It now holds the top position in terms of aggregated Mean Absolute Scaled Error and Continuous Ranked Probability Score, further validating its superior predictive capabilities.

The pre-trained model generates point forecasts based on a specified context length and prediction horizon. In our setup, we set the prediction horizon to one day while varying the look-back period (i.e., context length) across three values: 64, 128, and 512 points. The input window shifts daily in alignment with the prediction points.

\subsection{Incremental Fine-Tuning}
To better align the model with our objective, we conducted a \textbf{incremental fine-tuning procedure} using the realized variance for all the stocks separately. We fine-tuned the model using incremental learning. We adopted the code ``fine-tuning" in the TimesFM library. More specifically, fine-tuning was carried out using a PatchedDecoderFinetuneModel with specific adjustments tailored for our task. During fine-tuning, we applied linear probing where only the core layer parameters were updated, but the transformer layers were kept fixed. To prevent gradient explosion, we applied gradient clipping with a threshold of 100. The optimization process employed the Adam optimizer, following a cosine schedule for the learning rate that started at $1\times 10^{-3}$ and decayed to $1\times 10^{-4}$ over 40,000 steps. Further, an exponential moving average decay of 0.9999 was also included in the optimization. The training loop ran for up to 100 epochs, with a provision for early stopping tied to the loss on the validation set. The patience level was set to 5 epochs without improvement. 

\begin{figure}
    \centering
    \includegraphics[width=.8\linewidth]{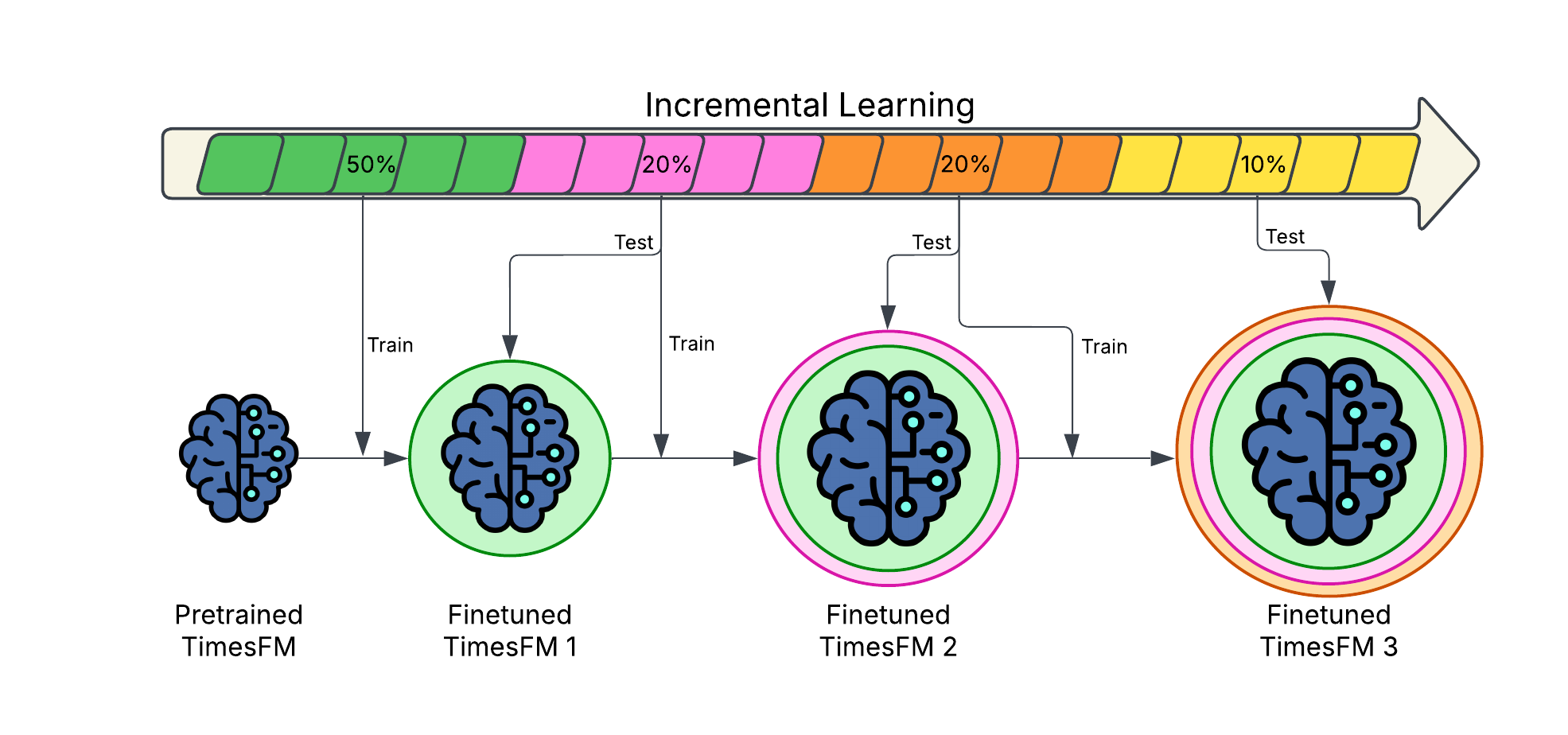}
    \caption{Fine-tuning procedure with incremental learning. An iterative fine-tuning process where a pre-trained model is progressively refined using sequential data splits. Initially, the model is fine-tuned on 50\% of the data and tested on the next 20\%. In each iteration, the previous test data becomes part of the training set, and the model is tested on the next segment. This process continues until the entire dataset is utilized, resulting in three rounds of fine-tuning.}
    \label{il}
\end{figure}

Throughout the fine-tuning procedure, checkpoints were saved, and finally, the best model was restored for evaluation purposes. To fine-tune the model in an incremental way using the most recent data available, we used transfer learning, where we train the previously fine-tuned model on the new data to obtain a new checkpoint. The process begins by fine-tuning the pre-trained model on 50\% of the data, with 40\% and 10\% of the data respectively for training and validation, and testing it on the subsequent 20\% of the data. In the next iteration, the previously fine-tuned model is further fine-tuned using the 20\% of the data from the previous test phase, with 16\% used as the training set and 4\% as the validation set. The updated model is then tested on the next 20\% of the data. This process is repeated iteratively, with each iteration using the fine-tuned model from the previous step, until the entire dataset is exhausted. In this way, we fine-tune the model three times (See Figure \ref{il}). For fine-tuning, we maintained a prediction horizon of one day and adopted the same three context lengths 64, 128, and 512 points, consistent with the pre-trained model\footnote{We also evaluated the performance of TimeGPT on the same task; however, its results did not surpass those of the TimesFM model. Furthermore, as TimeGPT is a proprietary model with restricted access to fine-tuning, we were unable to adapt it to the specific requirements of our study. This limitation underscores the advantage of open-source and fine-tunable models in domain-specific applications.}. The resulting models are denoted as {TFM64\(_{\text{PT}}\)}, {TFM128\(_{\text{PT}}\)}, and {TFM512\(_{\text{PT}}\)} for the pre-trained versions, while for fine-tuned models, we replace {PT} with {\text{IL}} representing incremental learning, resulting in {TFM64\(_{\text{IL}}\)}, {TFM128\(_{\text{IL}}\)}, and {TFM512\(_{\text{IL}}\)}. However, when forecasting log realized variance, we denote it using a superscript as \((\cdot)^{\log}\).

\section{Econometric Models for Comparison}

To compare the TimesFM model with standard econometric models, we consider the following models as the benchmarks: the Realized Generalized Autoregressive Conditional Heteroskedasticity (RGARCH) model, the Autoregressive Fractionally Integrated Moving Average (ARFIMA) model, the Heterogeneous Autoregressive model for realized variance (HAR), and the Continuous-HAR (CHAR) model.
In addition, we also considered these econometric models, except RGARCH, using log-transformed realized variance. The models incorporating log-transformed realized variance are represented with a superscript $^{\text{log}}$ as, e.g.,  HAR$^{\text{log}}$. 

The parameters of these models are estimated and updated iteratively within each data window. Initially, 50\% of the data is used for training, and the model is tested on the next 20\%. In the next step, the model is rebuilt using 70\% of the data and tested on the following 20\%. Finally, 90\% of the data is used for parameter estimation, and the model is tested on the remaining portion. This iterative process ensures that each model is refined three times, aligning with the incremental learning setup employed for the TimesFM model. Below, we formally introduce these models for completeness.

\subsubsection{Realized-GARCH Model}

The Realized-GARCH (RGARCH) model \citep{hansen2012realized}, is a widely used framework for modeling and forecasting volatility by incorporating realized measures of volatility. This approach extends traditional GARCH models by utilizing additional information from realized variance, enhancing both estimation efficiency and predictive accuracy. The model is expressed as:
\vspace{-0.1cm}
{\scriptsize
\begin{align*}
r_t &= \sqrt{h_t} \, z_t, \quad z_t \sim \mathcal{N}(0,1) \\
\log h_t &= \omega + \beta \log h_{t-1} + \alpha \log \text{RV}_{t-1} \\
\log \text{RV}_t &= \xi + \phi \log h_t + \tau z_t + u_t, \quad u_t \sim \mathcal{N}(0, \sigma_u^2)
\end{align*}
}
Here, $r_t$ is the return, $h_t$ is the latent conditional variance, and $\text{RV}_t$ is the realized variance observed at day $t$. The model extends standard GARCH dynamics by incorporating realized measures into the variance equation. The measurement equation links the realized variance to the latent variance, allowing the model to capture volatility clustering, leverage effects, and measurement noise in high-frequency data.

The model is estimated with maximum likelihood and can be tuned to incorporate different AR and GARCH lags by selecting the appropriate GARCH $(p,q)$ order using standard information criteria (we employ the AIC).

\subsubsection{ARFIMA Model}

The Autoregressive Fractionally Integrated Moving Average (ARFIMA) model is used to forecast realized variance by capturing long memory properties in the time series. Realized variance exhibits persistence and fractional integration, making ARFIMA particularly suitable for modeling its dynamics. The ARFIMA(\(p, d, q\)) model is specified as:
{\scriptsize
\begin{eqnarray*}
\Phi(B)(1 - B)^d \mathrm{RV}_t &= \Theta(B)\epsilon_t
\end{eqnarray*}
}

\vspace{-0.1cm}
where $B$ is the backshift operator ($\text{BRV}_t = \text{RV}_{t-1}$), $d$ is the fractional differencing parameter ($0 < d < 1$), $\Phi(B)$ is the autoregressive (AR) polynomial of order $p$, $\Theta(B)$ is the moving average (MA) polynomial of order $q$, and $\epsilon_t$ is a white noise process. Fractional differencing is applied to the time series \(\RV_t\) to account for its long-memory behaviour.

\subsubsection{HAR Model}

The Heterogeneous Autoregressive Model for realized variance (HAR), introduced by \cite{corsi2009simple}, is a widely-used approach for modeling realized variance. It is based on the idea of heterogeneous market participants operating over different time horizons—daily, weekly, and monthly—which is reflected in the model’s structure. The HAR model is specified as:

{\scriptsize
\begin{eqnarray*}
\text{RV}_t = \omega + \beta^{(d)} \text{RV}_{t-1}^{(d)} + \beta^{(w)} \text{RV}_{t-1}^{(w)} + \beta^{(m)} \text{RV}_{t-1}^{(m)} + \epsilon_t,
\end{eqnarray*}
}
where $\text{RV}_t$ is the realized variance at time \(t\), 
$\text{RV}_{t-1}^{(d)}$, $\text{RV}_{t-1}^{(w)}$, and $\text{RV}_{t-1}^{(m)}$ represent the daily, weekly, and monthly components of realized variance, respectively, 
$\omega$ is the intercept term, 
$\beta^{(d)}$, $\beta^{(w)}$, and $\beta^{(m)}$ are the coefficients corresponding to the daily, weekly, and monthly components, and 
$\epsilon_t$ is the error term. The weekly and monthly components are defined as:
{\scriptsize
\begin{eqnarray*}
\text{RV}_{t-1}^{(w)} = \frac{1}{5} \sum_{k=1}^{5} \text{RV}_{t-k}^{(d)},\qquad
\text{RV}_{t-1}^{(m)} = \frac{1}{22} \sum_{k=1}^{22} \text{RV}_{t-k}^{(d)},
\end{eqnarray*}
}
where $\text{RV}_{t-1}^{(w)}$ is the average realized variance over the past five days (weekly component), and $\text{RV}_{t-1}^{(m)}$ is the average over the past 22 days (monthly component).
The inclusion of lagged daily, weekly, and monthly components in the HAR model captures the long-memory dynamic dependence, that is, the persistence and heterogeneous nature of realized variance. Further, by incorporating multiple time horizons, the model reflects how market participants with different investment horizons process information, leading to varying volatility dynamics. The easy-to-estimate linear HAR model has shown remarkable forecasting performance \cite{corsi2009simple, izzeldin2019forecasting}.

\subsubsection{CHAR Model with Bi-Power Variation}

The CHAR (Continuous-HAR) model  \citep{andersen2007roughing}, extends the HAR framework by incorporating Bi-Power Variation (BPV) as a measure of quadratic variation. BPV decomposes the total variation of returns into a continuous component, mitigating the impact of jumps \citep{barndorff2004power}. The BPV, over a one-day period $[t-1,t]$, is defined as:
{\scriptsize
\begin{eqnarray*}
\BPV_t =  \frac{\pi}{2} \sum_{i=2}^M |r_i| |r_{i-1}|,
\end{eqnarray*}
}
with $r_i$ being the \(i\)-th intraday return.
Unlike to the realized volatility, the bipower variation is specifically designed to estimate the continuous part of the Quadratic Variation, the integrated variance: $\BPV_t\xrightarrow[]{p}\int_{t-1}^t \sigma^2 ds$. This makes the bipower variation preferable when  modeling or forecasting the continuous part of the volatility process (the one due to Brownian motion), leaving out the discontinuous jump component. 
The CHAR model replaces \(\operatorname{RV
}_t\) in the HAR specification with \(\BPV_t\), incorporating daily, weekly, and monthly components as follows:
{\scriptsize
\begin{eqnarray*}
{\BPV}_{t} = \omega + \beta^{(d)}{\BPV}_{t-1}^{(d)} + \beta^{(w)} {\BPV}_{t-1}^{(w)} + \beta^{(m)} {\BPV}_{t-1}^{(m)} + \epsilon_t,
\end{eqnarray*}
}
where 
\({BPV}_{t-1}^{(w)} = \frac{1}{5} \sum_{k=1}^5 \BPV_{t-k}\) is the weekly component, and 
\({\BPV}_{t-1}^{(m)} = \frac{1}{22} \sum_{k=1}^{22} \BPV_{t-k}\) is the monthly component, analogously to the HAR model.

\section{Evaluation Framework}

Next, we turn to the performance evaluation of forecasting models. It is well-established in the literature that forecast performance is sensitive to the choice of loss function. For instance, \cite{sharma2016forecasting} demonstrates that Realized-GARCH outperforms EGARCH when evaluated using  QLIKE (Quasi Likelihood), whereas EGARCH exhibits superior performance under MSE (Mean Squared Error). This underscores the importance of selecting an appropriate loss function tailored to the specific objectives of the forecasting task. As further emphasized by \cite{hansen2005forecast} and \cite{patton2011volatility}, the choice of loss function plays a crucial role in assessing the performance of volatility forecasting models. Accordingly, to ensure a comprehensive and unbiased evaluation, we employ six distinct loss functions, mitigating the risk of favoring any particular model based on a single metric. A robust forecasting model should perform well across all or most of the considered loss measures, demonstrating consistency in predictive accuracy across different evaluation criteria.

We follow  \cite{patton2011volatility} and choose MSE (Mean Squared Error) and QLIKE (Quasi Likelihood) loss functions in addition to MDA (Mean Directional Accuracy). MSE is a symmetric loss function that penalizes over-predictions and under-predictions equally. In contrast, QLIKE is asymmetric, imposing a heavier penalty on under-predictions than on over-predictions. This characteristic makes QLIKE particularly relevant in applications such as risk management, where under-predictions can lead to significant financial losses. As noted by \cite{patton2011volatility}, the MSE and QLIKE loss functions are robust to noise in the volatility proxy, and the rankings of different models are invariant to the choice of units of measurement. In contrast, MDA serves as a valuable complementary metric to these two primary measures, particularly effective in assessing and monitoring potential model overfitting. The other metrics that we use include Mean Absolute Error (MAE), Mean Absolute Percentage Error (MAPE) and Symmetric Mean Absolute Percentage Error (sMAPE). The statistical expressions of these measures are as follows:

\begin{itemize}
\scriptsize
\item Mean Squared Error (MSE)
$$
\text{MSE} = \frac{1}{P} \sum_{\ell=1}^{P} \left(\text{RV}_{t+\ell} - \widehat{\text{RV}}_{t+\ell} \right)^2 
$$
\item Mean Absolute Error (MAE)
$$
\text{MAE} = \frac{1}{P} \sum_{\ell=1}^{P} \left|\text{RV}_{t+\ell} - \widehat{\text{RV}}_{t+\ell} \right|
$$
\item Mean Absolute Percentage Error (MAPE)
$$
\text{MAPE} = \frac{100}{P} \sum_{\ell=1}^{P} \left| \frac{\text{RV}_{t+\ell} - \widehat{\text{RV}}_{t+\ell}}{\text{RV}_{t+\ell}} \right|
$$
\item Mean Directional Accuracy (MDA)
$$
\text{MDA} = \frac{100}{P} \sum_{\ell=1}^{P} \mathbf{1} \left( \operatorname{sign}(\text{RV}_{t+\ell} - \text{RV}_{t+\ell-1}) = \operatorname{sign}(\widehat{\text{RV}}_{t+\ell} - \widehat{\text{RV}}_{t+\ell-1}) \right)
$$
\item Quasi Likelihood (QLIKE)
$$
\text{QLIKE} = \frac{1}{P} \sum_{\ell=1}^{P} \left[ \frac{\text{RV}_{t+\ell}}{\widehat{\text{RV}}_{t+\ell}} - \log\left( \frac{\text{RV}_{t+\ell}}{\widehat{\text{RV}}_{t+\ell}} \right) - 1 \right]
$$
\item Symmetric Mean Absolute Percentage Error (sMAPE)
$$
\text{sMAPE} = \frac{100}{P} \sum_{\ell=1}^{P} \frac{2 \left|\text{RV}_{t+\ell} - \widehat{\text{RV}}_{t+\ell}\right|}{\left|\text{RV}_{t+\ell}\right| + \left|\widehat{\text{RV}}_{t+\ell}\right|}
$$
\end{itemize}

Here, $\widehat{\text{RV}}_{t+l}$ refers to the predicted value of the realized variance at a time $t+\ell$, opposed to its actual value $\text{RV}_{t+l}$. The above definitions can be analogously written in terms of Bipower variation or any other appropriate measure of volatility.
These measures reflect distinct dimensions of forecast performance, thereby providing a robust and comprehensive evaluation framework.

In addition to the performance measures for a particular model, we also compute skill scores, an intuitive measure of the relative skills of two forecasting models. More specifically, we compute the proportion of the forecasting error of a model relative to the error of the benchmark model.


Having analyzed the loss functions across different models, the next step is to formally assess whether the observed differences in predictive performance are statistically significant. To this end, we employ the \textit{Diebold-Mariano} (DM) test \cite{diebold2002comparing}, a widely used statistical test for forecast evaluation. This test evaluates the null hypothesis that the predictive accuracy of the two models is equal by analyzing the loss differential series:

Let \( d_{t, i, j} = L(y_t, \hat{y}_t^{(i)}) - L(y_t, \hat{y}_t^{(j)}) \) denote the loss differential at time \( t \) between models \( i \) and \( j \), where \( L(\cdot, \cdot) \) is a chosen loss function (e.g., MSE, QLIKE). With \( P \) denoting the number of out-of-sample forecast periods, the sample mean of the loss differentials is:
{\scriptsize
\[
\bar{d}_{i,j} = \frac{1}{P} \sum_{t=1}^{P} d_{t,i,j}.
\]
}
The DM test statistic is:
{\scriptsize
\[
\text{DM} = \frac{\bar{d}_{i,j}}{\sqrt{\widehat{\mathrm{Var}}(\bar{d}_{i,j})}},
\]}
where \( \widehat{\mathrm{Var}}(\bar{d}_{i,j}) \) is a consistent estimator of the variance of \( \bar{d}_{i,j} \), typically computed using a heteroskedasticity and autocorrelation consistent estimator such as Newey–West. Under the null hypothesis of equal predictive accuracy,
{\scriptsize
\[
H_0: \mathbb{E}[d_{t,i,j}] = 0,
\]}
and under standard regularity conditions, the DM statistic is asymptotically standard normal:
{\scriptsize
\[
\text{DM} \xrightarrow{d} \mathcal{N}(0,1) \quad \text{as } P \to \infty.
\]
}

We also employed the Giacomini and White (GW) test \citep{giacomini2006tests}, which accounts for parameter uncertainty and allows for the comparison of nested forecasting models. The GW test evaluates whether two competing models differ in predictive ability, even when one is nested within the other. Let \( d_{t,i,j} \) denote the difference in forecast losses (e.g., squared forecast errors) at time \( t \) between models \( i \) and \( j \), computed over a fixed forecast horizon \( h \). The null hypothesis of equal conditional predictive ability is:
{\scriptsize
\begin{equation}
    \operatorname{H_0} : \mathbb{E}[d_{t,i,j}] = 0.
\end{equation}
}
Let \( \bar{d}_{i,j} = \frac{1}{P} \sum_{t=1}^{P} d_{t,i,j} \) be the sample average of the loss differentials, where \( P \) denotes the number of out-of-sample forecasts. The GW test statistic is given by:
{\scriptsize
\begin{equation}
    \text{GW} = \bar{d}_{i,j}^\top \hat{V}^{-1} \bar{d}_{i,j} \sim \chi^2_k,
\end{equation}
}
where \( \hat{V} \) is a heteroskedasticity and autocorrelation consistent (HAC) estimator of the asymptotic variance of \( \bar{d}_{i,j} \), and \( k \) is the dimension of \( \bar{d}_{i,j} \) ($k=1$ in univariate loss comparison).

To formally assess the statistical significance of differences in predictive accuracy among a set $\mathcal{M}$ of competing models, we employ the \textit{Model Confidence Set} (MCS) procedure introduced by \citep{hansen2011model}. The MCS method involves a sequence of predictive accuracy tests to determine a subset of models, referred to as the \textit{Superior Set of Models} (SSM), that exhibit statistically indistinguishable forecasting performance at a given confidence level. This approach enables the identification of the most reliable models in a statistically rigorous manner.

The MCS procedure evaluates the differences in loss functions between competing models $i$ and $j$ over time, denoted as $d_{t,i,j}$. Assuming stationarity in the loss differentials, the null hypothesis of equal predictive accuracy is formulated as:
{\scriptsize
\begin{equation*} \operatorname{H_0}: \mathbb{E}[d_{t,i,j}] = 0, \quad \forall\, i, j \in \mathcal{M}. \end{equation*}
}
At a given confidence level, models for which the null hypothesis is rejected are sequentially excluded. The SSM consists of the models that remain after this elimination process, meaning they do not exhibit statistically significant differences in predictive performance. The $p$-values used for model selection within the MCS framework are computed using a stationary bootstrap procedure. Models with lower $p$-values provide stronger evidence of inferior predictive performance and are more likely to be eliminated from the SSM. For further details on the MCS methodology, refer to \citep{hansen2011model}.

\section{Empirical Analysis}

Table~\ref{tab:rv_summary_vol} presents the summary statistics for realized volatility across multiple indices considered, capturing key distributional characteristics such as the minimum (Min), mean, standard deviation (SD), median, and maximum (Max) values. The statistics are reported for the entire sample period and segmented into subsample periods: the first 50\%, next 20\%, next 20\%, and the last 10\% of observations. This segmentation allows for analyzing the periods used in incremental learning and assessing the impact of newly added data on model updates

A key observation is the substantial variation in volatility across indices. From the table, the indices with the highest mean realized volatility are BVSP (0.01091) and GDAXI (0.01063), indicating they experience the greatest market fluctuations. Conversely, the indices with the lowest mean realized volatility are AORD (0.00612) and STI (0.00687), suggesting more stable market conditions. Regarding market dispersion, as reflected in the standard deviation of realized volatility, GDAXI (0.00669) and STOXX50E (0.00669) exhibit the highest variability, indicating substantial volatility swings. On the other hand, STI (0.00274) and AORD (0.00399) show the lowest standard deviations, confirming that these indices have more stable volatility patterns.

When examining volatility over time, most indices experience higher volatility in the first half of the sample, followed by a decline. For instance, in AEX, the mean realized volatility in the first 50\% of the sample is 0.01031, decreasing to 0.00899 in the last 10\%. A similar pattern is observed in BVSP (from 0.01228 to 0.01023) and FTSE (from 0.01013 to 0.00960), reflecting market stabilization over time. However, some indices, such as RUT, exhibit an increase in volatility in the last 10\% (from 0.00737 to 0.01066), suggesting heightened market uncertainty in the later periods. The maximum realized volatility values provide insight into extreme market movements. The highest maximum volatility is observed in FTSE (0.10296), followed by SPX (0.08802) and BVSP (0.07689), indicating significant tail risk in these markets. Conversely, STI exhibits the lowest maximum volatility (0.04278), followed by IXIC (0.04573), highlighting relatively lower risk exposure.

Another important observation is the contrast between developed and emerging markets. Developed markets such as SPX, FTSE, and GDAXI show moderate fluctuations with relatively stable standard deviations across different periods. In contrast, emerging markets like MXX and SSEC exhibit larger variations, as seen in their higher standard deviations across different sub-periods. This suggests that emerging markets remain more susceptible to external shocks and policy changes, resulting in higher volatility clustering. These findings highlight the need for adaptive forecasting models that incorporate time-varying volatility dynamics, particularly for markets exhibiting higher fluctuations.

\begin{table}[htbp]
    \centering
    \caption{Summary Statistics for 5-Minute sub-sampled Realized Volatility}
    \label{tab:rv_summary_vol}
     \resizebox{.6\textwidth}{!}{%
    \begin{tabular}{lrrrrr|rrrrr}
        \toprule
      Period   & Min & Mean & SD & Median & Max & Min & Mean & SD & Median & Max \\ 
        \midrule
        &\multicolumn{5}{c}{AEX}  & \multicolumn{5}{c}{KS11}  \\ 
        \midrule
        Total     & 0.00126 & 0.00910 & 0.00573 & 0.00752 & 0.06481 & 0.00116 & 0.00919 & 0.00569 & 0.00756 & 0.07710 \\ 
        First 50\% & 0.00126 & 0.01031 & 0.00635 & 0.00857 & 0.06020 & 0.00315 & 0.01185 & 0.00619 & 0.01042 & 0.07710 \\ 
        Next 20\%  & 0.00150 & 0.00808 & 0.00417 & 0.00701 & 0.04576 & 0.00238 & 0.00638 & 0.00338 & 0.00561 & 0.04768 \\ 
        Next 20\%  & 0.00206 & 0.00714 & 0.00541 & 0.00583 & 0.06481 & 0.00269 & 0.00586 & 0.00341 & 0.00520 & 0.05107 \\ 
        Last 10\%  & 0.00169 & 0.00899 & 0.00404 & 0.00818 & 0.03104 & 0.00116 & 0.00816 & 0.00340 & 0.00724 & 0.03135 \\ 
        \midrule

        &\multicolumn{5}{c}{AORD}  & \multicolumn{5}{c}{KSE}  \\ 
        \midrule
        Total     & 0.00105 & 0.00612 & 0.00399 & 0.00510 & 0.06911 & 0.00014 & 0.00856 & 0.00527 & 0.00714 & 0.06079 \\ 
        First 50\% & 0.00105 & 0.00598 & 0.00383 & 0.00490 & 0.03905 & 0.00014 & 0.01034 & 0.00604 & 0.00860 & 0.04555 \\ 
        Next 20\%  & 0.00208 & 0.00600 & 0.00277 & 0.00534 & 0.03367 & 0.00214 & 0.00644 & 0.00295 & 0.00581 & 0.03010 \\ 
        Next 20\%  & 0.00181 & 0.00584 & 0.00494 & 0.00472 & 0.06911 & 0.00179 & 0.00699 & 0.00403 & 0.00605 & 0.06079 \\ 
        Last 10\%  & 0.00284 & 0.00765 & 0.00440 & 0.00647 & 0.03878 & 0.00236 & 0.00702 & 0.00372 & 0.00613 & 0.03707 \\ 
        \midrule
        
       & \multicolumn{5}{c}{BFX}  & \multicolumn{5}{c}{MXX}  \\ 
       \midrule
        Total     & 0.00200 & 0.00833 & 0.00478 & 0.00708 & 0.06074 & 0.00189 & 0.00782 & 0.00459 & 0.00662 & 0.07228 \\ 
        First 50\% & 0.00201 & 0.00876 & 0.00508 & 0.00749 & 0.05857 & 0.00189 & 0.00828 & 0.00537 & 0.00682 & 0.06675 \\ 
        Next 20\%  & 0.00219 & 0.00790 & 0.00366 & 0.00696 & 0.03930 & 0.00240 & 0.00716 & 0.00409 & 0.00609 & 0.07228 \\ 
        Next 20\%  & 0.00200 & 0.00723 & 0.00479 & 0.00620 & 0.06074 & 0.00295 & 0.00706 & 0.00323 & 0.00628 & 0.03730 \\ 
        Last 10\%  & 0.00244 & 0.00921 & 0.00474 & 0.00822 & 0.04879 & 0.00329 & 0.00833 & 0.00292 & 0.00763 & 0.02293 \\ 
        \midrule

        &\multicolumn{5}{c}{BVSP}  & \multicolumn{5}{c}{N225}  \\ 
        \midrule
        Total     & 0.00228 & 0.01091 & 0.00581 & 0.00966 & 0.07689 & 0.00144 & 0.00875 & 0.00485 & 0.00771 & 0.06204 \\ 
        First 50\% & 0.00243 & 0.01228 & 0.00651 & 0.01079 & 0.07689 & 0.00244 & 0.01011 & 0.00496 & 0.00924 & 0.05682 \\ 
        Next 20\%  & 0.00309 & 0.00973 & 0.00377 & 0.00895 & 0.05029 & 0.00229 & 0.00771 & 0.00389 & 0.00679 & 0.04103 \\ 
        Next 20\%  & 0.00228 & 0.00900 & 0.00458 & 0.00819 & 0.06361 & 0.00144 & 0.00687 & 0.00474 & 0.00563 & 0.06204 \\ 
        Last 10\%  & 0.00414 & 0.01023 & 0.00594 & 0.00901 & 0.06085 & 0.00213 & 0.00782 & 0.00417 & 0.00692 & 0.04356 \\ 
        \midrule

        &\multicolumn{5}{c}{DJI}  & \multicolumn{5}{c}{RUT}  \\ 
        \midrule
        Total     & 0.00139 & 0.00857 & 0.00613 & 0.00702 & 0.09287 & 2.27E-05 & 0.00760 & 0.00497 & 0.00626 & 0.05844 \\ 
        First 50\% & 0.00209 & 0.00987 & 0.00656 & 0.00829 & 0.09287 & 2.27E-05 & 0.00737 & 0.00492 & 0.00601 & 0.05844 \\ 
        Next 20\%  & 0.00177 & 0.00731 & 0.00485 & 0.00606 & 0.07719 & 0.00214 & 0.00740 & 0.00453 & 0.00618 & 0.05095 \\ 
        Next 20\%  & 0.00139 & 0.00639 & 0.00566 & 0.00496 & 0.06376 & 0.00171 & 0.00686 & 0.00497 & 0.00571 & 0.05552 \\ 
        Last 10\%  & 0.00243 & 0.00895 & 0.00525 & 0.00750 & 0.05334 & 0.00351 & 0.01066 & 0.00500 & 0.00940 & 0.03355 \\ 
         \midrule
        &\multicolumn{5}{c}{FCHI}  & \multicolumn{5}{c}{SPX}  \\ 
        \midrule
        Total     & 0.00166 & 0.00993 & 0.00589 & 0.00854 & 0.07157 & 0.00110 & 0.00850 & 0.00611 & 0.00689 & 0.08802 \\ 
        First 50\% & 0.00166 & 0.01100 & 0.00638 & 0.00969 & 0.07157 & 0.00213 & 0.00984 & 0.00651 & 0.00828 & 0.08802 \\ 
        Next 20\%  & 0.00209 & 0.00974 & 0.00486 & 0.00858 & 0.04751 & 0.00127 & 0.00736 & 0.00485 & 0.00596 & 0.06107 \\ 
        Next 20\%  & 0.00211 & 0.00769 & 0.00535 & 0.00663 & 0.06604 & 0.00110 & 0.00615 & 0.00559 & 0.00470 & 0.06444 \\ 
        Last 10\%  & 0.00265 & 0.00944 & 0.00483 & 0.00850 & 0.03951 & 0.00167 & 0.00880 & 0.00543 & 0.00753 & 0.05348 \\ 

        \midrule
        &\multicolumn{5}{c}{FTSE}  & \multicolumn{5}{c}{SSEC}  \\ 
        \midrule
        Total     & 0.00115 & 0.00917 & 0.00600 & 0.00756 & 0.10296 & 0.00188 & 0.01054 & 0.00660 & 0.00860 & 0.06541 \\ 
        First 50\% & 0.00211 & 0.01013 & 0.00650 & 0.00856 & 0.10296 & 0.00188 & 0.01194 & 0.00698 & 0.01023 & 0.06541 \\ 
        Next 20\%  & 0.00216 & 0.00814 & 0.00452 & 0.00704 & 0.04215 & 0.00327 & 0.01070 & 0.00749 & 0.00831 & 0.06439 \\ 
        Next 20\%  & 0.00115 & 0.00759 & 0.00590 & 0.00635 & 0.08166 & 0.00252 & 0.00821 & 0.00454 & 0.00691 & 0.04112 \\ 
        Last 10\%  & 0.00224 & 0.00960 & 0.00515 & 0.00828 & 0.03587 & 0.00213 & 0.00787 & 0.00322 & 0.00718 & 0.02558 \\ 
        \midrule

        &\multicolumn{5}{c}{GDAXI}  & \multicolumn{5}{c}{SSMI}  \\ 
        \midrule
        Total     & 0.00203 & 0.01063 & 0.00669 & 0.00887 & 0.07670 & 0.00246 & 0.00779 & 0.00495 & 0.00637 & 0.07448 \\ 
        First 50\% & 0.00208 & 0.01242 & 0.00763 & 0.01057 & 0.07670 & 0.00293 & 0.00873 & 0.00519 & 0.00710 & 0.05671 \\ 
        Next 20\%  & 0.00226 & 0.00984 & 0.00523 & 0.00856 & 0.04904 & 0.00271 & 0.00692 & 0.00404 & 0.00593 & 0.06495 \\ 
        Next 20\%  & 0.00203 & 0.00779 & 0.00479 & 0.00681 & 0.05583 & 0.00246 & 0.00671 & 0.00544 & 0.00558 & 0.07448 \\ 
        Last 10\%  & 0.00253 & 0.00893 & 0.00449 & 0.00798 & 0.03785 & 0.00268 & 0.00694 & 0.00307 & 0.00617 & 0.02651 \\ 
        \midrule

        &\multicolumn{5}{c}{HSI}  & \multicolumn{5}{c}{STI}  \\ 
        \midrule
        Total     & 0.00209 & 0.00876 & 0.00459 & 0.00761 & 0.06613 &  0 & 0.00687 & 0.00274 & 0.00628 & 0.04278 \\ 
        First 50\% & 0.00228 & 0.00999 & 0.00532 & 0.00871 & 0.06613 & 0.00254 & 0.00776 & 0.00263 & 0.00724 & 0.04278 \\ 
        Next 20\%  & 0.00209 & 0.00725 & 0.00324 & 0.00643 & 0.03499 &  0 & 0.00624 & 0.00252 & 0.00557 & 0.02974 \\ 
        Next 20\%  & 0.00218 & 0.00685 & 0.00283 & 0.00634 & 0.04866 & 0.00268 & 0.00604 & 0.00296 & 0.00539 & 0.03487 \\ 
        Last 10\%  & 0.00376 & 0.00948 & 0.00348 & 0.00880 & 0.03350 & 0.00287 & 0.00531 & 0.00137 & 0.00516 & 0.01577 \\ 
        \midrule

        &\multicolumn{5}{c}{IBEX}  & \multicolumn{5}{c}{STOXX50E}  \\ 
        \midrule
        Total     & 0.00202 & 0.01049 & 0.00559 & 0.00945 & 0.07423 &  0 & 0.01060 & 0.00669 & 0.00898 & 0.10405 \\ 
        First 50\% & 0.00202 & 0.01055 & 0.00585 & 0.00970 & 0.06345 & 0.00057 & 0.01169 & 0.00733 & 0.00988 & 0.10405 \\ 
        Next 20\%  & 0.00311 & 0.01213 & 0.00511 & 0.01113 & 0.04992 & 0.00022 & 0.01036 & 0.00520 & 0.00922 & 0.04946 \\ 
        Next 20\%  & 0.00305 & 0.00883 & 0.00547 & 0.00750 & 0.07423 & 0.00010 & 0.00799 & 0.00502 & 0.00693 & 0.07352 \\ 
        Last 10\%  & 0.00358 & 0.01026 & 0.00433 & 0.00942 & 0.03795 &  0 & 0.01090 & 0.00746 & 0.00937 & 0.06835 \\ 
        \midrule

        &\multicolumn{5}{c}{IXIC}  & &&&& \\ 
        \midrule
      Total     & 0.00163 & 0.00936 & 0.00619 & 0.00749 & 0.07722 & \\ 
        First 50\% & 0.00210 & 0.01115 & 0.00672 & 0.00935 & 0.06558 & \\ 
        Next 20\%  & 0.00215 & 0.00681 & 0.00359 & 0.00587 & 0.04573 & \\ 
        Next 20\%  & 0.00163 & 0.00699 & 0.00573 & 0.00549 & 0.07722 & \\ 
        Last 10\%  & 0.00298 & 0.01027 & 0.00499 & 0.00921 & 0.03054 & \\ 
        
        \bottomrule
    \end{tabular}}
\end{table}

\subsection{Skill Scores}

In Tables \ref{tab:lossa}-\ref{tab:lossc}, we present the out-of-sample errors (using the 6 loss functions defined previously) of each forecasting model relative to the benchmark specified in the selected row. Reading by column, the reported ratio represents the cross-sectional average of the relative one-day-ahead realized variance forecast errors across stocks. The first row, in particular, displays the out-of-sample error of each model relative to the basic ARFIMA model, averaged across stocks. Furthermore, for each stock and pairwise comparison, we conduct the Diebold–Mariano test to assess predictive accuracy. The number formatting in the table indicates whether the null hypothesis of equal predictive accuracy for a given pairwise comparison was rejected more than half of the time across stocks at different levels of statistical significance. To compare the forecasting accuracy of different models, we evaluate their performance across six loss functions: Mean Squared Error (MSE), Mean Absolute Deviation (MAD), Quasi-Likelihood (Qlike), Mean Absolute Percentage Error (MAPE), Mean Directional Accuracy (MDA), and Symmetric Mean Absolute Percentage Error (sMAPE). These loss functions allow for a comprehensive assessment of both error magnitude and directional accuracy across models.

\subsubsection{Comparison of Linear and Log Models}

Accurate volatility forecasting is essential in financial markets, where even small improvements in predictive accuracy can have significant implications for risk management and investment strategies. Given that financial time series often exhibit heteroskedasticity, log transformations are commonly employed to stabilize variance and improve model robustness. To evaluate their effectiveness, we compare the performance of linear and log-transformed models within this broader framework.

The results indicate that log-transformed models consistently outperform their linear counterparts across all categories. Specifically, ARFIMA$^{\text{log}}$, CHAR$^{\text{log}}$, and HAR$^{\text{log}}$ systematically achieve lower error scores in MSE, MAD, and Qlike, demonstrating the stabilizing effect of log transformations in realized volatility forecasting. In contrast, their linear counterparts (ARFIMA, CHAR, and HAR) struggle with capturing the distributional properties of realized volatility, leading to higher error metrics.

For TimesFM models, both pretrained (PT) and incremental learning (IL) variants benefit from log transformations, contributing to improved forecasting performance. TFM64$^{\text{log}}_{\text{IL}}$, TFM128$^{\text{log}}_{\text{IL}}$, and TFM512$^{\text{log}}_{\text{IL}}$, as well as their pretrained counterparts TFM64$^{\text{log}}_{\text{PT}}$, TFM128$^{\text{log}}_{\text{PT}}$, and TFM512$^{\text{log}}_{\text{PT}}$, outperform their linear versions in multiple loss functions. The advantage of log transformation is particularly evident in reducing MSE and Qlike errors. However, the improvement is relatively less pronounced than in econometric models, likely because TimesFM already integrates nonlinear relationships through its architecture. Notably, the gains from log transformation are more substantial in IL models compared to PT models, as the former continuously adjust to new data, leveraging the log transformation more effectively. Nevertheless, these findings indicate that log transformations remain an important component of financial time series modeling.

\subsubsection{Comparative Performance of Traditional Models and and the Incrementally Trained TimesFM Foundation Model}

The forecasting accuracy of traditional econometric models, including ARFIMA, CHAR, and HAR, is systematically lower compared to TimesFM models across all loss functions. A model is considered strong if its skill scores exceed 1 in its respective row, meaning it outperforms the benchmark across multiple loss functions. Conversely, models where all skill scores in a row are below 1 indicate weaker predictive capabilities. ARFIMA performs poorly in volatility forecasting, with skill scores exceeding 1 in at least five out of six loss functions, particularly in Qlike, where it struggles significantly in capturing tail risk. This suggests that ARFIMA is less effective at modeling realized volatility dynamics compared to more advanced models. This highlights its inefficiency in modeling realized volatility tails. CHAR and HAR show some improvements over ARFIMA, but they remain inconsistent across different loss functions. HAR exhibits skill scores greater than 1 in MSE, MAD, Qlike, and sMAPE, which suggests it captures some aspects of volatility but struggles with directional accuracy and percentage-based errors such as MDA and MAPE. CHAR, on the other hand, performs poorly in MSE and MAD, reinforcing its limitations in minimizing forecast errors, making it less reliable for volatility prediction. Notably, there exist models where all six skill scores are below 1, indicating weak performance in forecasting realized volatility, particularly in comparison to the stronger TimesFM models.

TimesFM models, particularly those fine-tuned with incremental learning (IL), achieve superior forecast accuracy across multiple loss functions. A key observation is that among the seven econometric models (ARFIMA, CHAR, HAR, and their log variants), none consistently achieve skill scores greater than 1 across all loss functions. In contrast, among the 12 TimesFM models (six with linear and six with log transformations), a majority demonstrate skill scores exceeding 1 in at least four loss functions, reinforcing their strong predictive performance. However, not all TimesFM models consistently outperform traditional econometric models. Specifically, some pretrained (PT) models, particularly those with shorter context lengths, exhibit relatively weaker performance in certain loss functions, such as MDA and MAPE. In contrast, PT models with longer context lengths generally perform better, suggesting that context length plays a crucial role in their ability to adapt to market dynamics. 

Among the TimesFM models, those utilizing incremental learning (IL) demonstrate superior adaptability to market conditions, as evidenced by consistently higher skill scores in MSE, MAD, and Qlike compared to their pretrained (PT) counterparts. In contrast, certain weaker models have skill scores below 1 across all loss functions, indicating their inferiority in forecasting realized volatility. Among the 19 evaluated models, TimesFM models with incremental learning (IL) outperform their pretrained (PT) counterparts and econometric models in most cases. Specifically, TFM64$_{\text{IL}}$, TFM128$_{\text{IL}}$, and TFM512$_{\text{IL}}$ exhibit skill scores exceeding 1 in key loss functions such as MSE, MAD, and Qlike, reinforcing their robustness in forecasting. However, some TimesFM PT models, particularly those with shorter context lengths, show weak performance, whereas PT models with longer context lengths generally perform better. This pattern indicates that static pretraining can achieve more competitive performance when combined with adequate context length. However, we observe that smaller context length leads to a better forecasting performance for the fine-tuned model. A likely explanation is that an increase in the context length is not necessarily contributing to predictive accuracy and might be contributing to the noise. For instance, TFM64$_{\text{IL}}$ achieves a skill score greater than 1 across multiple loss functions, indicating superior performance compared to traditional models. Similarly, TFM128$_{\text{IL}}$ and TFM512$_{\text{IL}}$ also exhibit strong performance, with skill scores exceeding 1 in key loss functions, reinforcing their robustness in forecasting. Furthermore, TimesFM models consistently achieve skill scores greater than 1 in Qlike for most assets, suggesting they better capture tail risks than econometric models. The log-transformed versions further enhance performance, particularly in Qlike and MSE. For example, TFM64$^{\text{log}}_{\text{IL}}$ has skill scores exceeding 1 across all loss functions, confirming its dominance as the most robust forecasting model. This underscores the compounding benefits of incremental learning and log transformations in stabilizing volatility estimation. Additionally, TimesFM models outperform all econometric models in directional accuracy (MDA), with at least three IL-based models achieving the best performance in this metric. This confirms their ability to capture volatility trends more effectively.

Overall, the results highlight the dominance of TimesFM IL models over traditional models in volatility forecasting. Notably, TimesFM models outperform all seven econometric models in at least four loss functions. However, only a subset of TimesFM models achieve skill scores greater than 1 across all loss functions, with TFM64$^{\text{log}}_{\text{IL}}$ standing out as the most consistently strong performer. Meanwhile, certain TimesFM PT models fail to achieve a competitive edge, particularly in MDA and MAPE. Among the econometric models, ARFIMA and its log variant struggle the most, with skill scores consistently below 1 across nearly all loss functions, confirming their weaker predictive capabilities. The systematic reduction in Qlike, MSE, and MAD errors, along with improved MDA scores, confirms that TFM64$^{\text{log}}_{\text{IL}}$ stands out as the strongest model, consistently outperforming traditional models across all six loss functions.

\subsubsection{Comparison of Zero-Shot and Incremental Learning}

The comparative analysis reveals that TimesFM models trained with incremental learning (IL) exhibit consistently better performance than their pretrained (PT) counterparts across the majority of loss functions. Notably, IL models outperform PT models in both short and long context settings, with performance differences becoming more pronounced in key metrics such as Qlike and MDA. When examining MSE, MAD, and Qlike, the IL models—TFM64$_{\text{IL}}$, TFM128$_{\text{IL}}$, and TFM512$_{\text{IL}}$, achieve skill scores exceeding 1 in nearly all cases, clearly demonstrating their ability to reduce forecast errors and adapt to dynamic financial environments. In particular, TFM64$_{\text{IL}}$ achieves skill scores greater than 1 in all six loss functions, indicating its robustness across multiple evaluation criteria.

In contrast, PT models show a mixed performance pattern. For instance, TFM512$_{\text{PT}}$ performs well in Qlike and sMAPE, where longer context lengths provide better stability, but its performance in MDA and MAPE remains below that of IL models. TFM128$_{\text{PT}}$ and TFM64$_{\text{PT}}$ generally fail to reach skill scores above 1 in key loss functions, reinforcing the limitations of static pretraining, particularly in capturing directional movements in volatility. The primary advantage of IL models lies in their continuous updating process, allowing them to incorporate new market information and adapt more effectively. This is particularly evident in directional accuracy, where TFM64$_{\text{IL}}$ and TFM128$_{\text{IL}}$ surpass their PT counterparts, demonstrating superior adaptability in volatile market conditions. Overall, TimesFM IL models establish themselves as the superior training approach, consistently achieving skill scores above 1 across the majority of loss functions. Their adaptability and responsiveness to new information enable them to provide more reliable forecasts compared to static PT models, reinforcing the importance of incremental learning in realized volatility forecasting.

\subsubsection{Identifying the Best Performing Model}

To determine the best-performing model, we analyze the relative skill scores across all loss functions. A model is considered superior if it consistently achieves skill scores above 1 across multiple loss functions, indicating higher predictive accuracy relative to the benchmark. The results clearly indicate that TFM64$^{\text{log}}_{\text{IL}}$ is the best-performing model across most loss functions. It achieves the lowest forecast errors in MSE, MAD, and Qlike while also maintaining strong performance in directional accuracy (MDA). The combination of incremental learning and log transformation enables it to effectively capture volatility dynamics while mitigating excessive historical dependencies. Although TFM128$^{\text{log}}_{\text{IL}}$ and TFM512$^{\text{log}}_{\text{IL}}$ also perform well, they do not consistently outperform the 64-context variant, reaffirming that excessive historical context can introduce noise.

Among PT models, TFM512$_{\text{PT}}$ achieves the best results, particularly in long-horizon loss functions such as Qlike and sMAPE. However, PT models fail to match the overall performance of IL models, further emphasizing the advantage of adaptation in financial time series forecasting. In general, empirical results suggest that TFM64$^{\text{log}}_{\text{IL}}$ provides the most robust prediction accuracy, balancing adaptability with historical dependency. This model offers a superior trade-off in volatility forecasting and should be considered the preferred approach for realized volatility modeling. To assess whether a small subset of stocks drives the superior forecast accuracy of TimesFM models or extends to the entire cross-section, we further analyze the aggregate results presented in Tables \ref{tab:lossa}-\ref{tab:lossc}. For further comparison and for brevity, we select a ``preferred" model from the econometric models, namely, CHAR$^{\text{log}}$.

\begin{figure}[h!]
\begin{center}
\subfigure[MSE]{%
\includegraphics[width=0.44\textwidth]{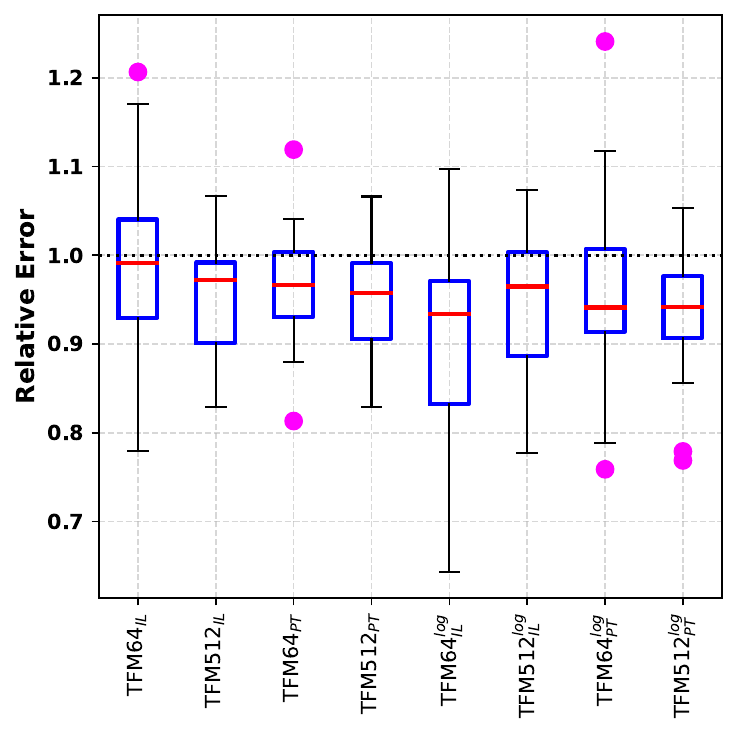}
\label{mse}}
\quad
\subfigure[MAE]{%
\includegraphics[width=0.44\textwidth]{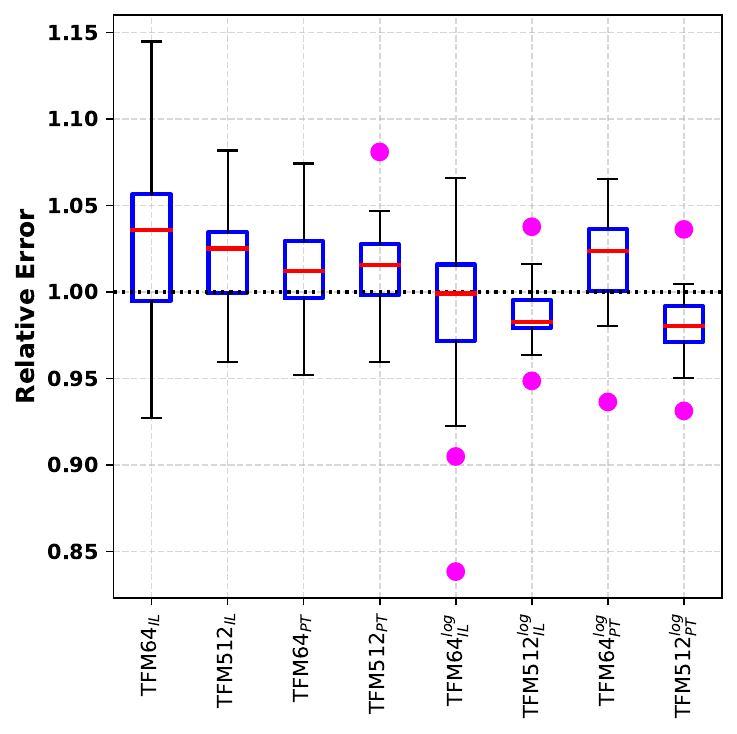}
\label{mad}}
\quad
\subfigure[MDA]{%
\includegraphics[width=0.44\textwidth]{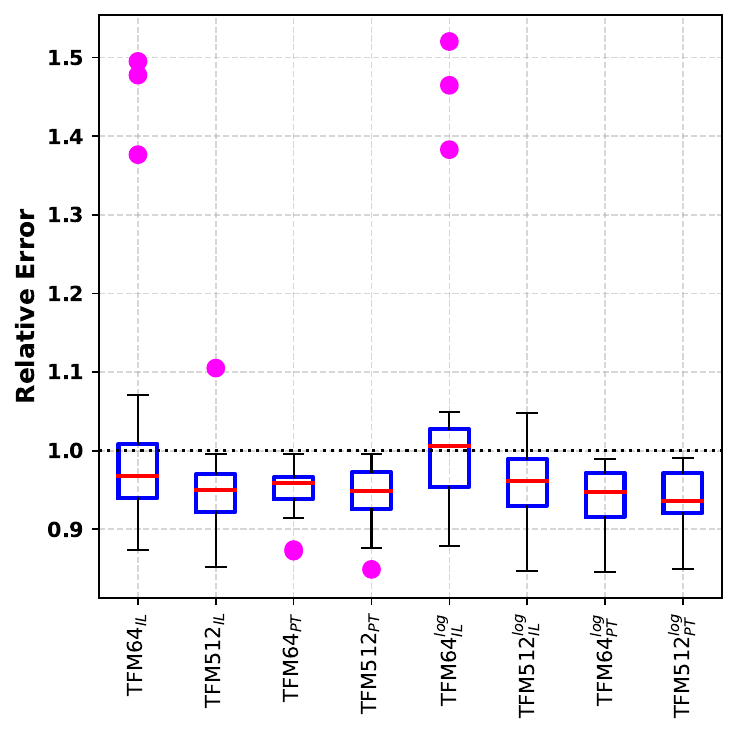}
\label{mda}}
\quad
\subfigure[MAPE]{%
\includegraphics[width=0.44\textwidth]{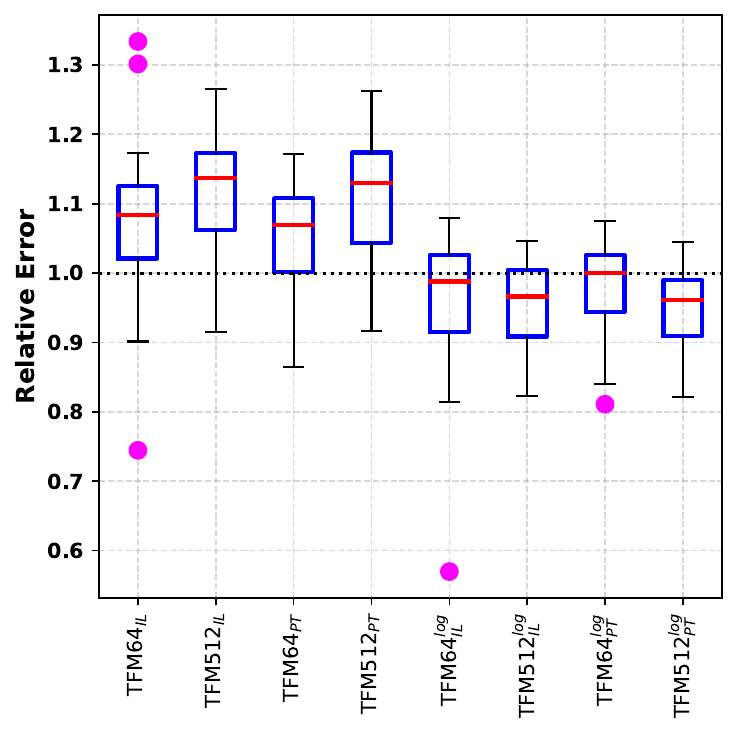}
\label{mape}}
\quad
\subfigure[sMAPE]{%
\includegraphics[width=0.44\textwidth]{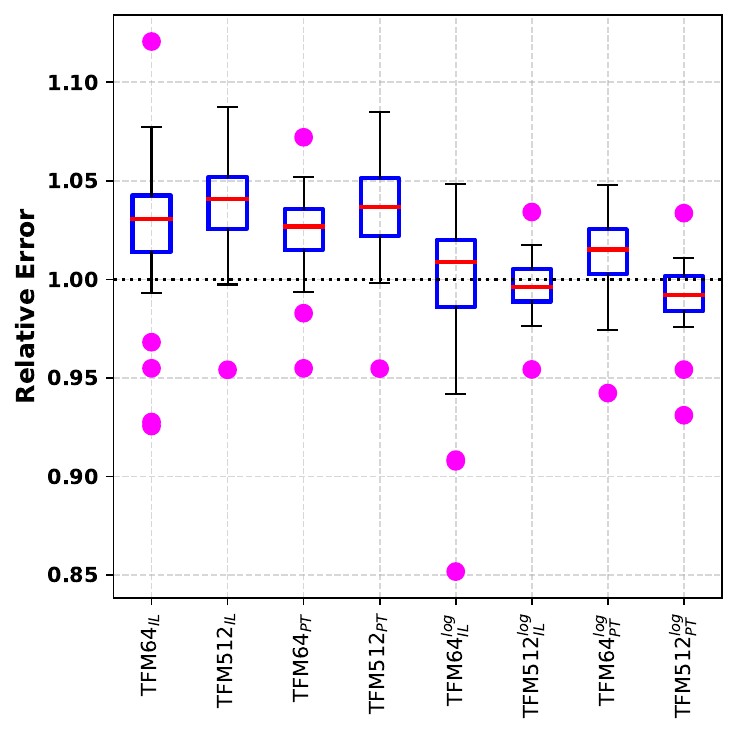}
\label{smape}}
\quad
\subfigure[Qlike]{%
\includegraphics[width=0.44\textwidth]{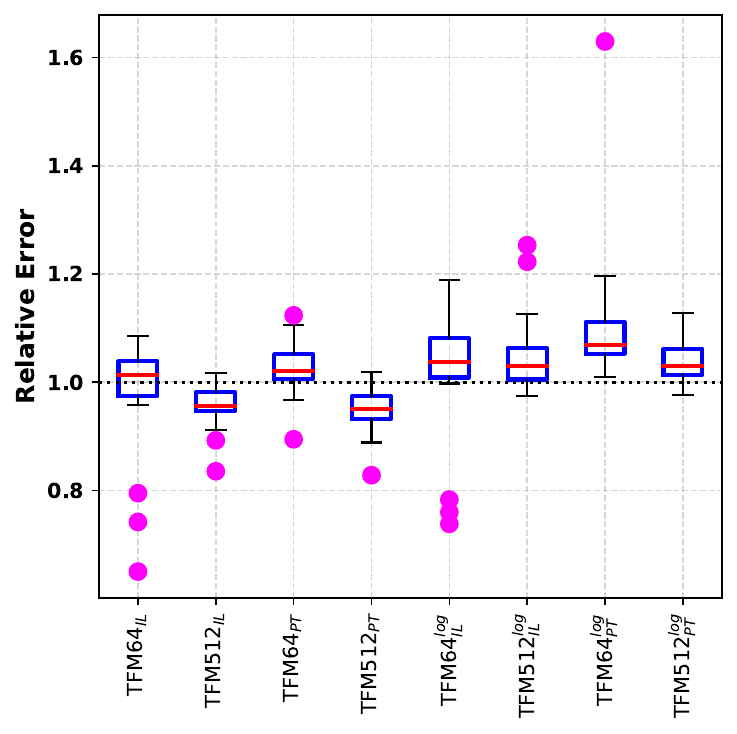}
\label{qlike}}\caption{{\small Visualization of Cross-Sectional Out-of-Sample Relative Error Distribution for six errors considered in the study: (a) MSE, (b) MAE, (c) MDA, (d) MAPE, (e) sMAPE and (f) Qlike. For each error, we construct a boxplot for the one-day-ahead out-of-sample forecast error of models ($\mathrm{TFM64}_{IL}$, $\mathrm{TFM512}_{IL}$, $\mathrm{TFM64}_{PT}$, $\mathrm{TFM512}_{PT}$, $\mathrm{TFM64}^{log}_{IL}$, $\mathrm{TFM512}^{log}_{IL}$, $\mathrm{TFM64}^{log}_{PT}$, $\mathrm{TFM512}^{log}_{PT}$)  relative to $\mathrm{CHAR}^{log}$ model. The sample includes 21 relative errors per model, corresponding to the number of stocks in our empirical analysis. The central line represents the median MSE, while the lower and upper edges of the box denote the inter-quartile range. The whiskers extend to the most extreme data points not identified as outliers, denoted by circles.}}
\label{boxplots}
\end{center}
\end{figure}

In Figure \ref{boxplots}, we construct a boxplot of the relative errors for the 6 loss functions. These boxplots visualize relative errors of different TimesFM-based models, with context lengths of 64 and 512, evaluated across 21 stocks, using the CHAR$^{\text{log}}$ model as the benchmark. The dotted line at 1 in each graph represents the benchmark; values below 1 indicate better performance than the benchmark. The results reveal no single model dominates across all criteria, but several models stand out in specific dimensions. Model performance varies across loss functions, highlighting distinct strengths of different TimesFM configurations. For MSE and MDA, most models \---across both pretrained (PT) and incremental learning (IL) setups\--- perform comparably well, with relative errors close to or below one. In contrast, for MAPE, sMAPE and MAD, log-transformed variants such as {TFM64$^{\log}_{\text{IL}}$} and {TFM512$^{\log}_{\text{PT}}$} offer clear improvements, suggesting that the log transformation helps stabilize percentage-based errors. For Qlike, linear models---particularly {TFM512$_{\text{PT}}$} and {TFM512$_{\text{IL}}$}---consistently outperform. These findings suggest that no single model dominates universally; instead, optimal performance depends on the loss function in focus, with log transformations proving especially effective for scale-sensitive metrics and longer context improving distributional calibration. However, the two models that outshine in most cases are {TFM512$^{\log}_{\text{PT}}$} and {TFM64$^{\log}_{\text{IL}}$}, combining strong performance across both scale and distribution sensitive metrics.

\begin{figure}[h!]
\begin{center}
\subfigure[MSE]{%
\includegraphics[width=0.47\textwidth]{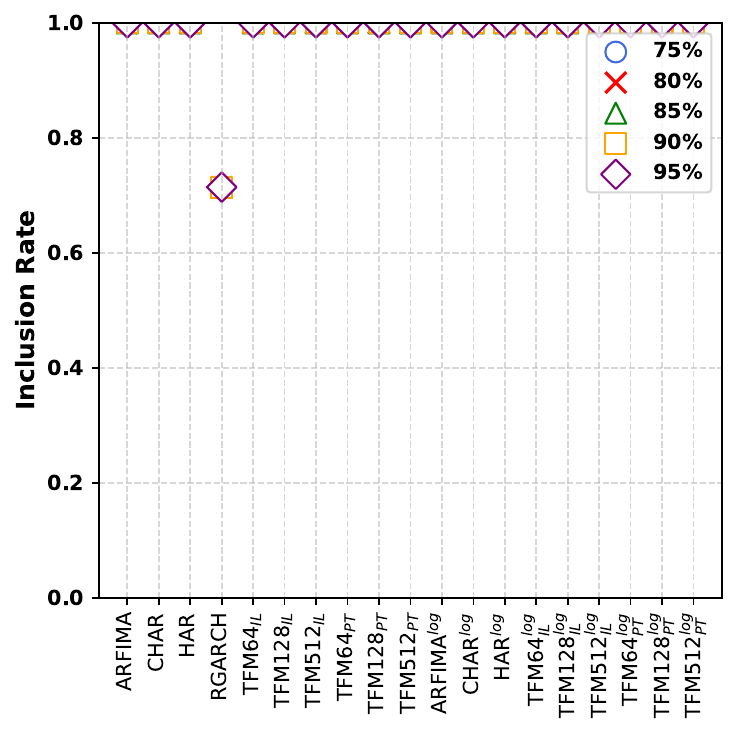}
\label{mse2}}
\quad
\subfigure[MAE]{%
\includegraphics[width=0.47\textwidth]{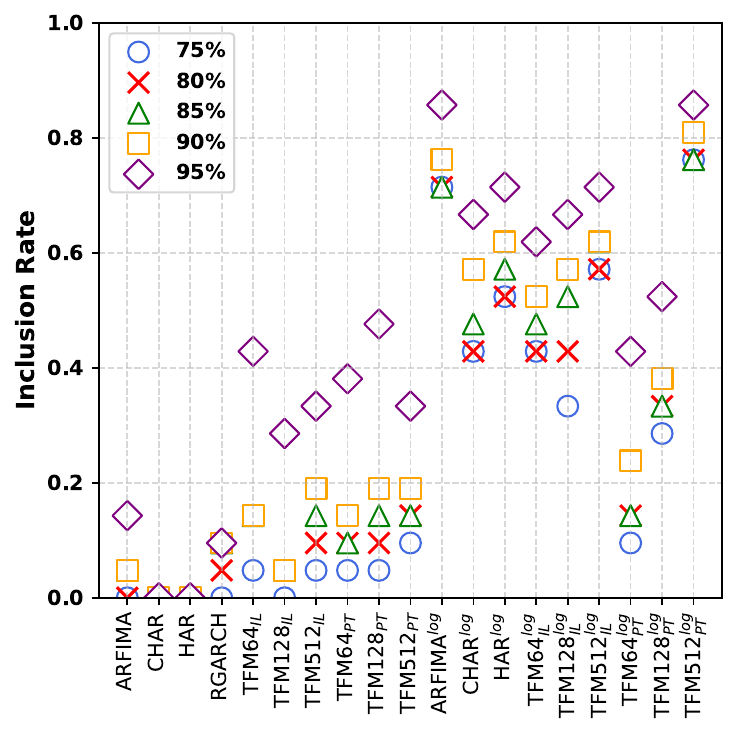}
\label{mad2}}
\quad
\subfigure[MDA]{%
\includegraphics[width=0.47\textwidth]{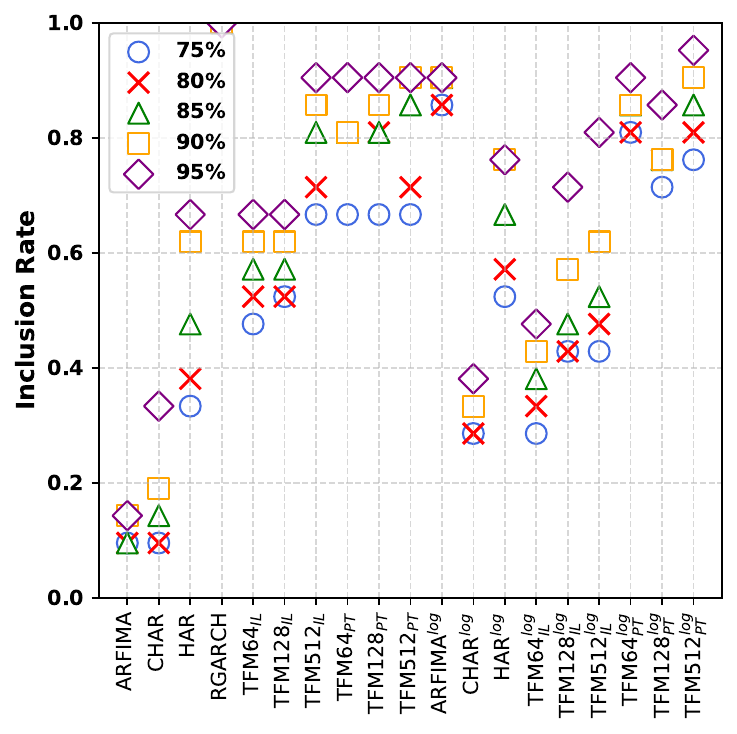}
\label{mda2}}
\quad
\subfigure[MAPE]{%
\includegraphics[width=0.47\textwidth]{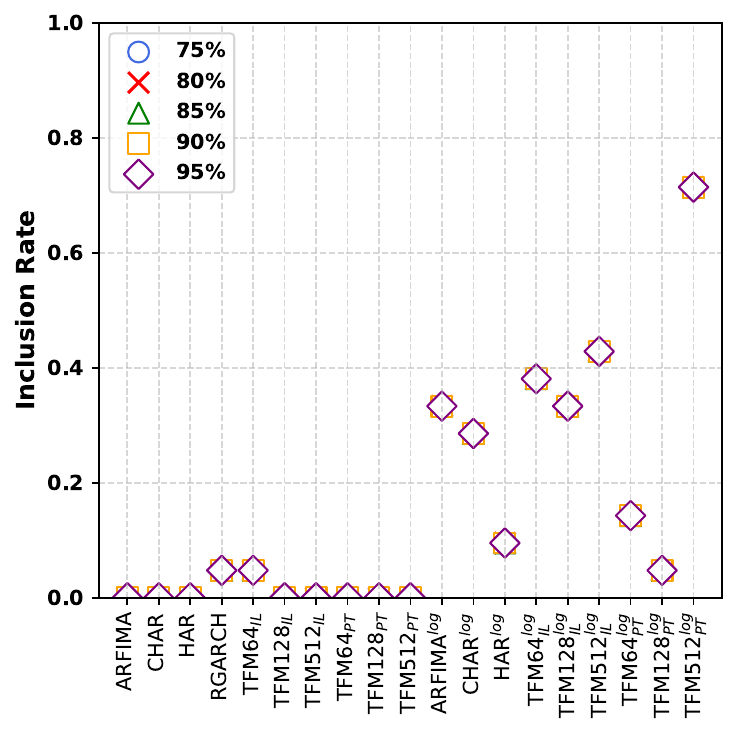}
\label{mape2}}
\quad
\subfigure[sMAPE]{%
\includegraphics[width=0.47\textwidth]{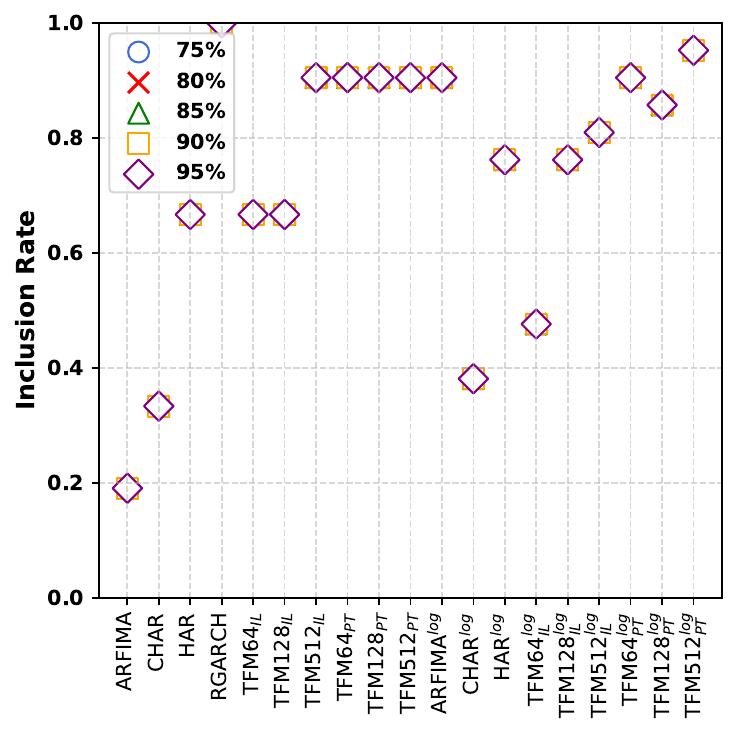}
\label{smape2}}
\quad
\subfigure[Qlike]{%
\includegraphics[width=0.47\textwidth]{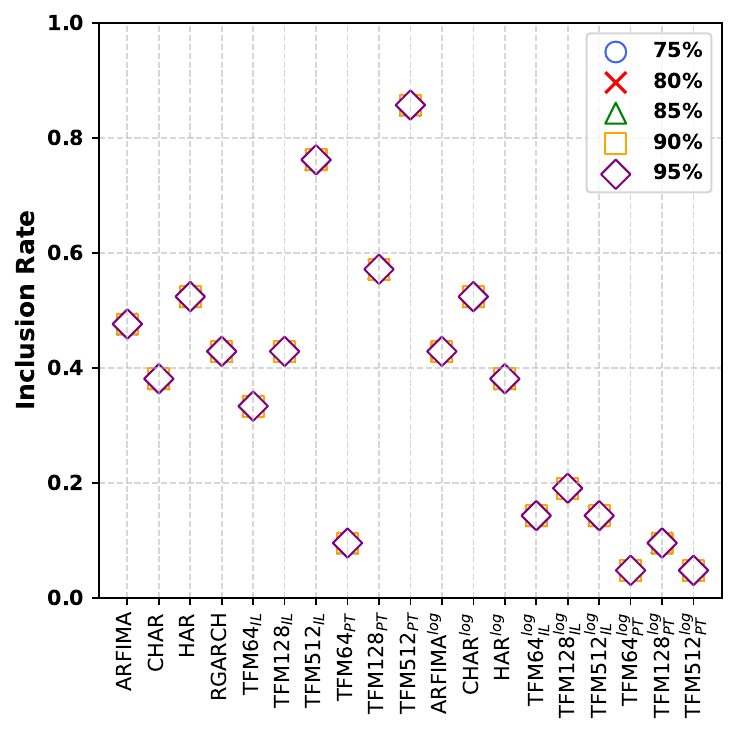}
\label{qlike2}}\caption{{\small Inclusion rate in the Model Inclusion Set (MCS): We present the results of the Model Confidence Set (MCS). This method identifies a subset of superior models by sequentially testing the null hypothesis of equal predictive accuracy, \(H_0: \mathrm{MSE}_{\text{model } i} = \mathrm{MSE}_{\text{model } j}.
\) We report the results for confidence levels $75\%,80\%,85\%,90\%$ and $95\%$, where the numbers on the y-axis indicate the percentage of times a given model is retained in the model confidence set.}}
\label{mcs}
\end{center}
\end{figure}

In Figure \ref{mcs}, we compute the Model Confidence Set (MCS) at different confidence levels $\in \{ 75\%, 80\%, 85\%, 90\% 95\%\}$. The figure illustrates the percentage of times each model was retained in the final set across the 21 stocks. The MCS results highlight that only a subset of models are consistently retained across stocks and loss functions. TFM64$^{\log}_{\text{IL}}$ and TFM512$^{\log}_{\text{PT}}$ stand out, achieving high inclusion rates \--- for MSE, MAD, MAPE, MDA and sMAPE. For Qlike, the inclusion rate is the highest for TFM512$^{_{\text{PT}}}$. Overall, TFM64$^{\log}_{\text{IL}}$ and TFM512$^{\log}_{\text{PT}}$ emerge as the most robust models across loss functions and confidence levels.

\begin{figure}[h!]
\begin{center}
\subfigure[Relative MSE]{%
\includegraphics[width=0.47\textwidth]{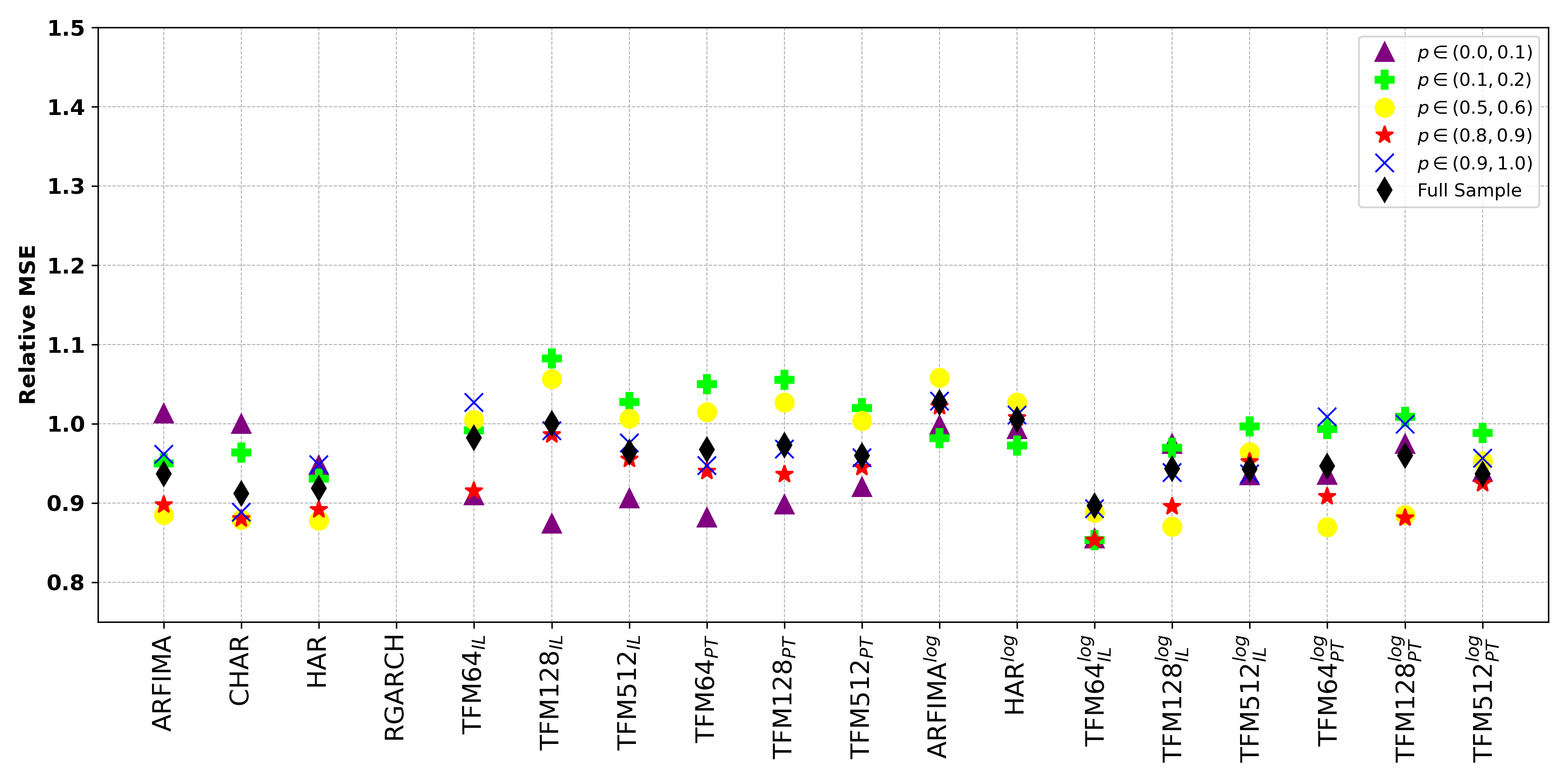}
\label{mse3}}
\quad
\subfigure[Relative MAE]{%
\includegraphics[width=0.47\textwidth]{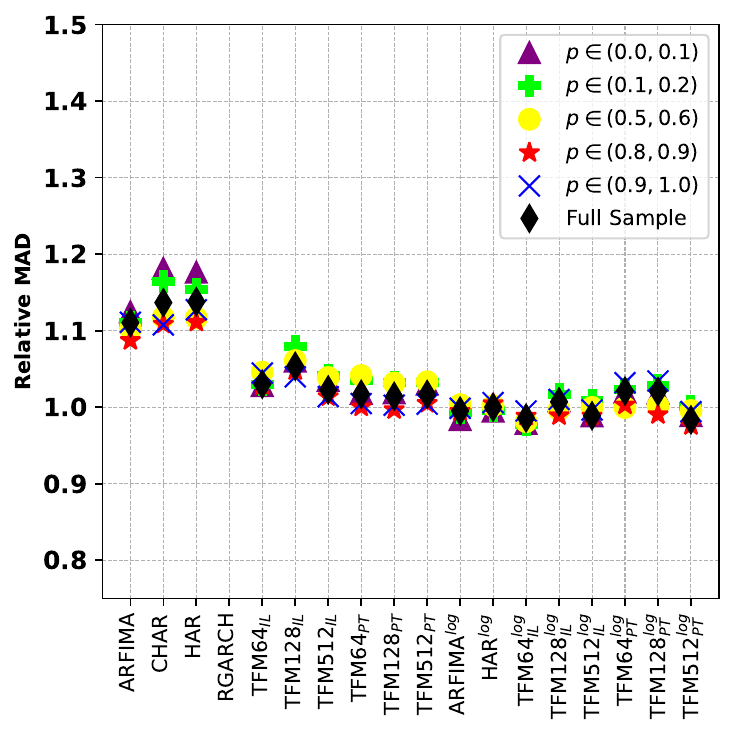}
\label{mad3}}
\quad
\subfigure[Relative MDA]{%
\includegraphics[width=0.47\textwidth]{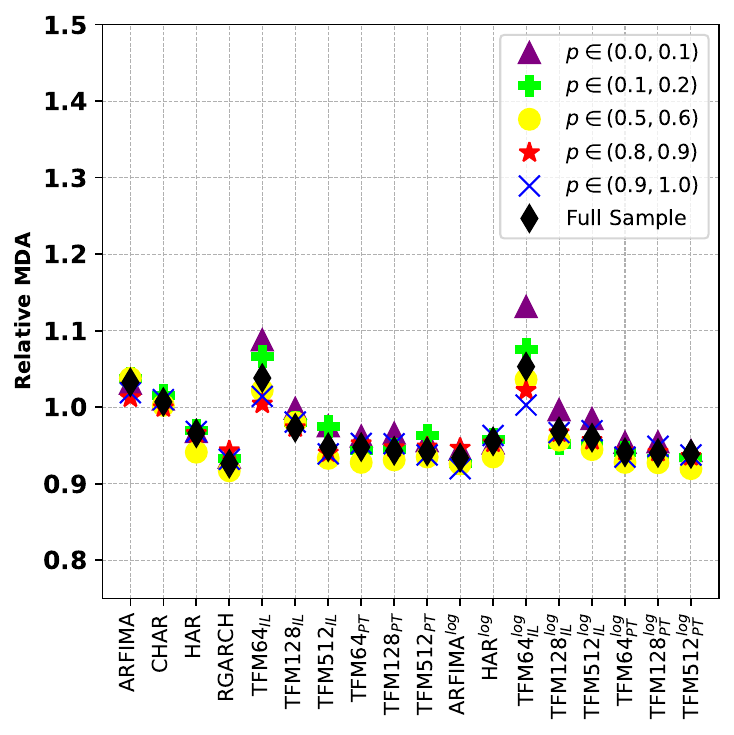}
\label{mda3}}
\quad
\subfigure[Relative MAPE]{%
\includegraphics[width=0.47\textwidth]{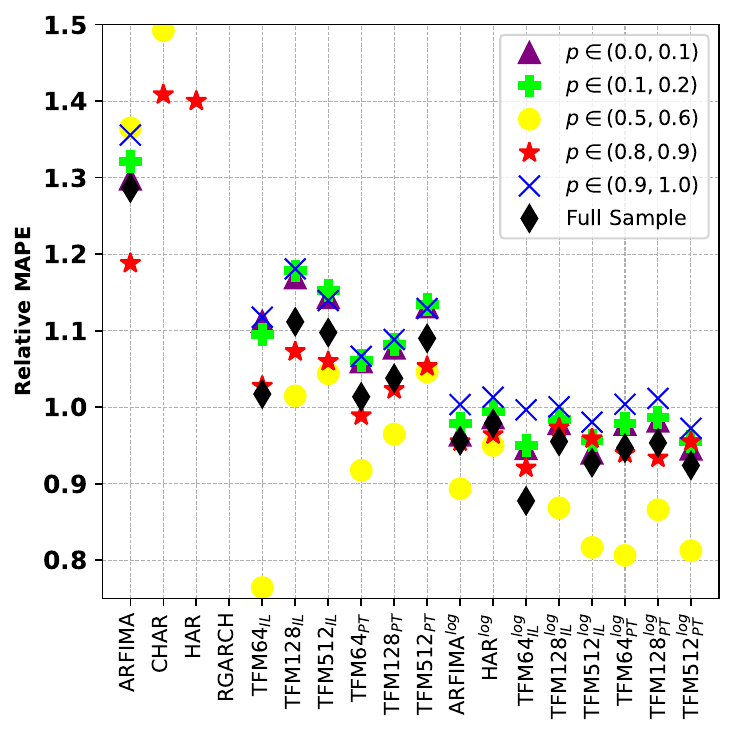}
\label{mape3}}
\quad
\subfigure[Relative sMAPE]{%
\includegraphics[width=0.47\textwidth]{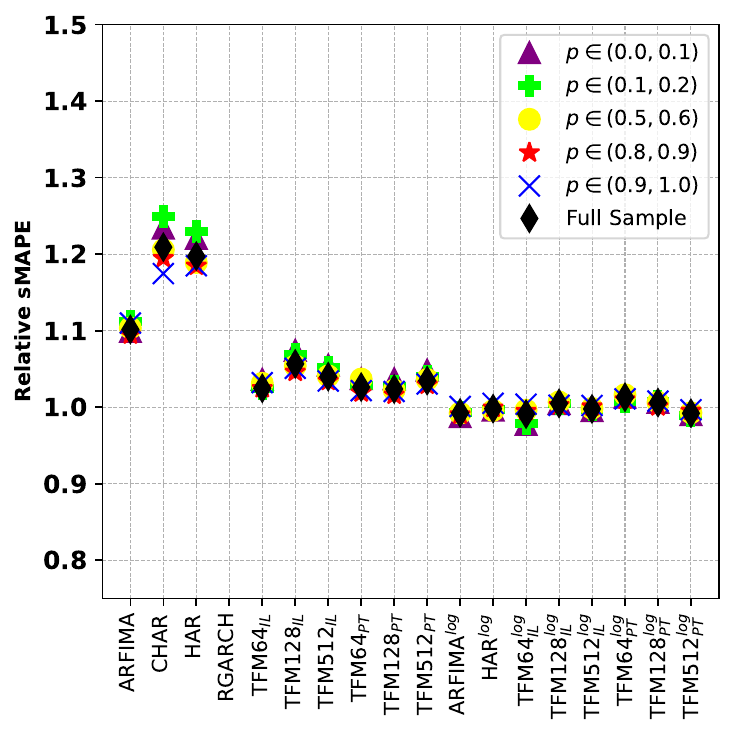}
\label{smape3}}
\quad
\subfigure[Relative Qlike]{%
\includegraphics[width=0.47\textwidth]{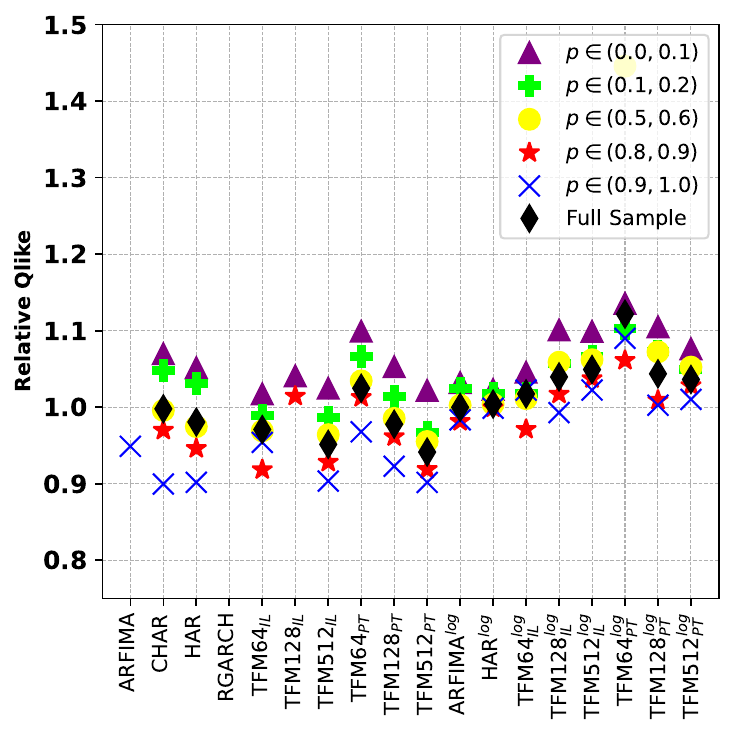}
\label{qlike3}}\caption{{\small We partition the test set into deciles based on the observed daily realized variance. Specifically, the first subsample contains the 10\% of observations with the lowest realized variance in the out-of-sample period and is labeled 
$(0.0,0.1)$. In this figure, we present the out-of-sample forecast errors for some deciles (denoted by $p$), reported relative to the $\mathrm{CHAR}^{\log}$ model, across all remaining 18 models.}}
\label{distribution}
\end{center}
\end{figure}

To gain insights into the forecasting performance of the models across different states of the volatility distribution, we partition the test set into deciles based on the observed daily realized variance. Specifically, the first subsample consists of the 10\% smallest realized variance values in the out-of-sample period, and is denoted as (0.0,0.1). In Figure \ref{distribution}, we present the results for selected low, medium, and high deciles for all 6 loss functions. All models are evaluated relative to the CHAR<sup>log</sup> benchmark, with the dotted line at one indicating equal performance. Values below one signify improved forecasting accuracy compared to CHAR$^{\text{log}}$. The figure reveals that the TimesFM models, particularly TFM64$^{\log}_{\text{IL}}$and TTFM512$^{_{\text{PT}}}$, consistently achieve lower relative errors across loss functions such as MSE, QLIKE, and sMAPE. These improvements are evident across volatility regimes, including both the lowest and highest deciles. Overall, the figure highlights that log-transformed and incrementally fine-tuned foundation models deliver stable and superior predictive performance relative to traditional approaches, especially in volatile environments.




\subsection{Average Losses}

The average losses across all assets, reported in Table \ref{tab:averageloss}, reinforce the insights derived from the skill score analysis. TimesFM models, particularly those employing incremental learning (IL), exhibit lower losses across key metrics such as MSE, MAD, and MAPE, confirming their superior forecasting accuracy. Notably, TFM64$^{\text{log}}_{\text{IL}}$ achieves the lowest MSE and MAPE values while also attaining the highest directional accuracy as measured by MDA, further supporting its identification as the top-performing model.  In contrast, traditional econometric models such as ARFIMA, CHAR, and HAR exhibit higher forecast errors across multiple loss functions. ARFIMA, in particular, struggles with capturing volatility patterns effectively, as evidenced by its higher Qlike values. CHAR and HAR models also underperform, showing higher MAPE and MAD values, reinforcing their limitations in minimizing absolute forecast errors. Among these models, the log-transformed variants slightly improve performance, but they remain inferior to TimesFM models. The results also highlight that the TimesFM model with zero-shot learning with longer context lengths, i.e., TFM512$_{\text{PT}}$, performs competitively in certain metrics, particularly Qlike, aligning with the observed trend that larger context lengths stabilize forecasts. These findings consistently validate the skill score-based ranking, demonstrating the robustness of TimesFM models in realized volatility forecasting.

\begin{table}[ht!]
\centering
\caption{Average Errors for One-day-ahead forecasts: We report cross-sectional (21 stocks) average of the the out-of-sample realized variance forecast error of each model. }\label{tab:averageloss}
\resizebox{.50\textwidth}{!}{
\begin{tabular}{@{}c | cccccc}
\toprule
Model  & \multicolumn{1}{c}{MSE} 
       & \multicolumn{1}{c}{MAD} 
       & \multicolumn{1}{c}{Qlike} 
       & \multicolumn{1}{c}{MAPE} 
       & \multicolumn{1}{c}{MDA}  
       & \multicolumn{1}{c}{sMAPE} \\
\midrule
ARFIMA            & 0.00120 & 0.01055 & 0.96190  & 75.005  & 37.888  & 44.828  \\
CHAR              & 0.00117 & 0.01081 & 0.20435  & 89.967  & 36.988  & 49.288  \\
HAR               & 0.00118 & 0.01082 & 0.20083  & 90.895  & 35.475  & 48.791  \\
RGARCH            & 0.00804 & 0.03359 & 0.53269  & 341.784 & 34.023  & 69.407  \\
TFM64$_{\text{IL}}$      & 0.00126 & 0.00979 & 0.19860  & 59.429  & 38.058  & 41.710  \\
TFM128$_{\text{IL}}$     & 0.00128 & 0.01001 & 0.90990  & 64.853  & 35.787  & 42.982  \\
TFM512$_{\text{IL}}$     & 0.00124 & 0.00972 & 0.19457  & 64.028  & 34.813  & 42.305  \\
TFM64$_{\text{PT}}$      & 0.00124 & 0.00966 & 0.20949  & 59.163  & 34.830  & 41.757  \\
TFM128$_{\text{PT}}$     & 0.00125 & 0.00964 & 0.19994  & 60.569  & 34.645  & 41.658  \\
TFM512$_{\text{PT}}$     & 0.00123 & 0.00966 & \textbf{0.19253}  & 63.584  & 34.618  & 42.096  \\
\midrule
ARFIMA$^{\text{log}}$    & 0.00132 & 0.00945 & 0.20441  & 55.814  & 34.287  & 40.394  \\
CHAR$^{\text{log}}$      & 0.00128 & 0.00950 & 0.20437  & 58.426  & 36.736  & 40.730  \\
HAR$^{\text{log}}$       & 0.00129 & 0.00950 & 0.20524  & 57.139  & 35.094  & 40.627  \\
TFM64$^{\text{log}}_{\text{IL}}$  & \textbf{0.00115} & 0.00937 & 0.20805  & \textbf{51.334}  & \textbf{38.592}  &\textbf{ 40.343}  \\
TFM128$^{\text{log}}_{\text{IL}}$ & 0.00121 & 0.00957 & 0.21256  & 55.754  & 35.583  & 40.895  \\
TFM512$^{\text{log}}_{\text{IL}}$ & 0.00121 & 0.00940 & 0.21460  & 54.163  & 35.323  & 40.615  \\
TFM64$^{\text{log}}_{\text{PT}}$  & 0.00121 & 0.00970 & 0.22898  & 55.265  & 34.565  & 41.220  \\
TFM128$^{\text{log}}_{\text{PT}}$ & 0.00123 & 0.00970 & 0.21330  & 55.669  & 34.556  & 40.939  \\
TFM512$^{\text{log}}_{\text{PT}}$ & 0.00120 & \textbf{0.00934} & 0.21180  & 53.937  & 34.494  & 40.370  \\
\bottomrule
\end{tabular}}
\end{table}

\subsection{Is there any statistically better model?}

To statistically assess the relative predictive performance of the models and determine whether any model significantly outperforms another—particularly whether TimesFM-based models outperform or at least match the econometric benchmarks—we employ the Diebold-Mariano (DM) and Giacomini-White (GW) tests. Tables \ref{tab:mse}-\ref{tab:smape} summarize the pairwise test results, where each model is represented by a block of two rows: the first contains p-values from the DM test and the second from the GW test. The test statistics are computed based on the mean difference in forecast errors between the model in the row and the model in the column. A low p-value indicates that the model in the row performs statistically worse than the model in the column. To illustrate how the table should be interpreted, consider the first row block corresponding to the ARFIMA model. For instance, in Table \ref{tab:mse}, the DM test reports p-values below 0.05 in two comparisons (and below 0.01 in one), indicating inferior predictive performance relative to the compared models in those cases. However, our goal is not to interpret every comparison individually, but to focus on whether the TimesFM-based models exhibit consistently better—or at least statistically comparable—performance when benchmarked against traditional econometric models.

Among the TimesFM models, TFM$_{\text{IL}}^{\log 64}$ stands out. In all three error metrics (MSE, MAPE, and SMAPE), it exhibits p-values greater than 0.70 in both the DM and GW tests when compared with every other model, indicating that it was never statistically worse. This stability suggests robustness in both point and distributional forecasting accuracy. When focusing on the Q-like loss function, the TimesFM-based models again demonstrate strong performance. In particular, TFM$_{\text{IL}}^{\log 64}$ achieves p-values greater than 0.20 in comparisons with all econometric models except the HAR model, for which the DM and GW tests yield p-values around 0.09. This suggests a marginally weaker performance relative to HAR, though not statistically significant at conventional thresholds. More notably, TFM$_{\text{PT}}^{512}$ and TFM$_{\text{IL}}^{512}$ consistently achieve p-values below 0.10 in all pairwise comparisons against econometric models under both DM and GW tests. These results imply that, at the 90\% confidence level, the null hypothesis of equal predictive accuracy is rejected in favor of these TimesFM models. In other words, both variants are statistically superior to all classical econometric models under the Q-like criterion.

Interestingly, when evaluating predictive direction using the Mean Directional Accuracy (MDA), RGARCH emerges as highly resilient. As seen in the table, none of the models achieve low p-values against RGARCH. This robustness is likely attributed to the model’s ability to capture volatility clustering and regime shifts, which are common in financial return series. RGARCH’s dynamic structure makes it particularly effective at identifying directional movements, hence its dominance under MDA tests.

\begin{sidewaystable}[ht!]
    \centering
    \caption{One-day-ahead relative error: We report the out-of-sample realized variance forecast error (MSE in Panel A and MAE in Panel B) of each model listed in the selected column, relative to the benchmark model indicated in the selected row. Each value represents the cross-sectional average of these pairwise relative errors across all stocks.}
    \resizebox{1\textwidth}{!}{%
    \begin{tabular}{@{\extracolsep{4pt}}l cccccccccccccccccccc}
    \toprule
            & \multicolumn{19}{c}{Panel A: Mean Squared Error (MSE)}\\
\hline
        \textbf{Model} & ARFIMA & CHAR & HAR & RGARCH & TFM64$_{\text{IL}}$ & TFM128$_{\text{IL}}$ & TFM512$_{\text{IL}}$ & TFM64$_{\text{PT}}$ & TFM128$_{\text{PT}}$ & TFM512$_{\text{PT}}$ & ARFIMA$^{\text{log}}$ & CHAR$^{\text{log}}$ & HAR$^{\text{log}}$ & TFM64$^{\text{log}}_{\text{IL}}$ & TFM128$^{\text{log}}_{\text{IL}}$ & TFM512$^{\text{log}}_{\text{IL}}$ & TFM64$^{\text{log}}_{\text{PT}}$ & TFM128$^{\text{log}}_{\text{PT}}$ & TFM512$^{\text{log}}_{\text{PT}}$ \\

\hline
ARFIMA & 1 & 0.984 & 0.994 & 13.033 & 1.066 & 1.072 & 1.030 & 1.037 & 1.041 & 1.028 & 1.109 & 1.079 & 1.085 & 0.979 & 1.012 & 1.013 & 1.023 & 1.032 & 1.003 \\
        CHAR & 1.020 & 1 & 1.013 & 12.634 & 1.084 & 1.091 & 1.048 & 1.056 & 1.060 & 1.046 & 1.130 & 1.099 & 1.106 & 0.997 & 1.030 & 1.031 & 1.042 & 1.050 & 1.021 \\
        HAR & 1.007 & 0.990 & 1 & 12.738 & 1.072 & 1.078 & 1.036 & 1.044 & 1.047 & 1.034 & 1.116 & 1.086 & 1.092 & 0.985 & 1.018 & 1.020 & 1.029 & 1.038 & 1.009 \\
        RGARCH & 0.802 & 0.782 & 0.793 & 1 & 0.852 & 0.857 & 0.824 & 0.829 & 0.834 & 0.822 & 0.887 & 0.863 & 0.869 & 0.776 & 0.808 & 0.805 & 0.824 & 0.825 & 0.805 \\
        TFM64$_{\text{IL}}$ & 0.951 & 0.933 & 0.944 & 12.274 & 1 & 1.016 & 0.976 & 0.981 & 0.986 & 0.974 & 1.050 & 1.022 & 1.028 & 0.923 & 0.958 & 0.961 & 0.970 & 0.977 & 0.952 \\
        TFM128$_{\text{IL}}$ & 0.938 & 0.921 & 0.931 & 11.350 & 0.996 & 1 & 0.963 & 0.970 & 0.973 & 0.961 & 1.037 & 1.011 & 1.016 & 0.917 & 0.947 & 0.948 & 0.958 & 0.965 & 0.939 \\
        TFM512$_{\text{IL}}$ & 0.974 & 0.956 & 0.967 & 12.676 & 1.034 & 1.041 & 1 & 1.007 & 1.011 & 0.998 & 1.078 & 1.049 & 1.055 & 0.952 & 0.983 & 0.985 & 0.995 & 1.002 & 0.975 \\
        TFM64$_{\text{PT}}$ & 0.968 & 0.950 & 0.961 & 12.474 & 1.026 & 1.034 & 0.994 & 1 & 1.004 & 0.992 & 1.070 & 1.042 & 1.048 & 0.945 & 0.976 & 0.979 & 0.987 & 0.995 & 0.968 \\
        TFM128$_{\text{PT}}$ & 0.964 & 0.947 & 0.958 & 12.538 & 1.024 & 1.030 & 0.990 & 0.997 & 1 & 0.988 & 1.066 & 1.038 & 1.044 & 0.943 & 0.973 & 0.976 & 0.984 & 0.991 & 0.965 \\
TFM512$_{\text{PT}}$ & 0.976 & 0.958 & 0.969 & 12.580 & 1.036 & 1.042 & 1.002 & 1.009 & 1.012 & 1 & 1.079 & 1.050 & 1.056 & 0.953 & 0.985 & 0.987 & 0.997 & 1.004 & 0.977 \\
ARFIMA$^{\text{log}}$ & 0.909 & 0.893 & 0.902 & 12.620 & 0.964 & 0.971 & 0.934 & 0.940 & 0.943 & 0.932 & 1 & 0.975 & 0.980 & 0.889 & 0.918 & 0.920 & 0.929 & 0.935 & 0.911 \\
CHAR$^{\text{log}}$ & 0.932 & 0.915 & 0.925 & 12.071 & 0.989 & 0.997 & 0.957 & 0.964 & 0.967 & 0.955 & 1.028 & 1 & 1.006 & 0.911 & 0.942 & 0.943 & 0.953 & 0.960 & 0.933 \\
HAR$^{\text{log}}$ & 0.927 & 0.911 & 0.920 & 12.727 & 0.984 & 0.991 & 0.953 & 0.959 & 0.962 & 0.951 & 1.021 & 0.996 & 1 & 0.906 & 0.936 & 0.938 & 0.948 & 0.954 & 0.929 \\

TFM64$^{\text{log}}_{\text{IL}}$ & 1.037 & 1.018 & 1.030 & 12.697 & 1.095 & 1.110 & 1.067 & 1.073 & 1.079 & 1.065 & 1.150 & 1.118 & 1.125 & 1 & 1.046 & 1.049 & 1.058 & 1.070 & 1.039 \\
TFM128$^{\text{log}}_{\text{IL}}$ & 0.993 & 0.974 & 0.986 & 12.952 & 1.053 & 1.061 & 1.020 & 1.026 & 1.030 & 1.018 & 1.099 & 1.070 & 1.075 & 0.968 & 1 & 1.004 & 1.013 & 1.020 & 0.994 \\
TFM512$^{\text{log}}_{\text{IL}}$ & 0.990 & 0.972 & 0.984 & 11.872 & 1.053 & 1.059 & 1.018 & 1.025 & 1.029 & 1.016 & 1.098 & 1.067 & 1.074 & 0.968 & 1.001 & 1 & 1.012 & 1.020 & 0.991 \\
TFM64$^{\text{log}}_{\text{PT}}$ & 0.983 & 0.966 & 0.976 & 13.143 & 1.045 & 1.051 & 1.011 & 1.017 & 1.021 & 1.009 & 1.090 & 1.061 & 1.067 & 0.959 & 0.992 & 0.995 & 1 & 1.012 & 0.984 \\
TFM128$^{\text{log}}_{\text{PT}}$ & 0.975 & 0.957 & 0.968 & 12.380 & 1.034 & 1.041 & 1.001 & 1.007 & 1.010 & 0.999 & 1.078 & 1.050 & 1.055 & 0.952 & 0.982 & 0.986 & 0.994 & 1 & 0.976 \\
TFM512$^{\text{log}}_{\text{PT}}$ & 0.999 & 0.982 & 0.993 & 12.830 & 1.063 & 1.069 & 1.027 & 1.034 & 1.038 & 1.025 & 1.108 & 1.078 & 1.084 & 0.977 & 1.010 & 1.011 & 1.020 & 1.030 & 1 \\
        \midrule
        & \multicolumn{19}{c}{Panel B: Mean Absolute Error (MAE)}\\
        \midrule
        ARFIMA & 1 & 1.040 & 1.034 & 3.597 & 0.931 & 0.952 & 0.924 & 0.918 & 0.915 & 0.918 & 0.899 & 0.906 & 0.903 & 0.895 & 0.910 & 0.894 & 0.922 & 0.922 & 0.888 \\
        CHAR & 0.968 & 1 & 0.997 & 3.254 & 0.899 & 0.920 & 0.893 & 0.888 & 0.885 & 0.887 & 0.869 & 0.875 & 0.873 & 0.864 & 0.880 & 0.864 & 0.891 & 0.891 & 0.859 \\
        HAR & 0.970 & 1.004 & 1 & 3.297 & 0.902 & 0.922 & 0.895 & 0.889 & 0.887 & 0.889 & 0.871 & 0.878 & 0.875 & 0.866 & 0.882 & 0.866 & 0.893 & 0.893 & 0.861 \\
        RGARCH & 0.792 & 0.817 & 0.814 & 1 & 0.729 & 0.750 & 0.729 & 0.726 & 0.723 & 0.725 & 0.710 & 0.714 & 0.714 & 0.703 & 0.720 & 0.705 & 0.732 & 0.730 & 0.703 \\
        TFM64$_{\text{IL}}$ & 1.078 & 1.119 & 1.114 & 3.719 & 1 & 1.025 & 0.995 & 0.988 & 0.986 & 0.988 & 0.968 & 0.976 & 0.973 & 0.962 & 0.980 & 0.962 & 0.993 & 0.993 & 0.957 \\
        TFM128$_{\text{IL}}$ & 1.052 & 1.093 & 1.087 & 3.674 & 0.978 & 1 & 0.971 & 0.965 & 0.962 & 0.965 & 0.945 & 0.953 & 0.950 & 0.940 & 0.957 & 0.940 & 0.970 & 0.970 & 0.934 \\
        TFM512$_{\text{IL}}$ & 1.084 & 1.125 & 1.120 & 3.777 & 1.007 & 1.030 & 1 & 0.994 & 0.991 & 0.993 & 0.973 & 0.981 & 0.978 & 0.968 & 0.985 & 0.967 & 0.998 & 0.998 & 0.962 \\
        TFM64$_{\text{PT}}$ & 1.091 & 1.133 & 1.127 & 3.910 & 1.014 & 1.037 & 1.007 & 1 & 0.997 & 1.000 & 0.979 & 0.987 & 0.984 & 0.974 & 0.992 & 0.974 & 1.005 & 1.005 & 0.968 \\
        TFM128$_{\text{PT}}$ & 1.094 & 1.136 & 1.130 & 3.916 & 1.016 & 1.040 & 1.010 & 1.003 & 1 & 1.003 & 0.982 & 0.990 & 0.987 & 0.977 & 0.994 & 0.976 & 1.007 & 1.007 & 0.971 \\
        TFM512$_{\text{PT}}$ & 1.091 & 1.133 & 1.127 & 3.890 & 1.014 & 1.037 & 1.007 & 1.000 & 0.997 & 1 & 0.979 & 0.988 & 0.984 & 0.975 & 0.992 & 0.974 & 1.005 & 1.005 & 0.968 \\
        ARFIMA$^{\text{log}}$ & 1.114 & 1.156 & 1.151 & 3.981 & 1.035 & 1.059 & 1.028 & 1.021 & 1.018 & 1.021 & 1 & 1.008 & 1.005 & 0.995 & 1.013 & 0.994 & 1.026 & 1.026 & 0.988 \\
        CHAR$^{\text{log}}$ & 1.105 & 1.146 & 1.141 & 3.885 & 1.027 & 1.051 & 1.020 & 1.013 & 1.011 & 1.013 & 0.992 & 1 & 0.997 & 0.988 & 1.005 & 0.986 & 1.018 & 1.018 & 0.981 \\
        HAR$^{\text{log}}$ & 1.108 & 1.150 & 1.145 & 3.960 & 1.030 & 1.054 & 1.023 & 1.016 & 1.013 & 1.016 & 0.995 & 1.003 & 1 & 0.990 & 1.008 & 0.989 & 1.021 & 1.021 & 0.983 \\
           TFM64$^{\text{log}}_{\text{IL}}$ & 1.122 & 1.164 & 1.159 & 3.937 & 1.041 & 1.066 & 1.036 & 1.029 & 1.026 & 1.028 & 1.007 & 1.015 & 1.013 & 1 & 1.020 & 1.001 & 1.034 & 1.034 & 0.995 \\
        TFM128$^{\text{log}}_{\text{IL}}$ & 1.100 & 1.142 & 1.136 & 3.904 & 1.022 & 1.046 & 1.016 & 1.009 & 1.006 & 1.009 & 0.988 & 0.996 & 0.993 & 0.983 & 1 & 0.982 & 1.013 & 1.013 & 0.976 \\
        TFM512$^{\text{log}}_{\text{IL}}$ & 1.120 & 1.163 & 1.157 & 3.944 & 1.041 & 1.066 & 1.034 & 1.027 & 1.025 & 1.027 & 1.006 & 1.014 & 1.011 & 1.001 & 1.019 & 1 & 1.032 & 1.032 & 0.994 \\
        TFM64$^{\text{log}}_{\text{PT}}$ & 1.086 & 1.127 & 1.122 & 3.902 & 1.009 & 1.033 & 1.002 & 0.996 & 0.993 & 0.995 & 0.975 & 0.983 & 0.980 & 0.970 & 0.987 & 0.969 & 1 & 1.000 & 0.964 \\
        TFM128$^{\text{log}}_{\text{PT}}$ & 1.086 & 1.127 & 1.122 & 3.879 & 1.009 & 1.033 & 1.003 & 0.996 & 0.993 & 0.996 & 0.975 & 0.983 & 0.980 & 0.971 & 0.987 & 0.969 & 1.000 & 1 & 0.964 \\
        TFM512$^{\text{log}}_{\text{PT}}$ & 1.127 & 1.170 & 1.164 & 4.029 & 1.047 & 1.072 & 1.040 & 1.033 & 1.030 & 1.033 & 1.012 & 1.020 & 1.017 & 1.007 & 1.025 & 1.006 & 1.038 & 1.038 & 1 \\
       
    \bottomrule
    \end{tabular}%
    }
    \label{tab:lossa}
\end{sidewaystable}

\begin{sidewaystable}[ht!]
    \centering
    \caption{One-day-ahead relative error: We report the out-of-sample realized variance forecast error (MDA in Panel A and MAPE in Panel B) of each model listed in the selected column, relative to the benchmark model indicated in the selected row. Each value represents the cross-sectional average of these pairwise relative errors across all stocks.}
    \resizebox{1\textwidth}{!}{%
    \begin{tabular}{@{\extracolsep{4pt}}l cccccccccccccccccccc}
    \toprule
                & \multicolumn{19}{c}{Panel A: Mean Directional Accuracy (MDA)}\\
\hline
        \textbf{Model} & ARFIMA & CHAR & HAR & RGARCH & TFM64$_{\text{IL}}$ & TFM128$_{\text{IL}}$ & TFM512$_{\text{IL}}$ & TFM64$_{\text{PT}}$ & TFM128$_{\text{PT}}$ & TFM512$_{\text{PT}}$ & ARFIMA$^{\text{log}}$ & CHAR$^{\text{log}}$ & HAR$^{\text{log}}$ & TFM64$^{\text{log}}_{\text{IL}}$ & TFM128$^{\text{log}}_{\text{IL}}$ & TFM512$^{\text{log}}_{\text{IL}}$ & TFM64$^{\text{log}}_{\text{PT}}$ & TFM128$^{\text{log}}_{\text{PT}}$ & TFM512$^{\text{log}}_{\text{PT}}$ \\
        \hline
        ARFIMA & 1 & 0.788 & 0.779 & 2.943 & 0.764 & 2.691 & 0.747 & 0.801 & 0.767 & 0.741 & 0.780 & 0.775 & 0.782 & 0.801 & 0.810 & 0.820 & 0.841 & 0.812 & 0.806 \\
        CHAR & 4.606 & 1 & 0.987 & 3.053 & 0.974 & 3.953 & 0.955 & 1.024 & 0.980 & 0.946 & 0.999 & 1.001 & 1.003 & 1.018 & 1.036 & 1.049 & 1.113 & 1.041 & 1.036 \\
        HAR & 4.641 & 1.015 & 1 & 3.050 & 0.988 & 4.001 & 0.966 & 1.036 & 0.991 & 0.957 & 1.010 & 1.012 & 1.014 & 1.031 & 1.048 & 1.062 & 1.123 & 1.053 & 1.048 \\
        RGARCH & 4.093 & 0.790 & 0.775 & 1 & 0.758 & 2.851 & 0.748 & 0.803 & 0.766 & 0.740 & 0.782 & 0.785 & 0.788 & 0.784 & 0.810 & 0.816 & 0.878 & 0.816 & 0.813 \\
        TFM64$_{\text{IL}}$ & 4.893 & 1.043 & 1.028 & 3.281 & 1 & 3.975 & 0.989 & 1.061 & 1.015 & 0.980 & 1.036 & 1.038 & 1.040 & 1.043 & 1.074 & 1.088 & 1.160 & 1.079 & 1.074 \\
        TFM128$_{\text{IL}}$ & 3.857 & 0.852 & 0.837 & 2.935 & 0.819 & 1 & 0.793 & 0.853 & 0.816 & 0.790 & 0.832 & 0.830 & 0.833 & 0.848 & 0.864 & 0.877 & 0.928 & 0.866 & 0.863 \\
        TFM512$_{\text{IL}}$ & 4.696 & 1.058 & 1.041 & 3.348 & 1.023 & 3.961 & 1 & 1.073 & 1.027 & 0.992 & 1.046 & 1.048 & 1.050 & 1.066 & 1.086 & 1.100 & 1.162 & 1.090 & 1.085 \\
        TFM64$_{\text{PT}}$ & 4.391 & 0.986 & 0.971 & 3.176 & 0.954 & 3.703 & 0.933 & 1 & 0.957 & 0.925 & 0.975 & 0.977 & 0.979 & 0.994 & 1.012 & 1.025 & 1.083 & 1.016 & 1.012 \\
        TFM128$_{\text{PT}}$ & 4.611 & 1.030 & 1.014 & 3.266 & 0.996 & 3.884 & 0.975 & 1.045 & 1 & 0.966 & 1.019 & 1.021 & 1.023 & 1.039 & 1.058 & 1.071 & 1.132 & 1.062 & 1.057 \\
        TFM512$_{\text{PT}}$ & 4.763 & 1.067 & 1.049 & 3.354 & 1.032 & 4.052 & 1.009 & 1.082 & 1.035 & 1 & 1.055 & 1.057 & 1.059 & 1.075 & 1.096 & 1.109 & 1.172 & 1.100 & 1.095 \\
        ARFIMA$^{\text{log}}$ & 4.688 & 1.037 & 1.021 & 3.345 & 1.003 & 3.965 & 0.983 & 1.053 & 1.007 & 0.972 & 1 & 1.002 & 1.004 & 1.020 & 1.038 & 1.051 & 1.113 & 1.042 & 1.037 \\
CHAR$^{\text{log}}$ & 4.737 & 1.049 & 1.033 & 3.381 & 1.015 & 4.006 & 0.995 & 1.065 & 1.019 & 0.984 & 1.011 & 1 & 1.002 & 1.031 & 1.050 & 1.063 & 1.126 & 1.054 & 1.048 \\
HAR$^{\text{log}}$ & 4.669 & 1.033 & 1.017 & 3.332 & 0.999 & 3.971 & 0.980 & 1.050 & 1.004 & 0.969 & 0.997 & 0.999 & 1 & 1.018 & 1.036 & 1.049 & 1.110 & 1.040 & 1.034 \\
TFM64$^{\text{log}}_{\text{IL}}$ & 4.983 & 1.109 & 1.093 & 3.533 & 1.074 & 4.187 & 1.055 & 1.126 & 1.080 & 1.045 & 1.081 & 1.083 & 1.085 & 1 & 1.103 & 1.116 & 1.187 & 1.108 & 1.103 \\
TFM128$^{\text{log}}_{\text{IL}}$ & 4.861 & 1.084 & 1.069 & 3.470 & 1.048 & 4.103 & 1.031 & 1.101 & 1.055 & 1.020 & 1.058 & 1.060 & 1.062 & 1.077 & 1 & 1.092 & 1.160 & 1.086 & 1.081 \\
TFM512$^{\text{log}}_{\text{IL}}$ & 4.810 & 1.072 & 1.057 & 3.447 & 1.037 & 4.060 & 1.019 & 1.089 & 1.043 & 1.008 & 1.047 & 1.049 & 1.051 & 1.065 & 1.083 & 1 & 1.148 & 1.074 & 1.069 \\
TFM64$^{\text{log}}_{\text{PT}}$ & 4.562 & 1.015 & 1.000 & 3.321 & 0.983 & 3.896 & 0.963 & 1.038 & 0.992 & 0.958 & 1.005 & 1.007 & 1.009 & 1.024 & 1.043 & 1.056 & 1 & 1.047 & 1.042 \\
TFM128$^{\text{log}}_{\text{PT}}$ & 4.705 & 1.046 & 1.030 & 3.374 & 1.013 & 4.003 & 0.993 & 1.065 & 1.019 & 0.984 & 1.037 & 1.039 & 1.041 & 1.056 & 1.075 & 1.088 & 1.159 & 1 & 1.074 \\
TFM512$^{\text{log}}_{\text{PT}}$ & 4.760 & 1.059 & 1.043 & 3.398 & 1.025 & 4.041 & 1.005 & 1.076 & 1.030 & 0.995 & 1.049 & 1.051 & 1.053 & 1.068 & 1.087 & 1.101 & 1.172 & 1.089 & 1 \\
 \midrule
        & \multicolumn{19}{c}{Panel B: Mean Absolute Percentage Error (MAPE)}\\
        \midrule
        ARFIMA & 1 & 1.226 & 1.206 & 4.957 & 0.836 & 0.888 & 0.868 & 0.811 & 0.825 & 0.860 & 0.757 & 0.781 & 0.770 & 0.740 & 0.761 & 0.741 & 0.760 & 0.763 & 0.737 \\
        CHAR & 0.841 & 1 & 0.998 & 3.663 & 0.701 & 0.746 & 0.727 & 0.680 & 0.693 & 0.722 & 0.634 & 0.652 & 0.645 & 0.620 & 0.639 & 0.621 & 0.638 & 0.640 & 0.618 \\
        HAR & 0.840 & 1.011 & 1 & 3.713 & 0.701 & 0.747 & 0.727 & 0.680 & 0.692 & 0.721 & 0.634 & 0.652 & 0.644 & 0.621 & 0.639 & 0.621 & 0.637 & 0.640 & 0.618 \\
        RGARCH & 0.794 & 0.973 & 0.954 & 1 & 0.648 & 0.696 & 0.684 & 0.641 & 0.652 & 0.683 & 0.597 & 0.612 & 0.607 & 0.581 & 0.603 & 0.586 & 0.604 & 0.605 & 0.583 \\
        TFM64$_{\text{IL}}$ & 1.211 & 1.479 & 1.459 & 5.381 & 1 & 1.069 & 1.047 & 0.977 & 0.995 & 1.038 & 0.912 & 0.942 & 0.929 & 0.886 & 0.917 & 0.892 & 0.915 & 0.919 & 0.887 \\
        TFM128$_{\text{IL}}$ & 1.133 & 1.386 & 1.367 & 5.281 & 0.942 & 1 & 0.981 & 0.916 & 0.933 & 0.974 & 0.855 & 0.883 & 0.871 & 0.834 & 0.860 & 0.837 & 0.859 & 0.863 & 0.833 \\
        TFM512$_{\text{IL}}$ & 1.156 & 1.411 & 1.391 & 5.260 & 0.962 & 1.024 & 1 & 0.934 & 0.951 & 0.992 & 0.872 & 0.900 & 0.888 & 0.852 & 0.877 & 0.854 & 0.876 & 0.879 & 0.849 \\
        TFM64$_{\text{PT}}$ & 1.238 & 1.514 & 1.492 & 6.077 & 1.030 & 1.097 & 1.072 & 1 & 1.018 & 1.062 & 0.934 & 0.964 & 0.950 & 0.911 & 0.939 & 0.913 & 0.937 & 0.941 & 0.908 \\
        TFM128$_{\text{PT}}$ & 1.216 & 1.487 & 1.464 & 5.938 & 1.012 & 1.078 & 1.053 & 0.982 & 1 & 1.043 & 0.917 & 0.946 & 0.933 & 0.895 & 0.922 & 0.897 & 0.920 & 0.924 & 0.892 \\
        TFM512$_{\text{PT}}$ & 1.165 & 1.424 & 1.403 & 5.655 & 0.971 & 1.033 & 1.009 & 0.942 & 0.959 & 1 & 0.879 & 0.908 & 0.895 & 0.859 & 0.884 & 0.861 & 0.883 & 0.887 & 0.856 \\
         ARFIMA$^{\text{log}}$ & 1.328 & 1.621 & 1.597 & 6.393 & 1.105 & 1.176 & 1.149 & 1.072 & 1.092 & 1.139 & 1 & 1.031 & 1.018 & 0.977 & 1.006 & 0.979 & 1.004 & 1.009 & 0.973 \\
CHAR$^{\text{log}}$ & 1.294 & 1.572 & 1.552 & 5.930 & 1.077 & 1.147 & 1.119 & 1.045 & 1.063 & 1.110 & 0.973 & 1 & 0.990 & 0.952 & 0.980 & 0.953 & 0.978 & 0.982 & 0.948 \\
HAR$^{\text{log}}$ & 1.305 & 1.591 & 1.568 & 6.176 & 1.086 & 1.157 & 1.129 & 1.054 & 1.073 & 1.120 & 0.983 & 1.013 & 1 & 0.961 & 0.989 & 0.962 & 0.987 & 0.991 & 0.957 \\
TFM64$^{\text{log}}_{\text{IL}}$ & 1.376 & 1.678 & 1.658 & 6.614 & 1.136 & 1.214 & 1.189 & 1.108 & 1.129 & 1.179 & 1.035 & 1.070 & 1.054 & 1 & 1.040 & 1.011 & 1.037 & 1.042 & 1.006 \\
TFM128$^{\text{log}}_{\text{IL}}$ & 1.321 & 1.617 & 1.592 & 6.511 & 1.099 & 1.170 & 1.143 & 1.066 & 1.086 & 1.133 & 0.995 & 1.026 & 1.013 & 0.972 & 1 & 0.974 & 0.999 & 1.003 & 0.968 \\
TFM512$^{\text{log}}_{\text{IL}}$ & 1.357 & 1.659 & 1.634 & 6.679 & 1.129 & 1.203 & 1.175 & 1.096 & 1.116 & 1.165 & 1.022 & 1.054 & 1.041 & 0.998 & 1.028 & 1 & 1.026 & 1.031 & 0.995 \\
TFM64$^{\text{log}}_{\text{PT}}$ & 1.323 & 1.617 & 1.593 & 6.578 & 1.101 & 1.172 & 1.145 & 1.068 & 1.087 & 1.135 & 0.997 & 1.028 & 1.014 & 0.972 & 1.002 & 0.975 & 1 & 1.004 & 0.969 \\
TFM128$^{\text{log}}_{\text{PT}}$ & 1.317 & 1.609 & 1.586 & 6.467 & 1.096 & 1.167 & 1.140 & 1.063 & 1.083 & 1.130 & 0.992 & 1.023 & 1.010 & 0.968 & 0.997 & 0.971 & 0.996 & 1 & 0.965 \\
TFM512$^{\text{log}}_{\text{PT}}$ & 1.365 & 1.668 & 1.643 & 6.695 & 1.135 & 1.209 & 1.181 & 1.102 & 1.122 & 1.171 & 1.028 & 1.060 & 1.046 & 1.003 & 1.034 & 1.005 & 1.032 & 1.036 & 1 \\
    \bottomrule
    \end{tabular}%
    }
    \label{tab:lossb}
\end{sidewaystable}

\begin{sidewaystable}[ht!]
    \centering
    \caption{One-day-ahead relative error: We report the out-of-sample realized variance forecast error (sMAPE in Panel A and QLIKE in Panel B) of each model listed in the selected column, relative to the benchmark model indicated in the selected row. Each value represents the cross-sectional average of these pairwise relative errors across all stocks.}
    \resizebox{1\textwidth}{!}{%
    \begin{tabular}{@{\extracolsep{4pt}}l cccccccccccccccccccc}
    \toprule
                & \multicolumn{19}{c}{Panel A: Symmetric Mean Absolute Percentage Error (sMAPE)}\\
\hline
        \textbf{Model} & ARFIMA & CHAR & HAR & RGARCH & TFM64$_{\text{IL}}$ & TFM128$_{\text{IL}}$ & TFM512$_{\text{IL}}$ & TFM64$_{\text{PT}}$ & TFM128$_{\text{PT}}$ & TFM512$_{\text{PT}}$ & ARFIMA$^{\text{log}}$ & CHAR$^{\text{log}}$ & HAR$^{\text{log}}$ & TFM64$^{\text{log}}_{\text{IL}}$ & TFM128$^{\text{log}}_{\text{IL}}$ & TFM512$^{\text{log}}_{\text{IL}}$ & TFM64$^{\text{log}}_{\text{PT}}$ & TFM128$^{\text{log}}_{\text{PT}}$ & TFM512$^{\text{log}}_{\text{PT}}$ \\
        \hline
        ARFIMA & 1 & 0.980 & 0.939 & 0.901 & 1.005 & 0.948 & 0.922 & 0.922 & 0.917 & 0.916 & 0.907 & 0.972 & 0.929 & 1.018 & 0.943 & 0.936 & 0.915 & 0.915 & 0.913 \\
        CHAR & 1.025 & 1 & 0.959 & 0.920 & 1.028 & 0.969 & 0.942 & 0.942 & 0.938 & 0.937 & 0.927 & 0.993 & 0.949 & 1.042 & 0.963 & 0.956 & 0.935 & 0.935 & 0.933 \\
        HAR & 1.068 & 1.044 & 1 & 0.959 & 1.071 & 1.010 & 0.982 & 0.982 & 0.977 & 0.976 & 0.966 & 1.036 & 0.989 & 1.086 & 1.004 & 0.996 & 0.975 & 0.975 & 0.973 \\
        RGARCH & 1.115 & 1.089 & 1.043 & 1 & 1.118 & 1.054 & 1.025 & 1.025 & 1.019 & 1.018 & 1.008 & 1.081 & 1.032 & 1.133 & 1.047 & 1.039 & 1.017 & 1.017 & 1.015 \\
        TFM64$_{\text{IL}}$ & 1.017 & 0.994 & 0.953 & 0.914 & 1 & 0.963 & 0.936 & 0.936 & 0.931 & 0.930 & 0.921 & 0.987 & 0.943 & 1.015 & 0.956 & 0.950 & 0.928 & 0.929 & 0.927 \\
        TFM128$_{\text{IL}}$ & 1.059 & 1.034 & 0.991 & 0.951 & 1.063 & 1 & 0.973 & 0.973 & 0.968 & 0.967 & 0.958 & 1.027 & 0.981 & 1.078 & 0.995 & 0.987 & 0.966 & 0.966 & 0.964 \\
        TFM512$_{\text{IL}}$ & 1.088 & 1.064 & 1.019 & 0.977 & 1.092 & 1.029 & 1 & 1.001 & 0.996 & 0.995 & 0.985 & 1.056 & 1.008 & 1.107 & 1.023 & 1.015 & 0.993 & 0.993 & 0.991 \\
        TFM64$_{\text{PT}}$ & 1.088 & 1.062 & 1.018 & 0.977 & 1.091 & 1.029 & 1.000 & 1 & 0.995 & 0.994 & 0.984 & 1.055 & 1.007 & 1.106 & 1.022 & 1.014 & 0.993 & 0.993 & 0.990 \\
        TFM128$_{\text{PT}}$ & 1.093 & 1.068 & 1.024 & 0.982 & 1.097 & 1.034 & 1.005 & 1.006 & 1 & 0.999 & 0.989 & 1.061 & 1.013 & 1.112 & 1.028 & 1.020 & 0.998 & 0.998 & 0.996 \\
        TFM512$_{\text{PT}}$ & 1.094 & 1.069 & 1.025 & 0.982 & 1.097 & 1.035 & 1.006 & 1.006 & 1.001 & 1 & 0.990 & 1.062 & 1.014 & 1.113 & 1.028 & 1.020 & 0.999 & 0.999 & 0.996 \\
ARFIMA$^{\text{log}}$ & 1.107 & 1.081 & 1.036 & 0.993 & 1.110 & 1.047 & 1.018 & 1.018 & 1.012 & 1.012 & 1 & 1.074 & 1.025 & 1.125 & 1.040 & 1.032 & 1.010 & 1.010 & 1.008 \\
CHAR$^{\text{log}}$ & 1.032 & 1.007 & 0.966 & 0.927 & 1.036 & 0.976 & 0.949 & 0.949 & 0.944 & 0.944 & 0.934 & 1 & 0.956 & 1.050 & 0.970 & 0.963 & 0.942 & 0.942 & 0.940 \\
HAR$^{\text{log}}$ & 1.080 & 1.055 & 1.011 & 0.969 & 1.083 & 1.022 & 0.993 & 0.993 & 0.988 & 0.987 & 0.977 & 1.048 & 1 & 1.098 & 1.015 & 1.007 & 0.986 & 0.986 & 0.983 \\
TFM64$^{\text{log}}_{\text{IL}}$ & 1.002 & 0.980 & 0.938 & 0.900 & 0.986 & 0.949 & 0.922 & 0.922 & 0.917 & 0.916 & 0.907 & 0.973 & 0.929 & 1 & 0.942 & 0.936 & 0.915 & 0.915 & 0.913 \\
TFM128$^{\text{log}}_{\text{IL}}$ & 1.065 & 1.040 & 0.997 & 0.956 & 1.068 & 1.007 & 0.979 & 0.979 & 0.974 & 0.973 & 0.963 & 1.033 & 0.986 & 1.083 & 1 & 0.993 & 0.971 & 0.972 & 0.969 \\
TFM512$^{\text{log}}_{\text{IL}}$ & 1.074 & 1.049 & 1.005 & 0.964 & 1.079 & 1.015 & 0.987 & 0.987 & 0.983 & 0.981 & 0.971 & 1.042 & 0.994 & 1.093 & 1.009 & 1 & 0.980 & 0.980 & 0.978 \\
TFM64$^{\text{log}}_{\text{PT}}$ & 1.096 & 1.071 & 1.026 & 0.984 & 1.099 & 1.036 & 1.008 & 1.008 & 1.003 & 1.002 & 0.991 & 1.063 & 1.015 & 1.114 & 1.030 & 1.022 & 1 & 1.000 & 0.998 \\
TFM128$^{\text{log}}_{\text{PT}}$ & 1.096 & 1.071 & 1.026 & 0.984 & 1.100 & 1.036 & 1.008 & 1.008 & 1.003 & 1.002 & 0.992 & 1.063 & 1.015 & 1.115 & 1.030 & 1.022 & 1.000 & 1 & 0.998 \\
TFM512$^{\text{log}}_{\text{PT}}$ & 1.098 & 1.073 & 1.028 & 0.986 & 1.102 & 1.039 & 1.010 & 1.010 & 1.005 & 1.004 & 0.993 & 1.065 & 1.017 & 1.117 & 1.032 & 1.024 & 1.002 & 1.002 & 1 \\
 \midrule
        & \multicolumn{19}{c}{Panel B: Quasi Likelihood (QLIKE)}\\
        \midrule
        ARFIMA & 1.000 & 1.103 & 1.091 & 1.573 & 0.932 & 0.959 & 0.944 & 0.933 & 0.930 & 0.940 & 0.903 & 0.911 & 0.908 & 0.903 & 0.914 & 0.908 & 0.921 & 0.915 & 0.902 \\
        CHAR & 0.916 & 1.000 & 0.993 & 1.431 & 0.853 & 0.878 & 0.864 & 0.853 & 0.852 & 0.860 & 0.826 & 0.833 & 0.830 & 0.826 & 0.836 & 0.831 & 0.843 & 0.837 & 0.826 \\
        HAR & 0.920 & 1.009 & 1.000 & 1.434 & 0.858 & 0.883 & 0.869 & 0.858 & 0.856 & 0.865 & 0.830 & 0.838 & 0.835 & 0.831 & 0.841 & 0.835 & 0.847 & 0.842 & 0.830 \\
        RGARCH & 0.808 & 0.889 & 0.878 & 1.000 & 0.747 & 0.773 & 0.761 & 0.753 & 0.751 & 0.759 & 0.728 & 0.733 & 0.732 & 0.725 & 0.737 & 0.731 & 0.744 & 0.739 & 0.728 \\
        TFM64$_{\text{IL}}$ & 1.075 & 1.185 & 1.173 & 1.671 & 1.000 & 1.030 & 1.014 & 1.002 & 1.000 & 1.010 & 0.970 & 0.979 & 0.976 & 0.968 & 0.982 & 0.976 & 0.990 & 0.983 & 0.970 \\
        TFM128$_{\text{IL}}$ & 1.044 & 1.151 & 1.139 & 1.638 & 0.972 & 1.000 & 0.985 & 0.973 & 0.971 & 0.980 & 0.942 & 0.951 & 0.947 & 0.942 & 0.953 & 0.947 & 0.961 & 0.955 & 0.941 \\
        TFM512$_{\text{IL}}$ & 1.060 & 1.168 & 1.156 & 1.663 & 0.988 & 1.016 & 1.000 & 0.988 & 0.986 & 0.996 & 0.956 & 0.966 & 0.962 & 0.956 & 0.968 & 0.962 & 0.976 & 0.969 & 0.956 \\
        TFM64$_{\text{PT}}$ & 1.073 & 1.183 & 1.170 & 1.687 & 1.000 & 1.029 & 1.012 & 1.000 & 0.998 & 1.008 & 0.968 & 0.977 & 0.974 & 0.968 & 0.980 & 0.974 & 0.988 & 0.981 & 0.968 \\
        TFM128$_{\text{PT}}$ & 1.075 & 1.185 & 1.173 & 1.689 & 1.002 & 1.031 & 1.015 & 1.002 & 1.000 & 1.010 & 0.970 & 0.980 & 0.976 & 0.970 & 0.982 & 0.976 & 0.990 & 0.983 & 0.970 \\
        TFM512$_{\text{PT}}$ & 1.065 & 1.174 & 1.161 & 1.674 & 0.992 & 1.021 & 1.005 & 0.992 & 0.990 & 1.000 & 0.961 & 0.970 & 0.966 & 0.961 & 0.972 & 0.966 & 0.980 & 0.974 & 0.960 \\
        ARFIMA$^{\text{log}}$ & 1.109 & 1.221 & 1.209 & 1.737 & 1.033 & 1.063 & 1.046 & 1.034 & 1.031 & 1.041 & 1.000 & 1.009 & 1.006 & 1.000 & 1.012 & 1.006 & 1.020 & 1.014 & 0.999 \\
CHAR$^{\text{log}}$ & 1.099 & 1.209 & 1.197 & 1.714 & 1.023 & 1.053 & 1.036 & 1.024 & 1.022 & 1.032 & 0.991 & 1.000 & 0.997 & 0.991 & 1.003 & 0.997 & 1.011 & 1.004 & 0.990 \\
HAR$^{\text{log}}$ & 1.102 & 1.214 & 1.202 & 1.727 & 1.027 & 1.057 & 1.040 & 1.027 & 1.025 & 1.035 & 0.994 & 1.004 & 1.000 & 0.994 & 1.007 & 1.000 & 1.014 & 1.008 & 0.994 \\
TFM64$^{\text{log}}_{\text{IL}}$ & 1.111 & 1.224 & 1.212 & 1.733 & 1.033 & 1.065 & 1.048 & 1.035 & 1.033 & 1.044 & 1.002 & 1.012 & 1.008 & 1.000 & 1.015 & 1.008 & 1.023 & 1.016 & 1.002 \\
TFM128$^{\text{log}}_{\text{IL}}$ & 1.095 & 1.207 & 1.194 & 1.719 & 1.020 & 1.050 & 1.033 & 1.021 & 1.018 & 1.028 & 0.988 & 0.997 & 0.994 & 0.988 & 1.000 & 0.994 & 1.008 & 1.001 & 0.987 \\
TFM512$^{\text{log}}_{\text{IL}}$ & 1.102 & 1.214 & 1.202 & 1.723 & 1.027 & 1.057 & 1.040 & 1.027 & 1.025 & 1.035 & 0.994 & 1.004 & 1.000 & 0.994 & 1.007 & 1.000 & 1.014 & 1.008 & 0.994 \\
TFM64$^{\text{log}}_{\text{PT}}$ & 1.087 & 1.197 & 1.185 & 1.710 & 1.012 & 1.042 & 1.025 & 1.013 & 1.010 & 1.020 & 0.980 & 0.990 & 0.986 & 0.980 & 0.992 & 0.986 & 1.000 & 0.993 & 0.980 \\
TFM128$^{\text{log}}_{\text{PT}}$ & 1.094 & 1.205 & 1.193 & 1.718 & 1.019 & 1.048 & 1.032 & 1.019 & 1.017 & 1.027 & 0.987 & 0.996 & 0.992 & 0.987 & 0.999 & 0.992 & 1.007 & 1.000 & 0.986 \\
TFM512$^{\text{log}}_{\text{PT}}$ & 1.109 & 1.222 & 1.210 & 1.743 & 1.033 & 1.063 & 1.046 & 1.034 & 1.031 & 1.042 & 1.001 & 1.010 & 1.006 & 1.001 & 1.013 & 1.006 & 1.021 & 1.014 & 1.000 \\
    \bottomrule
    \end{tabular}%
    }
    \label{tab:lossc}
\end{sidewaystable}

\section{Conclusion}

To the best of our knowledge, this study is the first to extensively explore the application of time series foundation models for volatility forecasting and systematically compare their performance against traditional econometric approaches. By leveraging the pre-trained TimesFM model and implementing an incremental fine-tuning procedure, we demonstrate the potential of foundation models to improve predictive accuracy while maintaining adaptability to evolving market conditions. 

Our results show that foundation models can match—and in several cases, statistically outperform—classical benchmarks across a range of error metrics. In particular, TFM$_\text{PT}^{512}$ and TFM$_\text{IL}^{512}$ consistently outperform traditional models under the Q-like loss function, with p-values below 0.10 in DM and GW tests. These findings suggest that the predictive accuracy of these foundation models is statistically superior. Meanwhile, TFM$_\text{IL}^{\log 64}$ stands out for its stability: across all comparisons, it was never found to be statistically worse, highlighting its robustness in both point and distributional forecasts.
The foundation models demonstrated broader strength in forecast accuracy across metrics. Importantly, these gains are achieved with minimal fine-tuning, showing the practical advantages of foundation models in adapting quickly to new data with limited computational effort. This has real implications for financial forecasting, where models often need to be updated frequently in response to shifting market conditions.

We envision that this study lays the groundwork for future research on the use of pre-trained time series models in financial forecasting, with a particular focus on volatility prediction. Our findings suggest that these models, especially when fine-tuned incrementally, offer a practical and scalable way to improve forecast accuracy while adapting to changing market conditions—without the need for frequent or full retraining. At the same time, we recognize that there is room for improvement. Further advances may come from better hyperparameter tuning, more targeted transfer learning strategies, and the use of alternative or higher-frequency data. Future work could also explore how architectural innovations, such as more dynamic model structures or multi-horizon forecasting techniques, might extend the capabilities of foundation models in finance. Improving model interpretability will be equally important to ensure these tools can be effectively trusted and adopted in financial decision-making environments.



\section{Funding} This work is supported by European Union's Horizon Europe programme under the Marie Skłodowska-Curie Actions (Grant Agreement No. 101150609, Project: HiddenTipChains).



\bibliographystyle{plain}
\bibliography{sample}

\newpage

\begin{sidewaystable}[ht!]
    \centering
    \caption{{\small Diebold-Mariano (DM) test and  Giacomini \& White (GW) test: We report the p-values of the DM and GW tests. The hypothesis being tested in DM test is \(H_0: \mathrm{MSE}_i = \mathrm{MSE}_j \quad \text{against the one-sided alternative} \quad H_1: \mathrm{MSE}_i > \mathrm{MSE}_j,\)
    The hypothesis being tested in GW test is \(H_0: \mathbb{E}[d_t \mid \mathcal{F}_{t-1}] = 0 \quad \text{against the one-sided alternative} \quad H_1: \mathbb{E}[d_t \mid \mathcal{F}_{t-1}] > 0,\)
where \( d_t = MSE_{i,t} - MSE_{j,t} \) is the loss differential, while model \( i \) corresponds to the label of the selected row, and model \( j \) corresponds to the label of the selected column. $\mathcal{F}_{t-1}$ denotes the information set available at time \( t-1 \). We applied these tests to cross-section (21 values) of average MSE obtained from different models for 21 stocks.}}
    \resizebox{1\textwidth}{!}{%
    \begin{tabular}{lrrrrrrrrrrrrrrrrrrr}
        \toprule
        & ARFIMA & CHAR & HAR & RGARCH & TFM$_{\text{IL}}^{64}$ & TFM$_{\text{IL}}^{128}$ & TFM$_{\text{IL}}^{512}$ & TFM$_{\text{PT}}^{64}$ & TFM$_{\text{PT}}^{128}$ & TFM$_{\text{PT}}^{512}$  
        & ARFIMA$^{\text{log}}$ & CHAR$^{\text{log}}$ & HAR$^{\text{log}}$ & TFM$_{\text{IL}}^{log 64}$ & TFM$_{\text{IL}}^{log 128}$ & TFM$_{\text{IL}}^{log 512}$ & TFM$_{\text{PT}}^{log 64}$ & TFM$_{\text{PT}}^{log 128}$ & TFM$_{\text{PT}}^{log 512}$ \\
        \midrule
        \multirow{2}{*}{ARFIMA} 
 & 0  & 0.0466 & 0.0093 & 0.8379 & 0.8783 & 0.9975 & 0.9806 & 0.9651 & 0.9907 & 0.9718 & 0.9990 & 0.9954 & 0.9995 & 0.1125 & 0.6284 & 0.7291 & 0.6449 & 0.8125 & 0.5291 \\
 & 0  & 0.0105 & 0.0005 & 0.8637 & 0.9645 & 1  & 0.8957 & 0.8968 & 0.9623 & 0.8913 & 1  & 0.9991 & 1  & 0.0288 & 0.6497 & 0.6892 & 0.6285 & 0.8099 & 0.5180 \\
        \midrule
        \multirow{2}{*}{CHAR} 
 & 0.9534 & 0  & 0.5813 & 0.8379 & 0.9933 & 0.9981 & 0.9996 & 0.9989 & 0.9994 & 0.9990 & 0.9978 & 0.9979 & 0.9980 & 0.3030 & 0.9371 & 0.9916 & 0.9741 & 0.9756 & 0.9758 \\
 & 0.9895 & 0  & 0.6412 & 0.8637 & 0.9996 & 1  & 1  & 1  & 1  & 1  & 1  & 1  & 0.9999 & 0.2376 & 0.9706 & 1  & 1  & 0.9885 & 1  \\
        \midrule
        \multirow{2}{*}{HAR} 
 & 0.9907 & 0.4187 & 0  & 0.8379 & 0.9732 & 0.9991 & 0.9952 & 0.9961 & 0.9985 & 0.9957 & 0.9997 & 0.9986 & 0.9999 & 0.2879 & 0.8939 & 0.9862 & 0.9598 & 0.9573 & 0.9773 \\
 & 0.9995 & 0.3588 & 0  & 0.8637 & 0.9987 & 1  & 0.9939 & 0.9860 & 0.9987 & 0.9937 & 1  & 0.9999 & 1  & 0.1704 & 0.9264 & 0.9814 & 0.9013 & 0.9722 & 0.9367 \\
        \midrule
        \multirow{2}{*}{RGARCH} 
 & 0.1621 & 0.1621 & 0.1621 & 0  & 0.1621 & 0.1622 & 0.1621 & 0.1621 & 0.1621 & 0.1621 & 0.1622 & 0.1622 & 0.1622 & 0.1620 & 0.1621 & 0.1621 & 0.1621 & 0.1621 & 0.1621 \\
 & 0.1363 & 0.1363 & 0.1363 & 0  & 0.1364 & 0.1364 & 0.1364 & 0.1364 & 0.1364 & 0.1364 & 0.1364 & 0.1364 & 0.1364 & 0.1363 & 0.1363 & 0.1363 & 0.1363 & 0.1364 & 0.1363 \\
        \midrule
        \multirow{2}{*}{TFM$_{\text{IL}}^{64}$} 
 & 0.1217 & 0.0067 & 0.0268 & 0.8379 & 0  & 0.8496 & 0.4082 & 0.4021 & 0.5192 & 0.3075 & 0.8748 & 0.6987 & 0.7662 & 0.0005 & 0.0763 & 0.1380 & 0.0932 & 0.2822 & 0.1055 \\
 & 0.0355 & 0.0004 & 0.0013 & 0.8636 & 0  & 0.9930 & 0.3712 & 0.3844 & 0.5293 & 0.2392 & 0.9781 & 0.7804 & 0.9358 & 0  & 0.0169 & 0.0665 & 0.0644 & 0.1256 & 0.0487 \\
        \midrule
        \multirow{2}{*}{TFM$_{\text{IL}}^{128}$} 
 & 0.0025 & 0.0019 & 0.0009 & 0.8378 & 0.1504 & 0  & 0.0311 & 0.0283 & 0.0289 & 0.0131 & 0.7436 & 0.3347 & 0.3695 & 0.0063 & 0.0107 & 0.0064 & 0.0087 & 0.0374 & 0.0032 \\
 & 0  & 0  & 0  & 0.8636 & 0.0070 & 0  & 0.0037 & 0.0072 & 0.0023 & 0.0002 & 0.8169 & 0.2760 & 0.3229 & 0  & 0  & 0.0001 & 0.0001 & 0.0015 & 0  \\
        \midrule
        \multirow{2}{*}{TFM$_{\text{IL}}^{512}$} 
 & 0.0194 & 0.0004 & 0.0048 & 0.8379 & 0.5918 & 0.9689 & 0  & 0.5341 & 0.8185 & 0.1191 & 0.9696 & 0.8829 & 0.9265 & 0.0231 & 0.1396 & 0.0200 & 0.0687 & 0.3581 & 0.0035 \\
 & 0.1043 & 0  & 0.0061 & 0.8636 & 0.6288 & 0.9963 & 0  & 0.5655 & 0.9002 & 0.0867 & 0.9898 & 0.9256 & 0.8996 & 0.0035 & 0.1064 & 0.0015 & 0.0033 & 0.3369 & 0.0002 \\
        \midrule
        \multirow{2}{*}{TFM$_{\text{PT}}^{64}$} 
 & 0.0349 & 0.0011 & 0.0039 & 0.8379 & 0.5979 & 0.9717 & 0.4659 & 0  & 0.8220 & 0.1621 & 0.9679 & 0.8384 & 0.9126 & 0.0109 & 0.1028 & 0.0177 & 0.0059 & 0.3311 & 0.0051 \\
 & 0.1032 & 0  & 0.0140 & 0.8636 & 0.6156 & 0.9928 & 0.4345 & 0  & 0.8792 & 0.0975 & 0.9558 & 0.9123 & 0.8492 & 0.0037 & 0.1090 & 0.0056 & 0.0001 & 0.2987 & 0.0018 \\
        \midrule
        \multirow{2}{*}{TFM$_{\text{PT}}^{128}$} 
 & 0.0093 & 0.0006 & 0.0015 & 0.8379 & 0.4808 & 0.9711 & 0.1815 & 0.1780 & 0  & 0.0135 & 0.9565 & 0.7769 & 0.8871 & 0.0111 & 0.0405 & 0.0107 & 0.0091 & 0.1791 & 0.0019 \\
 & 0.0377 & 0  & 0.0013 & 0.8636 & 0.4707 & 0.9977 & 0.0998 & 0.1208 & 0  & 0.0001 & 0.9566 & 0.8426 & 0.8080 & 0.0005 & 0.0148 & 0.0009 & 0  & 0.1290 & 0.0001 \\
        \midrule
        \multirow{2}{*}{TFM$_{\text{PT}}^{512}$} 
 & 0.0282 & 0.0010 & 0.0043 & 0.8379 & 0.6925 & 0.9869 & 0.8809 & 0.8379 & 0.9865 & 0  & 0.9844 & 0.9201 & 0.9668 & 0.0226 & 0.1852 & 0.0392 & 0.0887 & 0.4646 & 0.0060 \\
 & 0.1087 & 0  & 0.0063 & 0.8636 & 0.7608 & 0.9998 & 0.9133 & 0.9025 & 0.9999 & 0  & 0.9971 & 0.9711 & 0.9446 & 0.0025 & 0.1453 & 0.0099 & 0.0134 & 0.4581 & 0.0010 \\
        \midrule
        \multirow{2}{*}{ARFIMA$^{\text{log}}$} 
 & 0.0010 & 0.0022 & 0.0003 & 0.8378 & 0.1252 & 0.2564 & 0.0304 & 0.0321 & 0.0435 & 0.0156 & 0  & 0.0584 & 0.0171 & 0.0096 & 0.0319 & 0.0038 & 0.0113 & 0.0568 & 0.0022 \\
 & 0  & 0  & 0  & 0.8636 & 0.0219 & 0.1831 & 0.0102 & 0.0442 & 0.0434 & 0.0029 & 0  & 0  & 0.0063 & 0.0001 & 0.0042 & 0.0002 & 0.0030 & 0.0172 & 0.0002 \\
        \midrule
        \multirow{2}{*}{CHAR$^{\text{log}}$} 
 & 0.0046 & 0.0021 & 0.0014 & 0.8378 & 0.3013 & 0.6653 & 0.1171 & 0.1616 & 0.2231 & 0.0799 & 0.9416 & 0  & 0.6452 & 0.0252 & 0.1094 & 0.0137 & 0.0531 & 0.1885 & 0.0073 \\
 & 0.0009 & 0  & 0.0001 & 0.8636 & 0.2196 & 0.7240 & 0.0744 & 0.0877 & 0.1574 & 0.0289 & 1  & 0  & 0.6561 & 0.0007 & 0.0473 & 0.0002 & 0.0080 & 0.1404 & 0.0002 \\
        \midrule
        \multirow{2}{*}{HAR$^{\text{log}}$} 
 & 0.0005 & 0.0020 & 0.0001 & 0.8378 & 0.2338 & 0.6305 & 0.0735 & 0.0874 & 0.1129 & 0.0332 & 0.9829 & 0.3548 & 0  & 0.0128 & 0.0507 & 0.0058 & 0.0229 & 0.1026 & 0.0029 \\
 & 0  & 0.0001 & 0  & 0.8636 & 0.0642 & 0.6771 & 0.1004 & 0.1508 & 0.1920 & 0.0554 & 0.9937 & 0.3439 & 0  & 0  & 0.0135 & 0.0045 & 0.0191 & 0.0690 & 0.0069 \\
        \midrule
        \multirow{2}{*}{TFM$_{\text{IL}}^{log 64}$} 
 & 0.8875 & 0.6970 & 0.7121 & 0.8380 & 0.9995 & 0.9937 & 0.9769 & 0.9891 & 0.9889 & 0.9774 & 0.9904 & 0.9748 & 0.9872 & 0  & 0.9817 & 0.9289 & 0.9579 & 0.9801 & 0.9001 \\
 & 0.9712 & 0.7624 & 0.8296 & 0.8637 & 1  & 1  & 0.9965 & 0.9963 & 0.9995 & 0.9975 & 0.9999 & 0.9993 & 1  & 0  & 0.9982 & 0.9769 & 0.9754 & 0.9990 & 0.9567 \\
        \midrule
        \multirow{2}{*}{TFM$_{\text{IL}}^{log 128}$} 
 & 0.3716 & 0.0629 & 0.1061 & 0.8379 & 0.9237 & 0.9893 & 0.8604 & 0.8972 & 0.9595 & 0.8148 & 0.9681 & 0.8906 & 0.9493 & 0.0183 & 0  & 0.4764 & 0.4751 & 0.9134 & 0.3763 \\
 & 0.3503 & 0.0294 & 0.0736 & 0.8637 & 0.9831 & 1  & 0.8936 & 0.8910 & 0.9852 & 0.8547 & 0.9958 & 0.9527 & 0.9865 & 0.0018 & 0  & 0.4723 & 0.4691 & 0.9789 & 0.3536 \\
        \midrule
        \multirow{2}{*}{TFM$_{\text{IL}}^{log 512}$} 
 & 0.2709 & 0.0084 & 0.0138 & 0.8379 & 0.8620 & 0.9936 & 0.9800 & 0.9823 & 0.9893 & 0.9608 & 0.9962 & 0.9863 & 0.9942 & 0.0711 & 0.5236 & 0  & 0.5070 & 0.7458 & 0.1567 \\
 & 0.3108 & 0  & 0.0186 & 0.8637 & 0.9335 & 0.9999 & 0.9985 & 0.9944 & 0.9991 & 0.9901 & 0.9998 & 0.9998 & 0.9955 & 0.0231 & 0.5277 & 0  & 0.5138 & 0.7910 & 0.0029 \\
        \midrule
        \multirow{2}{*}{TFM$_{\text{PT}}^{log 64}$} 
 & 0.3551 & 0.0259 & 0.0402 & 0.8379 & 0.9068 & 0.9913 & 0.9313 & 0.9941 & 0.9909 & 0.9113 & 0.9887 & 0.9469 & 0.9771 & 0.0421 & 0.5249 & 0.4930 & 0  & 0.7873 & 0.3004 \\
 & 0.3715 & 0  & 0.0987 & 0.8637 & 0.9356 & 0.9999 & 0.9967 & 0.9999 & 1  & 0.9866 & 0.9970 & 0.9920 & 0.9809 & 0.0246 & 0.5309 & 0.4862 & 0  & 0.8741 & 0.2456 \\
        \midrule
        \multirow{2}{*}{TFM$_{\text{PT}}^{log 128}$} 
 & 0.1875 & 0.0244 & 0.0427 & 0.8379 & 0.7178 & 0.9626 & 0.6419 & 0.6689 & 0.8209 & 0.5354 & 0.9432 & 0.8115 & 0.8974 & 0.0199 & 0.0866 & 0.2542 & 0.2127 & 0  & 0.1808 \\
 & 0.1901 & 0.0115 & 0.0278 & 0.8636 & 0.8744 & 0.9985 & 0.6631 & 0.7013 & 0.8710 & 0.5419 & 0.9828 & 0.8596 & 0.9310 & 0.0010 & 0.0211 & 0.2090 & 0.1259 & 0  & 0.1355 \\
        \midrule
        \multirow{2}{*}{TFM$_{\text{PT}}^{log 512}$} 
 & 0.4709 & 0.0242 & 0.0227 & 0.8379 & 0.8945 & 0.9968 & 0.9965 & 0.9949 & 0.9981 & 0.9940 & 0.9978 & 0.9927 & 0.9971 & 0.0999 & 0.6237 & 0.8433 & 0.6996 & 0.8192 & 0  \\
 & 0.4820 & 0  & 0.0633 & 0.8637 & 0.9513 & 1  & 0.9998 & 0.9982 & 0.9999 & 0.9990 & 0.9998 & 0.9998 & 0.9931 & 0.0433 & 0.6464 & 0.9971 & 0.7544 & 0.8645 & 0  \\
        \midrule
    \end{tabular}%
    }
    \label{tab:mse}
\end{sidewaystable}

\begin{sidewaystable}[ht!]
    \centering
   \caption{{\small Diebold-Mariano (DM) test and  Giacomini \& White (GW) test: We report the p-values of the DM and GW tests. The hypothesis being tested in DM test is \(H_0: \mathrm{MAE}_i = \mathrm{MAE}_j \quad \text{against the one-sided alternative} \quad H_1: \mathrm{MAE}_i > \mathrm{MAE}_j,\)
    The hypothesis being tested in GW test is \(H_0: \mathbb{E}[d_t \mid \mathcal{F}_{t-1}] = 0 \quad \text{against the one-sided alternative} \quad H_1: \mathbb{E}[d_t \mid \mathcal{F}_{t-1}] > 0,\)
where \( d_t = MAE_{i,t} - MAE_{j,t} \) is the loss differential, while model \( i \) corresponds to the label of the selected row, and model \( j \) corresponds to the label of the selected column. $\mathcal{F}_{t-1}$ denotes the information set available at time \( t-1 \). We applied these tests to cross-section (21 values) of average MAE obtained from different models for 21 stocks.}}
    \resizebox{1\textwidth}{!}{%
    \begin{tabular}{lrrrrrrrrrrrrrrrrrrr}
        \toprule
        & ARFIMA & CHAR & HAR & RGARCH & TFM$_{\text{IL}}^{64}$ & TFM$_{\text{IL}}^{128}$ & TFM$_{\text{IL}}^{512}$ & TFM$_{\text{PT}}^{64}$ & TFM$_{\text{PT}}^{128}$ & TFM$_{\text{PT}}^{512}$  
        & ARFIMA$^{\text{log}}$ & CHAR$^{\text{log}}$ & HAR$^{\text{log}}$ & TFM$_{\text{IL}}^{log 64}$ & TFM$_{\text{IL}}^{log 128}$ & TFM$_{\text{IL}}^{log 512}$ & TFM$_{\text{PT}}^{log 64}$ & TFM$_{\text{PT}}^{log 128}$ & TFM$_{\text{PT}}^{log 512}$ \\
        \midrule
        \multirow{2}{*}{ARFIMA} 
 & 0 & 0.9322 & 0.9985 & 0.8561 & 0.0004 & 0 & 0 & 0 & 0 & 0 & 0 & 0 & 0 & 0 & 0 & 0 & 0 & 0 & 0 \\
 & 0 & 0.9521 & 0.9925 & 0.8813 & 0 & 0 & 0 & 0 & 0 & 0 & 0 & 0 & 0 & 0 & 0 & 0 & 0 & 0 & 0 \\
        \midrule
        \multirow{2}{*}{CHAR} 
 & 0.0678 & 0 & 0.6815 & 0.8560 & 0 & 0 & 0 & 0 & 0 & 0 & 0 & 0 & 0 & 0 & 0 & 0 & 0 & 0 & 0 \\
 & 0.0479 & 0 & 0.7657 & 0.8812 & 0 & 0 & 0 & 0 & 0 & 0 & 0 & 0 & 0 & 0 & 0 & 0 & 0 & 0 & 0 \\
        \midrule
        \multirow{2}{*}{HAR} 
 & 0.0015 & 0.3185 & 0 & 0.8560 & 0 & 0 & 0 & 0 & 0 & 0 & 0 & 0 & 0 & 0 & 0 & 0 & 0 & 0 & 0 \\
 & 0.0075 & 0.2343 & 0 & 0.8812 & 0 & 0 & 0 & 0 & 0 & 0 & 0 & 0 & 0 & 0 & 0 & 0 & 0 & 0 & 0 \\
        \midrule
        \multirow{2}{*}{RGARCH} 
 & 0.1439 & 0.1440 & 0.1440 & 0 & 0.1432 & 0.1434 & 0.1431 & 0.1431 & 0.1431 & 0.1431 & 0.1429 & 0.1429 & 0.1430 & 0.1428 & 0.1430 & 0.1429 & 0.1431 & 0.1431 & 0.1428 \\
 & 0.1187 & 0.1188 & 0.1188 & 0 & 0.1180 & 0.1182 & 0.1179 & 0.1179 & 0.1179 & 0.1179 & 0.1177 & 0.1177 & 0.1177 & 0.1176 & 0.1178 & 0.1177 & 0.1179 & 0.1179 & 0.1176 \\
        \midrule
        \multirow{2}{*}{TFM$_{\text{IL}}^{64}$} 
 & 0.9996 & 1 & 1 & 0.8568 & 0 & 0.9267 & 0.2882 & 0.1689 & 0.1421 & 0.1896 & 0.0223 & 0.0384 & 0.0402 & 0.0004 & 0.0494 & 0.0053 & 0.2577 & 0.2261 & 0.0055 \\
 & 1 & 1 & 1 & 0.8820 & 0 & 0.9895 & 0.1801 & 0.0547 & 0.0414 & 0.0732 & 0.0025 & 0.0126 & 0.0015 & 0 & 0.0062 & 0.0003 & 0.1315 & 0.0652 & 0.0003 \\
        \midrule
        \multirow{2}{*}{TFM$_{\text{IL}}^{128}$} 
 & 1 & 1 & 1 & 0.8566 & 0.0733 & 0 & 0.0003 & 0.0001 & 0 & 0.0001 & 0 & 0.0001 & 0 & 0.0002 & 0.0002 & 0 & 0.0010 & 0.0021 & 0 \\
 & 1 & 1 & 1 & 0.8818 & 0.0105 & 0 & 0 & 0 & 0 & 0 & 0 & 0 & 0 & 0 & 0 & 0 & 0 & 0 & 0 \\
        \midrule
        \multirow{2}{*}{TFM$_{\text{IL}}^{512}$} 
 & 1 & 1 & 1 & 0.8569 & 0.7118 & 0.9997 & 0 & 0.0816 & 0.0069 & 0.0203 & 0 & 0.0013 & 0 & 0.0078 & 0.0136 & 0 & 0.3231 & 0.3463 & 0 \\
 & 1 & 1 & 1 & 0.8821 & 0.8199 & 1 & 0 & 0.0017 & 0 & 0.0003 & 0 & 0 & 0 & 0 & 0 & 0 & 0.0009 & 0.3022 & 0 \\
        \midrule
        \multirow{2}{*}{TFM$_{\text{PT}}^{64}$} 
 & 1 & 1 & 1 & 0.8569 & 0.8311 & 0.9999 & 0.9184 & 0 & 0.0470 & 0.4814 & 0 & 0.0061 & 0.0003 & 0.0092 & 0.0356 & 0 & 0.7748 & 0.6826 & 0 \\
 & 1 & 1 & 1 & 0.8821 & 0.9453 & 1 & 0.9983 & 0 & 0.0025 & 0.4671 & 0 & 0.0045 & 0 & 0 & 0.0016 & 0 & 0.8945 & 0.7615 & 0 \\
        \midrule
        \multirow{2}{*}{TFM$_{\text{PT}}^{128}$} 
 & 1 & 1 & 1 & 0.8569 & 0.8579 & 1 & 0.9931 & 0.9530 & 0 & 0.9290 & 0.0002 & 0.0140 & 0.0009 & 0.0160 & 0.0668 & 0 & 0.8995 & 0.8161 & 0 \\
 & 1 & 1 & 1 & 0.8821 & 0.9586 & 1 & 1 & 0.9975 & 0 & 0.9655 & 0 & 0.0046 & 0 & 0 & 0.0054 & 0 & 0.9970 & 0.9112 & 0 \\
        \midrule
        \multirow{2}{*}{TFM$_{\text{PT}}^{512}$} 
 & 1 & 1 & 1 & 0.8569 & 0.8104 & 0.9999 & 0.9797 & 0.5186 & 0.0710 & 0 & 0 & 0.0057 & 0.0001 & 0.0137 & 0.0431 & 0 & 0.7805 & 0.6783 & 0 \\
 & 1 & 1 & 1 & 0.8821 & 0.9268 & 1 & 0.9997 & 0.5329 & 0.0345 & 0 & 0 & 0.0016 & 0 & 0 & 0.0029 & 0 & 0.8802 & 0.7231 & 0 \\
        \midrule
        \multirow{2}{*}{ARFIMA$^{\text{log}}$} 
 & 1 & 1 & 1 & 0.8571 & 0.9777 & 1 & 1 & 1 & 0.9998 & 1 & 0 & 0.8380 & 0.9994 & 0.2077 & 0.9432 & 0.0883 & 1 & 0.9996 & 0.0015 \\
 & 1 & 1 & 1 & 0.8823 & 0.9975 & 1 & 1 & 1 & 1 & 1 & 0 & 0.9246 & 0.9995 & 0.0948 & 0.9723 & 0 & 1 & 0.9998 & 0.0008 \\
        \midrule
        \multirow{2}{*}{CHAR$^{\text{log}}$} 
 & 1 & 1 & 1 & 0.8571 & 0.9616 & 0.9999 & 0.9987 & 0.9939 & 0.9860 & 0.9943 & 0.1620 & 0 & 0.6155 & 0.1560 & 0.7842 & 0.0319 & 0.9983 & 0.9860 & 0.0028 \\
 & 1 & 1 & 1 & 0.8823 & 0.9874 & 1 & 1 & 0.9955 & 0.9954 & 0.9984 & 0.0754 & 0 & 0.6501 & 0.0814 & 0.8669 & 0 & 1 & 0.9937 & 0 \\
        \midrule
        \multirow{2}{*}{HAR$^{\text{log}}$} 
 & 1 & 1 & 1 & 0.8570 & 0.9598 & 1 & 1 & 0.9997 & 0.9991 & 0.9999 & 0.0006 & 0.3845 & 0 & 0.1297 & 0.8225 & 0.0105 & 0.9999 & 0.9985 & 0.0001 \\
 & 1 & 1 & 1 & 0.8823 & 0.9985 & 1 & 1 & 1 & 1 & 1 & 0.0005 & 0.3499 & 0 & 0.0122 & 0.8932 & 0 & 1 & 0.9994 & 0 \\
        \midrule
        \multirow{2}{*}{TFM$_{\text{IL}}^{log 64}$} 
 & 1 & 1 & 1 & 0.8572 & 0.9996 & 0.9998 & 0.9922 & 0.9908 & 0.9840 & 0.9863 & 0.7923 & 0.8440 & 0.8703 & 0 & 0.9660 & 0.6891 & 0.9883 & 0.9920 & 0.4992 \\
 & 1 & 1 & 1 & 0.8824 & 1 & 1 & 1 & 1 & 1 & 1 & 0.9052 & 0.9186 & 0.9878 & 0 & 0.9996 & 0.7595 & 1 & 1 & 0.4989 \\
        \midrule
        \multirow{2}{*}{TFM$_{\text{IL}}^{log 128}$} 
 & 1 & 1 & 1 & 0.8570 & 0.9506 & 0.9998 & 0.9864 & 0.9644 & 0.9332 & 0.9569 & 0.0568 & 0.2158 & 0.1775 & 0.0340 & 0 & 0.0059 & 0.9768 & 0.9998 & 0.0009 \\
 & 1 & 1 & 1 & 0.8822 & 0.9938 & 1 & 1 & 0.9984 & 0.9946 & 0.9971 & 0.0277 & 0.1331 & 0.1068 & 0.0004 & 0 & 0.0005 & 0.9999 & 1 & 0.0001 \\
        \midrule
        \multirow{2}{*}{TFM$_{\text{IL}}^{log 512}$} 
 & 1 & 1 & 1 & 0.8571 & 0.9947 & 1 & 1 & 1 & 1 & 1 & 0.9117 & 0.9681 & 0.9895 & 0.3109 & 0.9941 & 0 & 1 & 0.9999 & 0.0478 \\
 & 1 & 1 & 1 & 0.8823 & 0.9997 & 1 & 1 & 1 & 1 & 1 & 1 & 1 & 1 & 0.2405 & 0.9995 & 0 & 1 & 1 & 0.0014 \\
        \midrule
        \multirow{2}{*}{TFM$_{\text{PT}}^{log 64}$} 
 & 1 & 1 & 1 & 0.8569 & 0.7423 & 0.9990 & 0.6769 & 0.2252 & 0.1005 & 0.2195 & 0 & 0.0017 & 0.0001 & 0.0117 & 0.0232 & 0 & 0 & 0.4672 & 0 \\
 & 1 & 1 & 1 & 0.8821 & 0.8685 & 1 & 0.9991 & 0.1055 & 0.0030 & 0.1198 & 0 & 0 & 0 & 0 & 0.0001 & 0 & 0 & 0.4614 & 0 \\
        \midrule
        \multirow{2}{*}{TFM$_{\text{PT}}^{log 128}$} 
 & 1 & 1 & 1 & 0.8569 & 0.7739 & 0.9979 & 0.6537 & 0.3174 & 0.1839 & 0.3217 & 0.0004 & 0.0140 & 0.0015 & 0.0080 & 0.0002 & 0.0001 & 0.5328 & 0 & 0 \\
 & 1 & 1 & 1 & 0.8821 & 0.9348 & 1 & 0.6978 & 0.2385 & 0.0888 & 0.2769 & 0.0002 & 0.0063 & 0.0006 & 0 & 0 & 0 & 0.5386 & 0 & 0 \\
        \midrule
        \multirow{2}{*}{TFM$_{\text{PT}}^{log 512}$} 
 & 1 & 1 & 1 & 0.8572 & 0.9945 & 1 & 1 & 1 & 1 & 1 & 0.9985 & 0.9972 & 0.9999 & 0.5008 & 0.9991 & 0.9522 & 1 & 1 & 0 \\
 & 1 & 1 & 1 & 0.8824 & 0.9997 & 1 & 1 & 1 & 1 & 1 & 0.9992 & 1 & 1 & 0.5011 & 0.9999 & 0.9986 & 1 & 1 & 0 \\
        \midrule
    \end{tabular}%
    }
    \label{tab:mad}
\end{sidewaystable}

\begin{sidewaystable}[ht!]
    \centering
    \caption{{\small Diebold-Mariano (DM) test and  Giacomini \& White (GW) test: We report the p-values of the DM and GW tests. The hypothesis being tested in DM test is \(H_0: \mathrm{MDA}_i = \mathrm{MDA}_j \quad \text{against the one-sided alternative} \quad H_1: \mathrm{MDA}_i > \mathrm{MDA}_j,\)
    The hypothesis being tested in GW test is \(H_0: \mathbb{E}[d_t \mid \mathcal{F}_{t-1}] = 0 \quad \text{against the one-sided alternative} \quad H_1: \mathbb{E}[d_t \mid \mathcal{F}_{t-1}] > 0,\)
where \( d_t = MDA_{i,t} - MDA_{j,t} \) is the loss differential, while model \( i \) corresponds to the label of the selected row, and model \( j \) corresponds to the label of the selected column. $\mathcal{F}_{t-1}$ denotes the information set available at time \( t-1 \). We applied these tests to cross-section (21 values) of average MDA obtained from different models for 21 stocks.}}
    \resizebox{1\textwidth}{!}{%
    \begin{tabular}{lrrrrrrrrrrrrrrrrrrr}
        \toprule
        & ARFIMA & CHAR & HAR & RGARCH & TFM$_{\text{IL}}^{64}$ & TFM$_{\text{IL}}^{128}$ & TFM$_{\text{IL}}^{512}$ & TFM$_{\text{PT}}^{64}$ & TFM$_{\text{PT}}^{128}$ & TFM$_{\text{PT}}^{512}$  
        & ARFIMA$^{\text{log}}$ & CHAR$^{\text{log}}$ & HAR$^{\text{log}}$ & TFM$_{\text{IL}}^{log 64}$ & TFM$_{\text{IL}}^{log 128}$ & TFM$_{\text{IL}}^{log 512}$ & TFM$_{\text{PT}}^{log 64}$ & TFM$_{\text{PT}}^{log 128}$ & TFM$_{\text{PT}}^{log 512}$ \\
        \midrule
        \multirow{2}{*}{ARFIMA} 
 & 0 & 0.0394 & 0 & 0 & 0.6601 & 0.0003 & 0 & 0 & 0 & 0 & 0 & 0.0144 & 0 & 0.7655 & 0.0003 & 0.0003 & 0 & 0 & 0 \\
 & 0 & 0.0099 & 0 & 0 & 0.6827 & 0 & 0 & 0 & 0 & 0 & 0 & 0.0022 & 0 & 0.8097 & 0 & 0 & 0 & 0 & 0 \\
        \midrule
        \multirow{2}{*}{CHAR} 
 & 0.9606 & 0 & 0 & 0 & 0.8246 & 0.0012 & 0 & 0 & 0 & 0 & 0 & 0.0079 & 0 & 0.8902 & 0.0001 & 0.0003 & 0 & 0 & 0 \\
 & 0.9901 & 0 & 0 & 0 & 0.8799 & 0.0001 & 0 & 0 & 0 & 0 & 0 & 0.0003 & 0 & 0.9580 & 0 & 0 & 0 & 0 & 0 \\
        \midrule
        \multirow{2}{*}{HAR} 
 & 1 & 1 & 0 & 0 & 0.9582 & 0.8158 & 0.0090 & 0.0009 & 0.0006 & 0 & 0 & 0.9997 & 0 & 0.9779 & 0.6939 & 0.3147 & 0 & 0.0001 & 0 \\
 & 1 & 1 & 0 & 0 & 0.9963 & 0.9784 & 0.0004 & 0 & 0 & 0 & 0 & 1 & 0 & 0.9996 & 0.7056 & 0.2453 & 0 & 0 & 0 \\
        \midrule
        \multirow{2}{*}{RGARCH} 
 & 1 & 1 & 1 & 0 & 0.9907 & 1 & 0.9960 & 0.9997 & 0.9933 & 0.9980 & 0.9097 & 1 & 1 & 0.9955 & 1 & 0.9999 & 0.9950 & 0.9882 & 0.9967 \\
 & 1 & 1 & 1 & 0 & 0.9999 & 1 & 0.9974 & 1 & 0.9972 & 0.9745 & 0.8982 & 1 & 1 & 1 & 1 & 1 & 0.9891 & 0.9990 & 0.9957 \\
        \midrule
        \multirow{2}{*}{TFM$_{\text{IL}}^{64}$} 
 & 0.3399 & 0.1754 & 0.0418 & 0.0093 & 0 & 0.0645 & 0.0222 & 0.0216 & 0.0174 & 0.0170 & 0.0136 & 0.1455 & 0.0296 & 0.9555 & 0.0480 & 0.0447 & 0.0158 & 0.0174 & 0.0149 \\
 & 0.3173 & 0.1201 & 0.0037 & 0.0001 & 0 & 0.0072 & 0.0013 & 0.0006 & 0.0002 & 0.0005 & 0.0004 & 0.0860 & 0.0014 & 0.9766 & 0.0063 & 0.0118 & 0.0004 & 0.0004 & 0.0009 \\
        \midrule
        \multirow{2}{*}{TFM$_{\text{IL}}^{128}$} 
 & 0.9997 & 0.9988 & 0.1842 & 0 & 0.9355 & 0 & 0.0001 & 0.0013 & 0 & 0 & 0.0010 & 0.9899 & 0.0274 & 0.9631 & 0.2497 & 0.1353 & 0.0001 & 0 & 0.0001 \\
 & 1 & 0.9999 & 0.0216 & 0 & 0.9928 & 0 & 0 & 0 & 0 & 0 & 0 & 0.9975 & 0 & 0.9993 & 0.1507 & 0.0502 & 0 & 0 & 0 \\
        \midrule
        \multirow{2}{*}{TFM$_{\text{IL}}^{512}$} 
 & 1 & 1 & 0.9910 & 0.0040 & 0.9778 & 0.9999 & 0 & 0.5262 & 0.2273 & 0.1384 & 0.0920 & 0.9999 & 0.8485 & 0.9885 & 0.9960 & 0.9226 & 0.1333 & 0.1030 & 0.0914 \\
 & 1 & 1 & 0.9996 & 0.0026 & 0.9987 & 1 & 0 & 0.5448 & 0.2183 & 0.0914 & 0.0581 & 1 & 0.9118 & 0.9998 & 1 & 0.9980 & 0.0620 & 0.0377 & 0.0761 \\
        \midrule
        \multirow{2}{*}{TFM$_{\text{PT}}^{64}$} 
 & 1 & 1 & 0.9991 & 0.0003 & 0.9784 & 0.9987 & 0.4738 & 0 & 0.1932 & 0.1387 & 0.0540 & 1 & 0.9206 & 0.9889 & 0.9994 & 0.9383 & 0.0661 & 0.0459 & 0.0218 \\
 & 1 & 1 & 1 & 0 & 0.9994 & 1 & 0.4552 & 0 & 0.0004 & 0.0083 & 0.0171 & 1 & 0.9871 & 1 & 1 & 0.9952 & 0.0041 & 0 & 0.0010 \\
        \midrule
        \multirow{2}{*}{TFM$_{\text{PT}}^{128}$} 
 & 1 & 1 & 0.9994 & 0.0067 & 0.9826 & 1 & 0.7727 & 0.8068 & 0 & 0.4291 & 0.1509 & 1 & 0.9674 & 0.9911 & 0.9994 & 0.9545 & 0.3269 & 0.2943 & 0.2345 \\
 & 1 & 1 & 1 & 0.0028 & 0.9998 & 1 & 0.7817 & 0.9996 & 0 & 0.4211 & 0.0910 & 1 & 0.9963 & 1 & 1 & 0.9944 & 0.1406 & 0.2314 & 0.0525 \\
        \midrule
        \multirow{2}{*}{TFM$_{\text{PT}}^{512}$} 
 & 1 & 1 & 1 & 0.0020 & 0.9830 & 1 & 0.8616 & 0.8613 & 0.5709 & 0 & 0.1479 & 1 & 0.9947 & 0.9914 & 1 & 0.9866 & 0.3306 & 0.3206 & 0.2065 \\
 & 1 & 1 & 1 & 0.0255 & 0.9995 & 1 & 0.9086 & 0.9917 & 0.5789 & 0 & 0.1508 & 1 & 0.9996 & 1 & 1 & 1 & 0.3356 & 0.1195 & 0.2714 \\
        \midrule
        \multirow{2}{*}{ARFIMA$^{\text{log}}$} 
 & 1 & 1 & 1 & 0.0903 & 0.9864 & 0.9990 & 0.9080 & 0.9460 & 0.8491 & 0.8521 & 0 & 1 & 0.9995 & 0.9934 & 0.9997 & 0.9978 & 0.8102 & 0.7539 & 0.7511 \\
 & 1 & 1 & 1 & 0.1018 & 0.9996 & 1 & 0.9419 & 0.9829 & 0.9090 & 0.8492 & 0 & 1 & 1 & 0.9999 & 1 & 1 & 0.8559 & 0.8248 & 0.8708 \\
        \midrule
        \multirow{2}{*}{CHAR$^{\text{log}}$} 
 & 0.9856 & 0.9921 & 0.0003 & 0 & 0.8545 & 0.0101 & 0.0001 & 0 & 0 & 0 & 0 & 0 & 0 & 0.9107 & 0.0022 & 0.0022 & 0 & 0 & 0 \\
 & 0.9978 & 0.9997 & 0 & 0 & 0.9140 & 0.0025 & 0 & 0 & 0 & 0 & 0 & 0 & 0 & 0.9740 & 0 & 0 & 0 & 0 & 0 \\
        \midrule
        \multirow{2}{*}{HAR$^{\text{log}}$} 
 & 1 & 1 & 1 & 0 & 0.9704 & 0.9726 & 0.1515 & 0.0794 & 0.0326 & 0.0053 & 0.0005 & 1 & 0 & 0.9847 & 0.9878 & 0.7994 & 0.0017 & 0.0075 & 0.0003 \\
 & 1 & 1 & 1 & 0 & 0.9986 & 1 & 0.0882 & 0.0129 & 0.0037 & 0.0004 & 0 & 1 & 0 & 0.9999 & 0.9929 & 0.8875 & 0.0006 & 0.0001 & 0 \\
        \midrule
        \multirow{2}{*}{TFM$_{\text{IL}}^{log 64}$} 
 & 0.2345 & 0.1098 & 0.0221 & 0.0045 & 0.0445 & 0.0369 & 0.0115 & 0.0111 & 0.0089 & 0.0086 & 0.0066 & 0.0893 & 0.0153 & 0 & 0.0260 & 0.0239 & 0.0079 & 0.0089 & 0.0074 \\
 & 0.1903 & 0.0420 & 0.0004 & 0 & 0.0234 & 0.0007 & 0.0002 & 0 & 0 & 0 & 0.0001 & 0.0260 & 0.0001 & 0 & 0.0013 & 0.0025 & 0 & 0 & 0.0001 \\
        \midrule
        \multirow{2}{*}{TFM$_{\text{IL}}^{log 128}$} 
 & 0.9997 & 0.9999 & 0.3061 & 0 & 0.9520 & 0.7503 & 0.0040 & 0.0006 & 0.0006 & 0 & 0.0003 & 0.9978 & 0.0122 & 0.9740 & 0 & 0.1887 & 0 & 0 & 0 \\
 & 1 & 1 & 0.2944 & 0 & 0.9937 & 0.8493 & 0 & 0 & 0 & 0 & 0 & 1 & 0.0071 & 0.9987 & 0 & 0.1310 & 0 & 0 & 0 \\
        \midrule
        \multirow{2}{*}{TFM$_{\text{IL}}^{log 512}$} 
 & 0.9997 & 0.9997 & 0.6853 & 0.0001 & 0.9553 & 0.8647 & 0.0774 & 0.0617 & 0.0455 & 0.0134 & 0.0022 & 0.9978 & 0.2006 & 0.9761 & 0.8113 & 0 & 0.0124 & 0.0100 & 0.0031 \\
 & 1 & 1 & 0.7547 & 0 & 0.9882 & 0.9498 & 0.0020 & 0.0048 & 0.0056 & 0 & 0 & 1 & 0.1125 & 0.9975 & 0.8690 & 0 & 0 & 0 & 0 \\
        \midrule
        \multirow{2}{*}{TFM$_{\text{PT}}^{log 64}$} 
 & 1 & 1 & 1 & 0.0050 & 0.9842 & 0.9999 & 0.8667 & 0.9339 & 0.6731 & 0.6694 & 0.1898 & 1 & 0.9983 & 0.9921 & 1 & 0.9876 & 0 & 0.4523 & 0.2855 \\
 & 1 & 1 & 1 & 0.0109 & 0.9996 & 1 & 0.9380 & 0.9959 & 0.8594 & 0.6644 & 0.1441 & 1 & 0.9994 & 1 & 1 & 1 & 0 & 0.4247 & 0.1784 \\
        \midrule
        \multirow{2}{*}{TFM$_{\text{PT}}^{log 128}$} 
 & 1 & 1 & 0.9999 & 0.0118 & 0.9826 & 1 & 0.8970 & 0.9541 & 0.7057 & 0.6794 & 0.2461 & 1 & 0.9925 & 0.9911 & 1 & 0.9900 & 0.5477 & 0 & 0.3517 \\
 & 1 & 1 & 1 & 0.0010 & 0.9996 & 1 & 0.9623 & 1 & 0.7686 & 0.8805 & 0.1752 & 1 & 0.9999 & 1 & 1 & 1 & 0.5753 & 0 & 0.3227 \\
        \midrule
        \multirow{2}{*}{TFM$_{\text{PT}}^{log 512}$} 
 & 1 & 1 & 1 & 0.0033 & 0.9851 & 0.9999 & 0.9086 & 0.9782 & 0.7655 & 0.7935 & 0.2489 & 1 & 0.9997 & 0.9926 & 1 & 0.9969 & 0.7145 & 0.6483 & 0 \\
 & 1 & 1 & 1 & 0.0043 & 0.9991 & 1 & 0.9239 & 0.9990 & 0.9475 & 0.7286 & 0.1292 & 1 & 1 & 0.9999 & 1 & 1 & 0.8216 & 0.6773 & 0 \\
        \midrule
    \end{tabular}%
    }
    \label{tab:mda}
\end{sidewaystable}

\begin{sidewaystable}[ht!]
    \centering
   \caption{{\small Diebold-Mariano (DM) test and  Giacomini \& White (GW) test: We report the p-values of the DM and GW tests. The hypothesis being tested in DM test is \(H_0: \mathrm{MAPE}_i = \mathrm{MAPE}_j \quad \text{against the one-sided alternative} \quad H_1: \mathrm{MAPE}_i > \mathrm{MAPE}_j,\)
    The hypothesis being tested in GW test is \(H_0: \mathbb{E}[d_t \mid \mathcal{F}_{t-1}] = 0 \quad \text{against the one-sided alternative} \quad H_1: \mathbb{E}[d_t \mid \mathcal{F}_{t-1}] > 0,\)
where \( d_t = MAPE_{i,t} - MAPE_{j,t} \) is the loss differential, while model \( i \) corresponds to the label of the selected row, and model \( j \) corresponds to the label of the selected column. $\mathcal{F}_{t-1}$ denotes the information set available at time \( t-1 \). We applied these tests to cross-section (21 values) of average MAPE obtained from different models for 21 stocks.}}
    \resizebox{1\textwidth}{!}{%
    \begin{tabular}{lrrrrrrrrrrrrrrrrrrr}
        \toprule
        & ARFIMA & CHAR & HAR & RGARCH & TFM$_{\text{IL}}^{64}$ & TFM$_{\text{IL}}^{128}$ & TFM$_{\text{IL}}^{512}$ & TFM$_{\text{PT}}^{64}$ & TFM$_{\text{PT}}^{128}$ & TFM$_{\text{PT}}^{512}$  
        & ARFIMA$^{\text{log}}$ & CHAR$^{\text{log}}$ & HAR$^{\text{log}}$ & TFM$_{\text{IL}}^{log 64}$ & TFM$_{\text{IL}}^{log 128}$ & TFM$_{\text{IL}}^{log 512}$ & TFM$_{\text{PT}}^{log 64}$ & TFM$_{\text{PT}}^{log 128}$ & TFM$_{\text{PT}}^{log 512}$ \\
        \midrule
        \multirow{2}{*}{ARFIMA} 
 & 0 & 0.9971 & 0.9396 & 0.8435 & 0.0946 & 0.0916 & 0.0723 & 0.0690 & 0.0648 & 0.0663 & 0.0502 & 0.0290 & 0.0446 & 0.0820 & 0.0572 & 0.0557 & 0.0622 & 0.0600 & 0.0552 \\
 & 0 & 0.9999 & 0.9820 & 0.8677 & 0.0350 & 0.0403 & 0.0160 & 0.0109 & 0.0084 & 0.0106 & 0.0029 & 0.0001 & 0.0016 & 0.0198 & 0.0054 & 0.0042 & 0.0071 & 0.0060 & 0.0045 \\
        \midrule
        \multirow{2}{*}{CHAR} 
 & 0.0029 & 0 & 0.7834 & 0.8427 & 0.0426 & 0.0307 & 0.0227 & 0.0276 & 0.0240 & 0.0209 & 0.0206 & 0.0104 & 0.0172 & 0.0428 & 0.0242 & 0.0245 & 0.0273 & 0.0258 & 0.0243 \\
 & 0.0001 & 0 & 0.7900 & 0.8670 & 0.0039 & 0.0028 & 0.0006 & 0.0009 & 0.0005 & 0.0005 & 0.0002 & 0 & 0.0001 & 0.0030 & 0.0005 & 0.0004 & 0.0006 & 0.0005 & 0.0004 \\
        \midrule
        \multirow{2}{*}{HAR} 
 & 0.0604 & 0.2166 & 0 & 0.8423 & 0.0766 & 0.0721 & 0.0645 & 0.0641 & 0.0620 & 0.0624 & 0.0553 & 0.0463 & 0.0529 & 0.0721 & 0.0586 & 0.0578 & 0.0610 & 0.0600 & 0.0576 \\
 & 0.0180 & 0.2100 & 0 & 0.8665 & 0.0222 & 0.0230 & 0.0143 & 0.0124 & 0.0112 & 0.0127 & 0.0070 & 0.0029 & 0.0058 & 0.0166 & 0.0091 & 0.0078 & 0.0100 & 0.0093 & 0.0081 \\
        \midrule
        \multirow{2}{*}{RGARCH} 
 & 0.1565 & 0.1573 & 0.1577 & 0 & 0.1553 & 0.1557 & 0.1558 & 0.1555 & 0.1556 & 0.1558 & 0.1554 & 0.1556 & 0.1555 & 0.1550 & 0.1554 & 0.1553 & 0.1553 & 0.1553 & 0.1553 \\
 & 0.1323 & 0.1330 & 0.1335 & 0 & 0.1310 & 0.1314 & 0.1314 & 0.1312 & 0.1313 & 0.1315 & 0.1311 & 0.1313 & 0.1312 & 0.1306 & 0.1310 & 0.1310 & 0.1310 & 0.1310 & 0.1310 \\
        \midrule
        \multirow{2}{*}{TFM$_{\text{IL}}^{64}$} 
 & 0.9054 & 0.9574 & 0.9234 & 0.8447 & 0 & 0.8978 & 0.8659 & 0.7506 & 0.8011 & 0.8551 & 0.6064 & 0.7524 & 0.6993 & 0.0455 & 0.5369 & 0.3359 & 0.3529 & 0.4764 & 0.3171 \\
 & 0.9650 & 0.9961 & 0.9778 & 0.8690 & 0 & 0.9625 & 0.9170 & 0.7402 & 0.7983 & 0.8980 & 0.5688 & 0.6771 & 0.6432 & 0.0018 & 0.5263 & 0.4019 & 0.3850 & 0.4838 & 0.3811 \\
        \midrule
        \multirow{2}{*}{TFM$_{\text{IL}}^{128}$} 
 & 0.9084 & 0.9693 & 0.9279 & 0.8443 & 0.1022 & 0 & 0.6340 & 0.0166 & 0.0012 & 0.5941 & 0.0006 & 0.2767 & 0 & 0.0730 & 0.0071 & 0.0097 & 0.0189 & 0.0127 & 0.0093 \\
 & 0.9597 & 0.9972 & 0.9770 & 0.8686 & 0.0375 & 0 & 0.6341 & 0.0001 & 0 & 0.5905 & 0 & 0.3906 & 0 & 0.0093 & 0 & 0 & 0 & 0 & 0 \\
        \midrule
        \multirow{2}{*}{TFM$_{\text{IL}}^{512}$} 
 & 0.9277 & 0.9773 & 0.9355 & 0.8442 & 0.1341 & 0.3660 & 0 & 0.0616 & 0.0379 & 0.3926 & 0.0136 & 0.0715 & 0.0011 & 0.0915 & 0.0318 & 0.0320 & 0.0466 & 0.0400 & 0.0311 \\
 & 0.9840 & 0.9994 & 0.9857 & 0.8686 & 0.0830 & 0.3659 & 0 & 0.0046 & 0.0005 & 0.3798 & 0 & 0.2863 & 0 & 0.0247 & 0.0004 & 0.0002 & 0.0015 & 0.0007 & 0.0002 \\
        \midrule
        \multirow{2}{*}{TFM$_{\text{PT}}^{64}$} 
 & 0.9310 & 0.9724 & 0.9359 & 0.8445 & 0.2494 & 0.9834 & 0.9384 & 0 & 0.8913 & 0.9233 & 0.0212 & 0.7520 & 0.5953 & 0.1084 & 0.0003 & 0.0047 & 0.0255 & 0.0089 & 0.0042 \\
 & 0.9891 & 0.9991 & 0.9876 & 0.8688 & 0.2598 & 0.9999 & 0.9954 & 0 & 0.9417 & 0.9878 & 0.1402 & 0.6544 & 0.5441 & 0.0468 & 0 & 0 & 0.0002 & 0 & 0 \\
        \midrule
        \multirow{2}{*}{TFM$_{\text{PT}}^{128}$} 
 & 0.9352 & 0.9760 & 0.9380 & 0.8444 & 0.1989 & 0.9988 & 0.9621 & 0.1087 & 0 & 0.9410 & 0.0019 & 0.6574 & 0.0125 & 0.1083 & 0.0276 & 0.0290 & 0.0534 & 0.0423 & 0.0277 \\
 & 0.9916 & 0.9995 & 0.9888 & 0.8687 & 0.2017 & 1 & 0.9995 & 0.0583 & 0 & 0.9964 & 0 & 0.5657 & 0.1307 & 0.0481 & 0.0004 & 0.0001 & 0.0039 & 0.0012 & 0.0002 \\
        \midrule
        \multirow{2}{*}{TFM$_{\text{PT}}^{512}$} 
 & 0.9337 & 0.9791 & 0.9376 & 0.8442 & 0.1449 & 0.4059 & 0.6074 & 0.0767 & 0.0590 & 0 & 0.0206 & 0.0553 & 0.0035 & 0.0973 & 0.0407 & 0.0396 & 0.0555 & 0.0491 & 0.0386 \\
 & 0.9894 & 0.9995 & 0.9873 & 0.8685 & 0.1020 & 0.4095 & 0.6202 & 0.0122 & 0.0036 & 0 & 0 & 0.1965 & 0 & 0.0318 & 0.0012 & 0.0006 & 0.0037 & 0.0020 & 0.0008 \\
        \midrule
        \multirow{2}{*}{ARFIMA$^{\text{log}}$} 
 & 0.9498 & 0.9794 & 0.9447 & 0.8446 & 0.3936 & 0.9994 & 0.9864 & 0.9788 & 0.9981 & 0.9794 & 0 & 0.8621 & 0.8870 & 0.1589 & 0.2173 & 0.1116 & 0.1856 & 0.2066 & 0.1033 \\
 & 0.9971 & 0.9998 & 0.9930 & 0.8689 & 0.4312 & 1 & 1 & 0.8598 & 1 & 1 & 0 & 0.8324 & 0.9232 & 0.1296 & 0.2872 & 0.0423 & 0.1884 & 0.2133 & 0.0546 \\
        \midrule
        \multirow{2}{*}{CHAR$^{\text{log}}$} 
 & 0.9710 & 0.9896 & 0.9537 & 0.8444 & 0.2476 & 0.7233 & 0.9285 & 0.2480 & 0.3426 & 0.9447 & 0.1379 & 0 & 0.1541 & 0.1515 & 0.1520 & 0.1294 & 0.1530 & 0.1548 & 0.1264 \\
 & 0.9999 & 1 & 0.9971 & 0.8687 & 0.3229 & 0.6094 & 0.7137 & 0.3456 & 0.4343 & 0.8035 & 0.1676 & 0 & 0.2227 & 0.1425 & 0.1962 & 0.1231 & 0.1770 & 0.1826 & 0.1291 \\
        \midrule
        \multirow{2}{*}{HAR$^{\text{log}}$} 
 & 0.9554 & 0.9828 & 0.9471 & 0.8445 & 0.3007 & 1 & 0.9989 & 0.4047 & 0.9875 & 0.9965 & 0.1130 & 0.8459 & 0 & 0.1510 & 0.1502 & 0.1122 & 0.1525 & 0.1558 & 0.1076 \\
 & 0.9984 & 0.9999 & 0.9942 & 0.8688 & 0.3568 & 1 & 1 & 0.4559 & 0.8693 & 1 & 0.0768 & 0.7773 & 0 & 0.1204 & 0.1620 & 0.0561 & 0.1381 & 0.1410 & 0.0640 \\
        \midrule
        \multirow{2}{*}{TFM$_{\text{IL}}^{log 64}$} 
 & 0.9180 & 0.9572 & 0.9279 & 0.8450 & 0.9545 & 0.9270 & 0.9085 & 0.8916 & 0.8917 & 0.9027 & 0.8411 & 0.8485 & 0.8490 & 0 & 0.8484 & 0.8258 & 0.8496 & 0.8509 & 0.8222 \\
 & 0.9802 & 0.9970 & 0.9834 & 0.8694 & 0.9982 & 0.9907 & 0.9753 & 0.9532 & 0.9519 & 0.9682 & 0.8704 & 0.8575 & 0.8796 & 0 & 0.8894 & 0.8372 & 0.8885 & 0.8870 & 0.8433 \\
        \midrule
        \multirow{2}{*}{TFM$_{\text{IL}}^{log 128}$} 
 & 0.9428 & 0.9758 & 0.9414 & 0.8446 & 0.4631 & 0.9929 & 0.9682 & 0.9997 & 0.9724 & 0.9593 & 0.7827 & 0.8480 & 0.8498 & 0.1516 & 0 & 0.0363 & 0.1595 & 0.1908 & 0.0294 \\
 & 0.9946 & 0.9995 & 0.9909 & 0.8690 & 0.4737 & 1 & 0.9996 & 1 & 0.9996 & 0.9988 & 0.7128 & 0.8038 & 0.8380 & 0.1106 & 0 & 0.0005 & 0.1099 & 0.1349 & 0.0006 \\
        \midrule
        \multirow{2}{*}{TFM$_{\text{IL}}^{log 512}$} 
 & 0.9443 & 0.9755 & 0.9422 & 0.8447 & 0.6641 & 0.9903 & 0.9680 & 0.9953 & 0.9710 & 0.9604 & 0.8884 & 0.8706 & 0.8878 & 0.1742 & 0.9637 & 0 & 0.7387 & 0.9997 & 0.0465 \\
 & 0.9958 & 0.9996 & 0.9922 & 0.8690 & 0.5981 & 1 & 0.9998 & 1 & 0.9999 & 0.9994 & 0.9577 & 0.8769 & 0.9439 & 0.1628 & 0.9995 & 0 & 0.6337 & 1 & 0.0395 \\
        \midrule
        \multirow{2}{*}{TFM$_{\text{PT}}^{log 64}$} 
 & 0.9378 & 0.9727 & 0.9390 & 0.8447 & 0.6471 & 0.9811 & 0.9534 & 0.9745 & 0.9466 & 0.9445 & 0.8144 & 0.8470 & 0.8475 & 0.1504 & 0.8405 & 0.2613 & 0 & 0.8634 & 0.1563 \\
 & 0.9929 & 0.9994 & 0.9900 & 0.8690 & 0.6150 & 1 & 0.9985 & 0.9998 & 0.9961 & 0.9963 & 0.8116 & 0.8230 & 0.8619 & 0.1115 & 0.8901 & 0.3663 & 0 & 0.8750 & 0.2763 \\
        \midrule
        \multirow{2}{*}{TFM$_{\text{PT}}^{log 128}$} 
 & 0.9400 & 0.9742 & 0.9400 & 0.8447 & 0.5236 & 0.9873 & 0.9600 & 0.9911 & 0.9577 & 0.9509 & 0.7934 & 0.8452 & 0.8442 & 0.1491 & 0.8092 & 0.0003 & 0.1366 & 0 & 0.0002 \\
 & 0.9940 & 0.9995 & 0.9907 & 0.8690 & 0.5162 & 1 & 0.9993 & 1 & 0.9988 & 0.9980 & 0.7867 & 0.8174 & 0.8590 & 0.1130 & 0.8651 & 0 & 0.1250 & 0 & 0 \\
        \midrule
        \multirow{2}{*}{TFM$_{\text{PT}}^{log 512}$} 
 & 0.9448 & 0.9757 & 0.9424 & 0.8447 & 0.6829 & 0.9907 & 0.9689 & 0.9958 & 0.9723 & 0.9614 & 0.8967 & 0.8736 & 0.8924 & 0.1778 & 0.9706 & 0.9535 & 0.8437 & 0.9998 & 0 \\
 & 0.9955 & 0.9996 & 0.9919 & 0.8690 & 0.6189 & 1 & 0.9998 & 1 & 0.9998 & 0.9992 & 0.9454 & 0.8709 & 0.9360 & 0.1567 & 0.9994 & 0.9605 & 0.7237 & 1 & 0 \\
        \midrule
    \end{tabular}%
    }
    \label{tab:mape}
\end{sidewaystable}

\begin{sidewaystable}[ht!]
    \centering
   \caption{{\small  Diebold-Mariano (DM) test and  Giacomini \& White (GW) test: We report the p-values of the DM and GW tests. The hypothesis being tested in DM test is \(H_0: \mathrm{sMAPE}_i = \mathrm{sMAPE}_j \quad \text{against the one-sided alternative} \quad H_1: \mathrm{sMAPE}_i > \mathrm{sMAPE}_j,\)
    The hypothesis being tested in GW test is \(H_0: \mathbb{E}[d_t \mid \mathcal{F}_{t-1}] = 0 \quad \text{against the one-sided alternative} \quad H_1: \mathbb{E}[d_t \mid \mathcal{F}_{t-1}] > 0,\)
where \( d_t = sMAPE_{i,t} - sMAPE_{j,t} \) is the loss differential, while model \( i \) corresponds to the label of the selected row, and model \( j \) corresponds to the label of the selected column. $\mathcal{F}_{t-1}$ denotes the information set available at time \( t-1 \). We applied these tests to cross-section (21 values) of average MAPE obtained from different models for 21 stocks.}}
    \resizebox{1\textwidth}{!}{%
    \begin{tabular}{lrrrrrrrrrrrrrrrrrrr}
        \toprule
        & ARFIMA & CHAR & HAR & RGARCH & TFM$_{\text{IL}}^{64}$ & TFM$_{\text{IL}}^{128}$ & TFM$_{\text{IL}}^{512}$ & TFM$_{\text{PT}}^{64}$ & TFM$_{\text{PT}}^{128}$ & TFM$_{\text{PT}}^{512}$  
        & ARFIMA$^{\text{log}}$ & CHAR$^{\text{log}}$ & HAR$^{\text{log}}$ & TFM$_{\text{IL}}^{log 64}$ & TFM$_{\text{IL}}^{log 128}$ & TFM$_{\text{IL}}^{log 512}$ & TFM$_{\text{PT}}^{log 64}$ & TFM$_{\text{PT}}^{log 128}$ & TFM$_{\text{PT}}^{log 512}$ \\
        \midrule
        \multirow{2}{*}{ARFIMA} 
 & 0 & 0.9997 & 1 & 0.9881 & 0 & 0.0001 & 0 & 0 & 0 & 0 & 0 & 0 & 0 & 0 & 0 & 0 & 0 & 0 & 0 \\
 & 0 & 0.9999 & 0.9999 & 0.9951 & 0 & 0 & 0 & 0 & 0 & 0 & 0 & 0 & 0 & 0 & 0 & 0 & 0 & 0 & 0 \\
        \midrule
        \multirow{2}{*}{CHAR} 
 & 0.0003 & 0 & 0.1576 & 0.9809 & 0 & 0 & 0 & 0 & 0 & 0 & 0 & 0 & 0 & 0 & 0 & 0 & 0 & 0 & 0 \\
 & 0.0001 & 0 & 0.1066 & 0.9917 & 0 & 0 & 0 & 0 & 0 & 0 & 0 & 0 & 0 & 0 & 0 & 0 & 0 & 0 & 0 \\
        \midrule
        \multirow{2}{*}{HAR} 
 & 0 & 0.8424 & 0 & 0.9825 & 0 & 0 & 0 & 0 & 0 & 0 & 0 & 0 & 0 & 0 & 0 & 0 & 0 & 0 & 0 \\
 & 0.0001 & 0.8934 & 0 & 0.9923 & 0 & 0 & 0 & 0 & 0 & 0 & 0 & 0 & 0 & 0 & 0 & 0 & 0 & 0 & 0 \\
        \midrule
        \multirow{2}{*}{RGARCH} 
 & 0.0119 & 0.0191 & 0.0175 & 0 & 0.0080 & 0.0096 & 0.0088 & 0.0085 & 0.0084 & 0.0088 & 0.0073 & 0.0074 & 0.0075 & 0.0071 & 0.0077 & 0.0074 & 0.0080 & 0.0078 & 0.0074 \\
 & 0.0049 & 0.0083 & 0.0077 & 0 & 0.0033 & 0.0039 & 0.0034 & 0.0032 & 0.0032 & 0.0034 & 0.0025 & 0.0026 & 0.0026 & 0.0027 & 0.0028 & 0.0026 & 0.0030 & 0.0028 & 0.0026 \\
        \midrule
        \multirow{2}{*}{TFM$_{\text{IL}}^{64}$} 
 & 1 & 1 & 1 & 0.9920 & 0 & 0.9959 & 0.9152 & 0.5430 & 0.4502 & 0.8007 & 0.0033 & 0.0183 & 0.0109 & 0 & 0.0298 & 0.0065 & 0.1393 & 0.0422 & 0.0033 \\
 & 1 & 1 & 1 & 0.9967 & 0 & 1 & 0.9947 & 0.5731 & 0.4242 & 0.9152 & 0 & 0.0021 & 0.0002 & 0 & 0.0014 & 0.0001 & 0.0459 & 0.0039 & 0 \\
        \midrule
        \multirow{2}{*}{TFM$_{\text{IL}}^{128}$} 
 & 0.9999 & 1 & 1 & 0.9904 & 0.0041 & 0 & 0.0167 & 0.0011 & 0.0005 & 0.0074 & 0 & 0 & 0 & 0 & 0 & 0 & 0.0001 & 0 & 0 \\
 & 1 & 1 & 1 & 0.9961 & 0 & 0 & 0.0012 & 0 & 0 & 0.0004 & 0 & 0 & 0 & 0 & 0 & 0 & 0 & 0 & 0 \\
        \midrule
        \multirow{2}{*}{TFM$_{\text{IL}}^{512}$} 
 & 1 & 1 & 1 & 0.9912 & 0.0848 & 0.9833 & 0 & 0.0014 & 0.0004 & 0.0531 & 0 & 0 & 0 & 0.0007 & 0 & 0 & 0 & 0 & 0 \\
 & 1 & 1 & 1 & 0.9966 & 0.0053 & 0.9988 & 0 & 0 & 0 & 0.0188 & 0 & 0 & 0 & 0 & 0 & 0 & 0 & 0 & 0 \\
        \midrule
        \multirow{2}{*}{TFM$_{\text{PT}}^{64}$} 
 & 1 & 1 & 1 & 0.9915 & 0.4570 & 0.9989 & 0.9986 & 0 & 0.0145 & 0.9999 & 0 & 0.0001 & 0 & 0.0053 & 0 & 0 & 0 & 0 & 0 \\
 & 1 & 1 & 1 & 0.9968 & 0.4269 & 1 & 1 & 0 & 0.0055 & 1 & 0 & 0 & 0 & 0 & 0 & 0 & 0 & 0 & 0 \\
        \midrule
        \multirow{2}{*}{TFM$_{\text{PT}}^{128}$} 
 & 1 & 1 & 1 & 0.9916 & 0.5498 & 0.9995 & 0.9996 & 0.9855 & 0 & 1 & 0 & 0.0002 & 0 & 0.0077 & 0 & 0 & 0 & 0 & 0 \\
 & 1 & 1 & 1 & 0.9968 & 0.5758 & 1 & 1 & 0.9945 & 0 & 1 & 0 & 0 & 0 & 0 & 0 & 0 & 0 & 0 & 0 \\
        \midrule
        \multirow{2}{*}{TFM$_{\text{PT}}^{512}$} 
 & 1 & 1 & 1 & 0.9912 & 0.1993 & 0.9926 & 0.9469 & 0.0001 & 0 & 0 & 0 & 0 & 0 & 0.0018 & 0 & 0 & 0 & 0 & 0 \\
 & 1 & 1 & 1 & 0.9966 & 0.0848 & 0.9996 & 0.9812 & 0 & 0 & 0 & 0 & 0 & 0 & 0 & 0 & 0 & 0 & 0 & 0 \\
        \midrule
        \multirow{2}{*}{ARFIMA$^{\text{log}}$} 
 & 1 & 1 & 1 & 0.9927 & 0.9967 & 1 & 1 & 1 & 1 & 1 & 0 & 0.9782 & 1 & 0.4497 & 1 & 0.9810 & 1 & 1 & 0.3523 \\
 & 1 & 1 & 1 & 0.9975 & 1 & 1 & 1 & 1 & 1 & 1 & 0 & 0.9979 & 1 & 0.4346 & 1 & 1 & 1 & 1 & 0.3353 \\
        \midrule
        \multirow{2}{*}{CHAR$^{\text{log}}$} 
 & 1 & 1 & 1 & 0.9926 & 0.9817 & 1 & 1 & 0.9999 & 0.9998 & 1 & 0.0218 & 0 & 0.3072 & 0.2346 & 0.8729 & 0.2349 & 0.9957 & 0.9200 & 0.0323 \\
 & 1 & 1 & 1 & 0.9974 & 0.9979 & 1 & 1 & 1 & 1 & 1 & 0.0021 & 0 & 0.2545 & 0.1819 & 0.8982 & 0.1642 & 0.9970 & 0.9387 & 0.0076 \\
        \midrule
        \multirow{2}{*}{HAR$^{\text{log}}$} 
 & 1 & 1 & 1 & 0.9925 & 0.9891 & 1 & 1 & 1 & 1 & 1 & 0 & 0.6928 & 0 & 0.2761 & 0.9992 & 0.3949 & 1 & 0.9997 & 0.0061 \\
 & 1 & 1 & 1 & 0.9974 & 0.9998 & 1 & 1 & 1 & 1 & 1 & 0 & 0.7455 & 0 & 0.2133 & 0.9996 & 0.0146 & 1 & 1 & 0 \\
        \midrule
        \multirow{2}{*}{TFM$_{\text{IL}}^{log 64}$} 
 & 1 & 1 & 1 & 0.9929 & 1 & 1 & 0.9993 & 0.9947 & 0.9923 & 0.9982 & 0.5503 & 0.7654 & 0.7239 & 0 & 0.8792 & 0.7196 & 0.9551 & 0.8901 & 0.5269 \\
 & 1 & 1 & 1 & 0.9973 & 1 & 1 & 1 & 1 & 1 & 1 & 0.5654 & 0.8181 & 0.7867 & 0 & 0.9661 & 0.7842 & 0.9964 & 0.9667 & 0.5358 \\
        \midrule
        \multirow{2}{*}{TFM$_{\text{IL}}^{log 128}$} 
 & 1 & 1 & 1 & 0.9923 & 0.9702 & 1 & 1 & 1 & 1 & 1 & 0 & 0.1271 & 0.0008 & 0.1208 & 0 & 0.0112 & 0.9999 & 0.7422 & 0 \\
 & 1 & 1 & 1 & 0.9972 & 0.9986 & 1 & 1 & 1 & 1 & 1 & 0 & 0.1018 & 0.0004 & 0.0339 & 0 & 0.0002 & 1 & 0.8196 & 0 \\
        \midrule
        \multirow{2}{*}{TFM$_{\text{IL}}^{log 512}$} 
 & 1 & 1 & 1 & 0.9926 & 0.9935 & 1 & 1 & 1 & 1 & 1 & 0.0190 & 0.7651 & 0.6051 & 0.2804 & 0.9888 & 0 & 0.9999 & 0.9979 & 0.0169 \\
 & 1 & 1 & 1 & 0.9974 & 0.9999 & 1 & 1 & 1 & 1 & 1 & 0 & 0.8358 & 0.9854 & 0.2158 & 0.9998 & 0 & 1 & 1 & 0.0008 \\
        \midrule
        \multirow{2}{*}{TFM$_{\text{PT}}^{log 64}$} 
 & 1 & 1 & 1 & 0.9920 & 0.8607 & 0.9999 & 1 & 1 & 1 & 1 & 0 & 0.0043 & 0 & 0.0449 & 0.0001 & 0.0001 & 0 & 0 & 0 \\
 & 1 & 1 & 1 & 0.9970 & 0.9541 & 1 & 1 & 1 & 1 & 1 & 0 & 0.0030 & 0 & 0.0036 & 0 & 0 & 0 & 0 & 0 \\
        \midrule
        \multirow{2}{*}{TFM$_{\text{PT}}^{log 128}$} 
 & 1 & 1 & 1 & 0.9922 & 0.9578 & 1 & 1 & 1 & 1 & 1 & 0 & 0.0800 & 0.0003 & 0.1099 & 0.2578 & 0.0021 & 1 & 0 & 0 \\
 & 1 & 1 & 1 & 0.9972 & 0.9961 & 1 & 1 & 1 & 1 & 1 & 0 & 0.0613 & 0 & 0.0333 & 0.1804 & 0 & 1 & 0 & 0 \\
        \midrule
        \multirow{2}{*}{TFM$_{\text{PT}}^{log 512}$} 
 & 1 & 1 & 1 & 0.9926 & 0.9967 & 1 & 1 & 1 & 1 & 1 & 0.6477 & 0.9677 & 0.9939 & 0.4731 & 1 & 0.9831 & 1 & 1 & 0 \\
 & 1 & 1 & 1 & 0.9974 & 1 & 1 & 1 & 1 & 1 & 1 & 0.6647 & 0.9924 & 1 & 0.4642 & 1 & 0.9992 & 1 & 1 & 0 \\
        \midrule
    \end{tabular}%
    }
    \label{tab:smape}
\end{sidewaystable}

\begin{sidewaystable}[ht!]
    \centering
    \caption{{\small Diebold-Mariano (DM) test and  Giacomini \& White (GW) test: We report the p-values of the DM and GW tests. The hypothesis being tested in DM test is \(H_0: \mathrm{Qlike}_i = \mathrm{Qlike}_j \quad \text{against the one-sided alternative} \quad H_1: \mathrm{Qlike}_i > \mathrm{Qlike}_j,\)
    The hypothesis being tested in GW test is \(H_0: \mathbb{E}[d_t \mid \mathcal{F}_{t-1}] = 0 \quad \text{against the one-sided alternative} \quad H_1: \mathbb{E}[d_t \mid \mathcal{F}_{t-1}] > 0,\)
where \( d_t = Qlike_{i,t} - Qlike_{j,t} \) is the loss differential, while model \( i \) corresponds to the label of the selected row, and model \( j \) corresponds to the label of the selected column. $\mathcal{F}_{t-1}$ denotes the information set available at time \( t-1 \). We applied these tests to cross-section (21 values) of average Qlike obtained from different models for 21 stocks.}}
    \resizebox{1\textwidth}{!}{%
    \begin{tabular}{lrrrrrrrrrrrrrrrrrrr}
        \toprule
        & ARFIMA & CHAR & HAR & RGARCH & TFM$_{\text{IL}}^{64}$ & TFM$_{\text{IL}}^{128}$ & TFM$_{\text{IL}}^{512}$ & TFM$_{\text{PT}}^{64}$ & TFM$_{\text{PT}}^{128}$ & TFM$_{\text{PT}}^{512}$  
        & ARFIMA$^{\text{log}}$ & CHAR$^{\text{log}}$ & HAR$^{\text{log}}$ & TFM$_{\text{IL}}^{log 64}$ & TFM$_{\text{IL}}^{log 128}$ & TFM$_{\text{IL}}^{log 512}$ & TFM$_{\text{PT}}^{log 64}$ & TFM$_{\text{PT}}^{log 128}$ & TFM$_{\text{PT}}^{log 512}$ \\
        \midrule
        \multirow{2}{*}{ARFIMA} 
 & 0 & 0.0495 & 0.0494 & 0.1045 & 0.0495 & 0.4982 & 0.0493 & 0.0496 & 0.0494 & 0.0493 & 0.0495 & 0.0494 & 0.0495 & 0.0496 & 0.0496 & 0.0497 & 0.0496 & 0.0496 & 0.0496 \\
 & 0 & 0.0409 & 0.0408 & 0.0106 & 0.0408 & 0.4978 & 0.0407 & 0.0410 & 0.0408 & 0.0406 & 0.0409 & 0.0408 & 0.0409 & 0.0410 & 0.0411 & 0.0411 & 0.0412 & 0.0411 & 0.0410 \\
        \midrule
        \multirow{2}{*}{CHAR} 
 & 0.9505 & 0 & 0.0474 & 0.9437 & 0.2186 & 0.9398 & 0.0160 & 0.9341 & 0.1433 & 0.0061 & 0.5914 & 0.5518 & 0.6562 & 0.7593 & 0.9694 & 0.9781 & 0.9671 & 0.9671 & 0.9460 \\
 & 0.9591 & 0 & 0.0992 & 0.9718 & 0.2100 & 0.9856 & 0.0327 & 0.9133 & 0.1639 & 0.0056 & 0.5799 & 0.5543 & 0.6353 & 0.7662 & 0.9776 & 0.9901 & 0.9973 & 0.9733 & 0.9373 \\
        \midrule
        \multirow{2}{*}{HAR} 
 & 0.9506 & 0.9526 & 0 & 0.9441 & 0.4450 & 0.9399 & 0.0219 & 0.9984 & 0.5073 & 0.0045 & 0.9477 & 0.8837 & 0.9571 & 0.9066 & 0.9990 & 0.9988 & 0.9858 & 0.9990 & 0.9984 \\
 & 0.9592 & 0.9008 & 0 & 0.9721 & 0.4358 & 0.9856 & 0.0199 & 0.9966 & 0.5084 & 0 & 0.9488 & 0.9240 & 0.9566 & 0.9027 & 0.9998 & 0.9998 & 1 & 0.9999 & 0.9996 \\
        \midrule
        \multirow{2}{*}{RGARCH} 
 & 0.8955 & 0.0563 & 0.0559 & 0 & 0.0559 & 0.8839 & 0.0555 & 0.0570 & 0.0560 & 0.0552 & 0.0565 & 0.0566 & 0.0567 & 0.0568 & 0.0573 & 0.0573 & 0.0598 & 0.0574 & 0.0572 \\
 & 0.9894 & 0.0282 & 0.0279 & 0 & 0.0281 & 0.9608 & 0.0276 & 0.0287 & 0.0280 & 0.0274 & 0.0282 & 0.0282 & 0.0284 & 0.0287 & 0.0288 & 0.0288 & 0.0295 & 0.0289 & 0.0287 \\
        \midrule
        \multirow{2}{*}{TFM$_{\text{IL}}^{64}$} 
 & 0.9505 & 0.7814 & 0.5550 & 0.9441 & 0 & 0.9400 & 0.1795 & 0.9814 & 0.5730 & 0.0831 & 0.8461 & 0.8171 & 0.8719 & 0.9970 & 0.9924 & 0.9923 & 0.9733 & 0.9889 & 0.9845 \\
 & 0.9592 & 0.7900 & 0.5642 & 0.9719 & 0 & 0.9857 & 0.1627 & 0.9986 & 0.5889 & 0.0682 & 0.9107 & 0.8590 & 0.9347 & 1 & 1 & 1 & 0.9997 & 1 & 0.9999 \\
        \midrule
        \multirow{2}{*}{TFM$_{\text{IL}}^{128}$} 
 & 0.5018 & 0.0602 & 0.0601 & 0.1161 & 0.0600 & 0 & 0.0599 & 0.0602 & 0.0601 & 0.0599 & 0.0602 & 0.0601 & 0.0602 & 0.0602 & 0.0603 & 0.0604 & 0.0608 & 0.0603 & 0.0603 \\
 & 0.5022 & 0.0144 & 0.0144 & 0.0392 & 0.0143 & 0 & 0.0143 & 0.0144 & 0.0144 & 0.0143 & 0.0144 & 0.0144 & 0.0144 & 0.0144 & 0.0145 & 0.0145 & 0.0147 & 0.0145 & 0.0145 \\
        \midrule
        \multirow{2}{*}{TFM$_{\text{IL}}^{512}$} 
 & 0.9507 & 0.9840 & 0.9781 & 0.9445 & 0.8205 & 0.9401 & 0 & 1 & 0.9998 & 0.0562 & 0.9999 & 0.9990 & 1 & 0.9907 & 1 & 1 & 0.9932 & 1 & 1 \\
 & 0.9593 & 0.9673 & 0.9801 & 0.9724 & 0.8373 & 0.9857 & 0 & 1 & 0.9984 & 0.0165 & 1 & 1 & 1 & 0.9926 & 1 & 1 & 1 & 1 & 1 \\
        \midrule
        \multirow{2}{*}{TFM$_{\text{PT}}^{64}$} 
 & 0.9504 & 0.0659 & 0.0016 & 0.9430 & 0.0186 & 0.9398 & 0 & 0 & 0 & 0 & 0.0109 & 0.0560 & 0.0437 & 0.4645 & 0.9693 & 0.9618 & 0.9536 & 0.9611 & 0.8476 \\
 & 0.9590 & 0.0867 & 0.0034 & 0.9713 & 0.0014 & 0.9856 & 0 & 0 & 0 & 0 & 0.0425 & 0.0244 & 0.0575 & 0.4628 & 0.9917 & 0.9779 & 0.9979 & 0.9821 & 0.8202 \\
        \midrule
        \multirow{2}{*}{TFM$_{\text{PT}}^{128}$} 
 & 0.9506 & 0.8567 & 0.4927 & 0.9440 & 0.4270 & 0.9399 & 0.0002 & 1 & 0 & 0 & 0.9952 & 0.9501 & 0.9946 & 0.9484 & 1 & 0.9999 & 0.9866 & 1 & 1 \\
 & 0.9592 & 0.8361 & 0.4916 & 0.9720 & 0.4111 & 0.9856 & 0.0016 & 1 & 0 & 0 & 1 & 0.9981 & 1 & 0.9536 & 1 & 1 & 1 & 1 & 1 \\
        \midrule
        \multirow{2}{*}{TFM$_{\text{PT}}^{512}$} 
 & 0.9507 & 0.9939 & 0.9955 & 0.9448 & 0.9169 & 0.9401 & 0.9438 & 1 & 1 & 0 & 0.9999 & 0.9996 & 0.9999 & 0.9954 & 1 & 1 & 0.9949 & 1 & 1 \\
 & 0.9594 & 0.9944 & 1 & 0.9726 & 0.9318 & 0.9857 & 0.9835 & 1 & 1 & 0 & 1 & 1 & 1 & 0.9977 & 1 & 1 & 1 & 1 & 1 \\
        \midrule
        \multirow{2}{*}{ARFIMA$^{\text{log}}$} 
 & 0.9505 & 0.4086 & 0.0523 & 0.9435 & 0.1539 & 0.9398 & 0.0001 & 0.9891 & 0.0048 & 0.0001 & 0 & 0.4406 & 0.8266 & 0.7657 & 0.9999 & 0.9996 & 0.9834 & 1 & 1 \\
 & 0.9591 & 0.4201 & 0.0512 & 0.9718 & 0.0893 & 0.9856 & 0 & 0.9575 & 0 & 0 & 0 & 0.3832 & 0.8773 & 0.7704 & 1 & 1 & 1 & 1 & 1 \\
        \midrule
        \multirow{2}{*}{CHAR$^{\text{log}}$} 
 & 0.9506 & 0.4482 & 0.1163 & 0.9434 & 0.1829 & 0.9399 & 0.0010 & 0.9440 & 0.0499 & 0.0004 & 0.5594 & 0 & 0.7649 & 0.7604 & 0.9952 & 0.9918 & 0.9866 & 0.9994 & 0.9981 \\
 & 0.9592 & 0.4457 & 0.0760 & 0.9718 & 0.1410 & 0.9856 & 0 & 0.9756 & 0.0019 & 0 & 0.6168 & 0 & 0.8983 & 0.7531 & 0.9995 & 0.9995 & 1 & 1 & 1 \\
        \midrule
        \multirow{2}{*}{HAR$^{\text{log}}$} 
 & 0.9505 & 0.3438 & 0.0429 & 0.9433 & 0.1281 & 0.9398 & 0 & 0.9563 & 0.0054 & 0.0001 & 0.1734 & 0.2351 & 0 & 0.7110 & 0.9986 & 0.9952 & 0.9815 & 0.9999 & 0.9997 \\
 & 0.9591 & 0.3647 & 0.0434 & 0.9716 & 0.0653 & 0.9856 & 0 & 0.9425 & 0 & 0 & 0.1227 & 0.1017 & 0 & 0.7095 & 0.9993 & 0.9986 & 1 & 1 & 0.9994 \\
        \midrule
        \multirow{2}{*}{TFM$_{\text{IL}}^{log 64}$} 
 & 0.9504 & 0.2407 & 0.0934 & 0.9432 & 0.0030 & 0.9398 & 0.0093 & 0.5355 & 0.0516 & 0.0046 & 0.2343 & 0.2396 & 0.2890 & 0 & 0.7860 & 0.8442 & 0.9280 & 0.7911 & 0.6873 \\
 & 0.9590 & 0.2338 & 0.0973 & 0.9713 & 0 & 0.9856 & 0.0074 & 0.5372 & 0.0464 & 0.0023 & 0.2296 & 0.2469 & 0.2905 & 0 & 0.8028 & 0.9196 & 0.9789 & 0.8231 & 0.7285 \\
        \midrule
        \multirow{2}{*}{TFM$_{\text{IL}}^{log 128}$} 
 & 0.9504 & 0.0306 & 0.0010 & 0.9427 & 0.0076 & 0.9397 & 0 & 0.0307 & 0 & 0 & 0.0001 & 0.0048 & 0.0014 & 0.2140 & 0 & 0.7815 & 0.9420 & 0.6997 & 0.1823 \\
 & 0.9589 & 0.0224 & 0.0002 & 0.9712 & 0 & 0.9855 & 0 & 0.0083 & 0 & 0 & 0 & 0.0005 & 0.0007 & 0.1972 & 0 & 0.8530 & 0.9975 & 0.6809 & 0.1205 \\
        \midrule
        \multirow{2}{*}{TFM$_{\text{IL}}^{log 512}$} 
 & 0.9503 & 0.0219 & 0.0012 & 0.9427 & 0.0077 & 0.9396 & 0 & 0.0382 & 0.0001 & 0 & 0.0004 & 0.0082 & 0.0048 & 0.1558 & 0.2185 & 0 & 0.9279 & 0.3206 & 0.0813 \\
 & 0.9589 & 0.0099 & 0.0002 & 0.9712 & 0 & 0.9855 & 0 & 0.0221 & 0 & 0 & 0 & 0.0005 & 0.0014 & 0.0804 & 0.1470 & 0 & 0.9902 & 0.3001 & 0.0419 \\
        \midrule
        \multirow{2}{*}{TFM$_{\text{PT}}^{log 64}$} 
 & 0.9504 & 0.0329 & 0.0142 & 0.9402 & 0.0267 & 0.9392 & 0.0068 & 0.0464 & 0.0134 & 0.0051 & 0.0166 & 0.0134 & 0.0185 & 0.0720 & 0.0580 & 0.0721 & 0 & 0.0535 & 0.0445 \\
 & 0.9588 & 0.0027 & 0 & 0.9705 & 0.0003 & 0.9853 & 0 & 0.0021 & 0 & 0 & 0 & 0 & 0 & 0.0211 & 0.0025 & 0.0098 & 0 & 0.0031 & 0.0014 \\
        \midrule
        \multirow{2}{*}{TFM$_{\text{PT}}^{log 128}$} 
 & 0.9504 & 0.0329 & 0.0010 & 0.9426 & 0.0111 & 0.9397 & 0 & 0.0389 & 0 & 0 & 0 & 0.0006 & 0.0001 & 0.2089 & 0.3003 & 0.6794 & 0.9465 & 0 & 0.0475 \\
 & 0.9589 & 0.0267 & 0.0001 & 0.9711 & 0 & 0.9855 & 0 & 0.0179 & 0 & 0 & 0 & 0 & 0 & 0.1769 & 0.3191 & 0.6999 & 0.9969 & 0 & 0 \\
        \midrule
        \multirow{2}{*}{TFM$_{\text{PT}}^{log 512}$} 
 & 0.9504 & 0.0540 & 0.0016 & 0.9428 & 0.0155 & 0.9397 & 0 & 0.1524 & 0 & 0 & 0 & 0.0019 & 0.0003 & 0.3127 & 0.8177 & 0.9187 & 0.9555 & 0.9525 & 0 \\
 & 0.9590 & 0.0627 & 0.0004 & 0.9713 & 0.0001 & 0.9855 & 0 & 0.1798 & 0 & 0 & 0 & 0 & 0.0006 & 0.2715 & 0.8795 & 0.9581 & 0.9986 & 1 & 0 \\
        \midrule
    \end{tabular}%
    }
    \label{tab:qlike}
\end{sidewaystable}

\end{document}